\documentclass[journal]{IEEEtran} 
\usepackage{units}
\usepackage{cite}
\usepackage{graphicx}
\usepackage[tight,footnotesize]{subfigure} 
\usepackage{url}
\usepackage[cmex10]{amsmath}
\usepackage{amsfonts}
\usepackage{bm}
\usepackage{array}
\usepackage{algorithmic}
\usepackage{booktabs}
\usepackage{multirow}
\newcommand{\condmid}{\,|\,} 
\newcommand{\e}{e} 
\newcommand{\seq}[1]{\bm{#1}}
\newcommand{\seqs}[1]{\bm{#1}}
\newcommand{\subseq}[3]{\seq{#1}_{#2}^{#3}}

\newcommand{\ct}{\tilde{c}} 
\newcommand{\ut}{\tilde{u}} 
\newcommand{\yt}{\tilde{y}} 
\newcommand{\cck}{\check{c}} 
\newcommand{\uck}{\check{u}} 
\newcommand{\yck}{\check{y}} 
\newcommand{\yest}{\hat{y}}
\newcommand{\Fest}{\hat{F}}
\newcommand{\Gest}{\hat{G}}

\newcommand{\mat}[1]{\bm{#1}}
\newcommand{\abs}[1]{\left\lvert #1 \right\rvert}
\newcommand{\avg}[1]{\left\langle #1 \right\rangle}
\newcommand{\avga}[1]{\avg{#1}}

\newcommand{\cardinal}[1]{\left\lvert #1 \right\rvert}
\newcommand{\norm}[1]{\left\lVert #1 \right\rVert}
\newcommand{\asymptO}{O} 
\newcommand{\asympteq}{\doteq}
\newcommand{\SetR}{\mathbb{R}}
\newcommand{\SetZ}{\mathbb{Z}}

\DeclareMathOperator*{\argmax}{arg\,max}
\DeclareMathOperator*{\sgn}{sgn}
\newcommand{\bmodI}{_{\mathcal{I}}}
\newcommand{\bmodIn}{_{\mathcal{I}^n}}
\newcommand{\pz}{p_{\mathsf{z}}} 
\newcommand{\pyq}[1]{p_{\mathsf{y},#1}}
\newcommand{\pyqest}[1]{\hat{p}_{\mathsf{y},#1}}
\newcommand{\Fyq}[1]{F_{\mathsf{y},#1}}
\newcommand{\Fyqest}[1]{\Fest_{\mathsf{y},#1}}
\newcommand{\Hc}{H_{\node{c}}}
\newcommand{\Hcx}[1]{H_{\node{c},#1}}
\newcommand{\db}{d_{\node{b}}}
\newcommand{\dc}{d_{\node{c}}}
\newcommand{\dcmax}{d_{\node{c}}^{\mathrm{max}}}
\newcommand{\vc}[1]{v_{\node{c}#1}}
\newcommand{\wc}[1]{w_{\node{c}#1}}
\newcommand{\degp}{d_{\node{p}}}
\newcommand{\vq}[1]{v_{\node{q}#1}}
\newcommand{\node}[1]{\ensuremath{\mathsf{#1}}}

\newcommand{\msg}[2]{\mu^{\node{#1}}_{#2}} 
\newcommand{\msgx}[3]{\mu^{\node{#1}}_{#2}(#3)} 
\newcommand{\priprob}[2]{\lambda^{\node{#1}}_{#2}}
\newcommand{\priprobx}[3]{\priprob{#1}{#2}(#3)}
\newcommand{\extprob}[2]{\nu^{\node{#1}}_{#2}}
\newcommand{\extprobx}[3]{\extprob{#1}{#2}(#3)}
\newcommand{\trueextprob}[2]{\nu^{\node{#1}*}_{#2}}
\newcommand{\trueextprobx}[3]{\trueextprob{#1}{#2}(#3)}

\newcommand{\constmsg}[1]{\overline{#1}}
\newcommand{\neighbor}[2]{\mathcal{N}^{\node{#1}}_{#2}} 
\newcommand{\Info}[1]{I_{\node{#1}}} 
\newcommand{\Infox}[2]{I_{\node{#1},#2}}
\newcommand{\nextInfo}[1]{I^{+}_{\node{#1}}}
\newcommand{\Ibext}{\Infox{b}{\mathrm{ext}}}
\newcommand{\Ibfix}{\Info{b}^*}
\newcommand{\Infoiter}[2]{I_{\node{#1}}^{(#2)}}
\newcommand{\Ibextiter}[1]{I_{\node{b},\mathrm{ext}}^{(#1)}}
\newcommand{\Ibfixiter}[1]{I_{\node{b}}^{*(#1)}}
\newcommand{\deltaIb}{\Delta\Info{b}}
\newcommand{\deltaIbiter}[1]{\deltaIb^{(#1)}}

\newcommand{\tthr}{t^{\mathrm{thr}}}
\newcommand{\Icthr}{\Info{c}^{\mathrm{thr}}}
\newcommand{\IcthrL}{\Info{c}^{\mathrm{thr},L}}
\newcommand{\IcthrorL}{\Info{c}^{\mathrm{thr}(,L)}}
\newcommand{\IcDE}{\Info{c}^{\mathrm{de}}}
\newcommand{\sL}{s^{(L)}}
\newcommand{\sDE}{s^{\mathrm{de}}}
\newcommand{\sDEL}{s^{\mathrm{de},(L)}}
\newcommand{\rL}{r^{(L)}}
\newcommand{\smax}{s_{\mathrm{max}}}
\newcommand{\assign}{\Leftarrow}
\newcommand{\deltamax}{\delta_{\mathrm{max}}}
\newcommand{\excluding}[1]{\backslash \{#1\}}
\newcommand{\minprog}{\Delta\Info{bc}^{\mathrm{min}}} 
\newcommand{\nofootnote}[1]{} 
\newcommand{\vpp}[1]{#1^{\vphantom{\prime}}} 

\newcommand{\nb}{n_{\mathrm{b}}}
\newcommand{\nbx}[1]{n_{\mathrm{b},#1}}
\newcommand{\nc}{n_{\mathrm{c}}}
\newcommand{\nne}{n_{\mathrm{ne}}}
\newcommand{\Ane}{A_{\mathrm{ne}}}
\newcommand{\nr}{n_{\mathrm{r}}}
\newcommand{\nguess}{n_{\mathrm{g}}}
\newcommand{\Aguess}{A_{\mathrm{g}}}
\newcommand{\Adet}{A_{\mathrm{d}}}
\newcommand{\Adetiter}[1]{\Adet^{(#1)}}
\newcommand{\nignore}{n_{\mathrm{i}}}
\newcommand{\Aignore}{A_{\mathrm{i}}}
\newcommand{\conjL}{\tilde{L}}
\newcommand{\Gne}{\mat{G}_{\mathrm{ne}}}
\newcommand{\yne}{\seq{y}_{\mathrm{ne}}}
\newcommand{\qseqy}{q_{\seqs{y}}}
\newcommand{\Qseqy}{Q_{\seqs{y}}}
\newcommand{\Qseqyta}{Q_{\seqs{\yt} + \seqs{a}}}
\newcommand{\pseqy}{p_{\seqs{y}}}

\newcommand{\pseqa}{p_{\seqs{a}}}
\newcommand{\Qsum}{Q_{\Sigma}}
\newcommand{\GF}{\mathrm{GF}}

\newcommand{\IbvsIbext}{$\Info{b}$-versus-$\Ibext$}
\newcommand{\oneif}{\mathbf{1}} 
\newcommand{\dB}{\ensuremath{\mathrm{dB}}}
\newcommand{\qed}{\hfill \ensuremath{\Box}}

\newcommand{\mrow}[2]{\multirow{#1}{*}{#2}}
\newcommand{\mrowa}[2]{\multirow{#1}{*}[-1.5mm]{#2}}
\newcommand{\mcol}[2]{\multicolumn{#1}{c}{#2}}
\newcommand{\mytitle}{Design and Analysis of LDGM-Based Codes for MSE
  Quantization}

\urldef{\ouremail}\url|{r6144,chenhe}@sjtu.edu.cn| \title{\mytitle}%
\author{Qingchuan~Wang,~\IEEEmembership{Student Member,~IEEE,} Chen
  He,~\IEEEmembership{Member, IEEE,}%
  \thanks{The authors are with Department of Electronic Engineering,
    Shanghai Jiao Tong University, Shanghai, 200240, China.  E-mail:
    \ouremail.  This paper was supported in part by National Natural
    Science Foundation of China Grant No.~60772100 and in part by
    Science \& Technology Committee of Shanghai Municipality Grant
    No.~06DZ15013.  Part of the material in this paper has been
    presented in \cite{ldgm-vq-globecom07} at IEEE Global
    Communications Conference, Washington, DC, November 2007.}%
} 
\begin{document}
\maketitle

\begin{abstract}
  Approaching the 1.5329-dB shaping (granular) gain limit in
  mean-squared error (MSE) quantization of $\SetR^n$ is important in a
  number of problems, notably dirty-paper coding.  For this purpose,
  we start with a binary low-density generator-matrix (LDGM) code,
  and construct the quantization codebook by periodically repeating
  its set of binary codewords, or them mapped to $m$-ary ones with
  Gray mapping.  The quantization algorithm is based on belief
  propagation, and it uses a decimation procedure to do the guessing
  necessary for convergence.  Using the results of a true typical
  decimator (TTD) as reference, it is shown that the asymptotic
  performance of the proposed quantizer can be characterized by
  certain monotonicity conditions on the code's fixed point
  properties, which can be analyzed with density evolution, and degree
  distribution optimization can be carried out accordingly.  When the
  number of iterations is finite, the resulting loss is made amenable
  to analysis through the introduction of a recovery algorithm from
  ``bad'' guesses, and the results of such analysis enable further
  optimization of the pace of decimation and the degree distribution.
  Simulation results show that the proposed LDGM-based quantizer can
  achieve a shaping gain of \unit[1.4906]{dB}, or \unit[0.0423]{dB}
  from the limit, and significantly outperforms trellis-coded
  quantization (TCQ) at a similar computational complexity.
\end{abstract}


\section{Introduction}
\IEEEPARstart{T}{he} mean-squared error (MSE) quantization problem of
$\SetR^n$ \cite[Sec.~II-C]{lattices-good-everything} can be formulated
as follows:\footnote{Notational conventions: $\SetZ$ and $\SetR$ are
  respectively the set of integers and real numbers.  $\norm{\cdot}$
  is the Euclidean norm.  $\cardinal{\mathcal{A}}$ is the cardinality
  of set $\mathcal{A}$\@.  $\asympteq$ denotes asymptotic equality,
  usually with respect to block length $n\to\infty$.  $\log(\cdot)$,
  entropy and mutual information are computed in base-2, while
  $\ln(\cdot)$ and $\exp(\cdot)$ are base-$\e$.  Bold letters denote
  sequences or vectors whose elements are indicated by subscripts,
  e.g.\ $\seq{y}=(y_1,\dotsc,y_n)$, and sub-sequences are denoted by $
  \subseq{y}{i}{j} = (y_i,y_{i+1},\dotsc,y_j)$ or
  $\seq{y}_{\mathcal{S}} = (y_i)_{i\in\mathcal{S}}$.  Addition and
  multiplication on sets apply element-by-element, e.g.\
  $\mathcal{U}+2\SetZ^n = \left\{ \seq{u} + (2d_1,\dotsc,2d_n) \mid
    \seq{u} \in \mathcal{U}, d_i \in \SetZ \right\}$.  $x \bmod [a,b)$
  (or simply $(x)_{[a,b)}$) is defined as the unique element of
  $(x-(b-a)\SetZ) \cap [a,b)$, and similarly $\seq{x} \bmod [a,b)^n$
  or $(\seq{x})_{[a,b)^n}$ is the unique element of
  $(\seq{x}-(b-a)\SetZ^n) \cap [a,b)^n$.  The unit ``\unit{b/s}''
  means ``bits per symbol''.}  let $\Lambda$ be a discrete subset of
$\SetR^n$ (the \emph{quantization codebook}, or simply
\emph{code})\footnote{Ref.~\cite{lattices-good-everything} assumes
  that $\Lambda$ is a lattice, but in practice neither the trellis in
  TCQ nor the non-binary codebooks proposed here are lattices.
  Therefore, we allow $\Lambda$ to be any discrete set, and
  definitions are modified accordingly.} and $Q_{\Lambda}:
\SetR^n\to\Lambda$ be a quantizer that maps each $\seq{y} \in \SetR^n$
to a nearby codeword $Q_{\Lambda}(\seq{y}) \in \Lambda$.  The
mean-square quantization error, averaged over $\seq{y}$, is given by
\begin{equation}
  \label{eq:mse}
  \sigma^2 = \limsup_{M\to\infty} \frac{1}{(2M)^n} \cdot \frac{1}{n} \int_{[-M,M]^n} \norm{\seq{y}
  - Q_{\Lambda}(\seq{y})}^2 \, d\seq{y}.
\end{equation}
The objective is to design $\Lambda$ and a practical quantizer
$Q_{\Lambda}(\cdot)$ such that the scale-normalized MSE
$G(\Lambda)=\sigma^2\rho^{2/n}$ is minimized,\footnote{This agrees with the
  definition of $G(\Lambda)$ for lattices in
  \cite{lattices-good-everything}.} where $\rho$ is the codeword density
\begin{equation}
  \label{eq:rho}
  \rho = \limsup_{M\to\infty} \frac{1}{(2M)^n} \cardinal{\Lambda \cap
    [-M,M]^n}.
\end{equation}

In this paper we consider asymptotically large dimensionality $n$.  By
a volume argument, it is easy to find a lower bound $G^* =
\frac{1}{2\pi\e}$ for $G(\Lambda)$.  This bound can be approached by
the nearest-neighbor quantizer with a suitable random codebook e.g.\
in \cite{lattices-good-everything}, whose codewords' Voronoi regions
are asymptotically spherical, but such a quantizer has exponential
complexity in $n$ and is thus impractical.  The simplest scalar
quantizer $\Lambda_1=\SetZ^n$, on the other hand, has the 1.5329-dB
larger $G_1=G(\Lambda_1)=\frac{1}{12}$, which corresponds to the
well-known 1.53-dB loss of scalar quantization.  In general, we call
$10 \log_{10} (G(\Lambda) / G^*)$ the \emph{shaping loss} of a
quantizer, and it is also the gap of the \emph{granular gain} and
\emph{shaping gain} defined in \cite{lattice-trellis-quant-highrate},
for source and channel coding respectively, toward the 1.53-dB limit.

MSE quantizers with near-zero shaping losses are important in both
source and channel coding.  In lossy source coding, the shaping loss
naturally dictates rate-distortion performance at high rates
\cite{lattice-trellis-quant-highrate}.  In channel coding on Gaussian
channels, MSE quantizers can be used for \emph{shaping} to make the
channel input closer to the optimal Gaussian distribution
\cite{trellis-shaping}.  Basically, instead of transmitting the
channel-coded and QAM-modulated signal $\seq{u}$ (each element of
$\seq{u}$ corresponding to one symbol in the code block), we transmit
$\seq{x} = \seq{u} - \seq{a}$ with $\seq{a} = Q_{\Lambda}(\seq{u}) \in
\Lambda$, which should be closer to Gaussian.  $\seq{u}$ and $\seq{a}$
are separated at the receiver side, and the shaping loss determines
the achievable gap from channel capacity at high SNRs.  Shaping is
particularly important in \emph{dirty-paper coding} (DPC)
\cite{writing-on-dirty-paper} on the channel
\begin{equation}
  \label{eq:dpc-model}
  \seq{y} = \seq{x} + \seq{s} + \seq{z},
\end{equation}
where $\seq{x}$ is the transmitted signal, $\seq{s}$ is the
interference known only at the transmitter, and $\seq{z}$ is the
``MMSE-adjusted'' noise.  Using an MSE quantizer, arbitrarily large
$\seq{s}$ can be pre-cancelled without significantly increasing signal
power by transmitting
\begin{equation}
  \label{eq:dpc-tx}
  \seq{x} = \seq{u} - \seq{s} - \seq{a},\ \text{with } \seq{a} = Q_{\Lambda}(\seq{u} - \seq{s}),
\end{equation}
so that the received signal
\begin{equation}
  \label{eq:dpc-rx}
  \seq{y} = \seq{u} - \seq{a} + \seq{z}.  
\end{equation}
Again, the receiver must separate $\seq{u}$ and $\seq{a}$, and the
shaping loss determines the achievable gap from channel capacity.  In
this case, however, due to the lack of receiver-side knowledge of
$\seq{s}$, the rate loss caused by non-ideal shaping is most
significant at \emph{low} SNRs and can be a significant fraction of
channel capacity
\cite{sup-coding-side-info-chan,close-to-cap-dpc,trellis-conv-prec-tx-intf-presub,near-cap-dpc-tcq-ira}.
For example, the shaping quantizer in \cite{near-cap-dpc-tcq-ira} has
\unit[0.15]{dB} shaping loss, corresponding to a rate loss of
\unit[0.025]{b/s}, yet in the 0.25-b/s DPC system this is already 10\%
of the rate and is responsible for \unit[0.49]{dB} of its 0.83-dB gap
from capacity.  Apart from its obvious application in steganography
\cite{img-data-hiding-cap-appr-dpc}, DPC and its extension to vector
channels (similar in principle to vector precoding
\cite{vector-perturbation-mod-precoding1} but done in both time and
spatial domains) are also essential in approaching the capacity of
vector Gaussian broadcast channels such as MIMO downlink, therefore
the design of near-ideal MSE quantizers is of great interest in these
applications.

Currently, near-optimal MSE quantizers usually employ trellis-coded
quantization (TCQ) \cite{tcq-memoryless-gauss-markov}, in which
$\Lambda=\mathcal{U} + 2\SetZ^n$ or $\mathcal{U} + 4\SetZ^n$ with
$\mathcal{U}$ being respectively the codeword set of a binary
convolution code or a 4-ary trellis code.  The number of required
trellis states increases very rapidly as the shaping gain approaches
the 1.53-dB limit, and the computational complexity and memory
requirement are thus very high.  This is particularly bad at the
receiver side of DPC systems, where the BCJR
(Bahl-Cocke-Jelinek-Raviv) algorithm must be run many times on the
trellis in an iterative fashion to separate $\seq{u}$ and $\seq{a}$
\cite{near-cap-dpc-tcq-ira}, resulting in a time complexity
proportional to both the number of trellis states and the outer
iteration count.

Inspired by the effectiveness of Turbo and low-density parity-check
(LDPC) codes in channel coding, it is natural to consider the use of
sparse-graph codes in quantization.  In \cite{turbo-trellis-vq} Turbo
codes are used in quantization of uniform sources, but convergence
issues make the scheme usable only for very small block sizes $n$, and
the shaping loss is thus unsatisfactory.  In
\cite{it-quant-codes-graphs, analysis-ldgm-loss-compression,
  ld-ach-wz-gp-bounds}, it is shown that low-density generator matrix
(LDGM) codes, being the duals of LDPC codes, are good for lossy
compression of binary sources, and practical quantization algorithms
based on belief propagation (BP) and survey propagation (SP) have also
been proposed in \cite{lossy-src-enc-msgpass-dec-ldgm} and \cite
{binary-quant-bp-dec-ldgm}, but these works consider binary sources
only.  Practical algorithms for the MSE quantization of $\SetR^n$ with
LDGM codes have not received much attention before.  Even in the
binary case, little has been done in the analysis of the BP
quantizer's behavior and the optimization of the LDGM code for it.

In \cite{ldgm-vq-globecom07}, we have addressed the problem
of MSE quantization using LDGM-based codes of structure $\Lambda =
\mathcal{U} + m\SetZ^n$, known as \emph{$m$-ary codes}, where each
$\seq{u} \in \mathcal{U}$ is a codeword of a binary LDGM code when
$m=2$, and is the combination of two codewords, each from a binary
LDGM code, by Gray mapping when $m=4$.  The degree distributions of
the codes are optimized under the erasure approximation, and
shaping losses as low as \unit[0.056]{dB} have been demonstrated.

In this paper, we will improve upon the results in
\cite{ldgm-vq-globecom07} by using better analytical techniques and
more accurate methods for code optimization.  We start in
Section~\ref{sec:basics} by analyzing the minimum shaping loss
achievable by this $m$-ary structure using random-coding arguments.
Although binary quantization codes have significant random-coding
loss, they are analyzed first due to their simplicity.  In
Section~\ref{sec:binary}, we present the quantization algorithm for
binary codes, which consists, like \cite{binary-quant-bp-dec-ldgm}, of
BP and a guessing (``decimation'') procedure to aid convergence.

Like LDPC, degree distribution plays an important role in the
performance of LDGM quantization codes, but the use of decimation
makes direct analysis difficult.  To solve this problem, we propose
the \emph{typical decimator} (TD) as a suboptimal but analytically
more tractable version of the decimation algorithm, and analyze first
its use in the simpler binary erasure quantization (BEQ) problem in
Section~\ref{sec:deg-opt-erasure}, which also forms the basis for the
erasure approximation in \cite{ldgm-vq-globecom07}.  We find that the
TD can obtain asymptotically correct extrinsic information for
decimation, and a solution to BEQ can be found with such information,
as long as the code's extended BP (EBP) extrinsic information transfer
(EXIT) curve \cite{maxwell-constr} characterizing the fixed points of
the BP process satisfies certain \emph{monotonicity conditions}.  For
a given LDGM code, the most difficult BEQ problem it can solve is then
parametrized by a \emph{monotonicity threshold} $\Icthr$, and the
degree distribution can be optimized by maximizing this $\Icthr$.

In Section~\ref{sec:deg-opt-mse}, these arguments are extended to our
MSE quantization problem, and similar monotonicity conditions are
obtained, which can be checked by quantized density evolution (DE).
These DE results can be visualized with modified EXIT curves, and a
similar method to the BEQ case can then be used for degree
distribution optimization.

We have assumed iteration counts $L\to\infty$ in the above analysis.
In Section~\ref{sec:decimation}, we proceed to analyze the impact of
finite $L$.  We will show that a finite $L$ causes ``bad'' guesses in
decimation, and a \emph{recovery algorithm} is sometimes required for
BP to continue normally afterwards.  With recovery, the loss due to
finite $L$ can be characterized by the \emph{delta-area} $\Aignore$
between the EBP curve and the actual trajectory, which will be used in
the subsequent optimization of the pace of decimation as well as the
degree distribution.

All these results are extended to $m$-ary codes (where $m=2^K$) in a
straightforward manner in Section~\ref{sec:nonbinary}\@.  Numerical
results on MSE performance in Section~\ref{sec:numerical} shows that
LDGM quantization codes optimized with the aforementioned methods have
the expected good performance and can achieve shaping losses of
\unit[0.2676]{dB} at 99 iterations, \unit[0.0741]{dB} at 1022 and
\unit[0.0423]{dB} at 8356 iterations, the latter two of which are far
better than what TCQ can reasonably offer and are also significantly
better than the results in \cite{ldgm-vq-globecom07}.  Indeed, a
heuristical analysis on the asymptotic loss-complexity tradeoff carried
out in Section~\ref{sec:discussion} indicates that LDGM quantization
codes can achieve the same shaping loss with far lower complexity than
TCQ.  We conclude the paper in Section~\ref{sec:conclusion}.

\section{Performance Bounds of $m$-ary Quantizers}
\label{sec:basics}
In this paper, we consider $\Lambda$ with a periodic structure
$\Lambda = \mathcal{U} + m\SetZ^n$, where $\mathcal{U}$ is a set of
$2^{nR}$ codewords from $\{0,1,\dotsc,m-1\}^n$ with each $\seq{u} =
\seq{u}(\seq{b}) \in \mathcal{U}$ labeled by a binary sequence
$\seq{b} \in \{0,1\}^{nR}$.  We call $\Lambda$ an \emph{$m$-ary
  rate-$R$ quantization code}.  In this section, we will analyze the
achievable shaping loss by this periodic structure.

Given the source sequence $\seq{y} \in \SetR^n$, for each
$\seq{u} = \seq{u}(\seq{b}) \in \mathcal{U}$ the nearest sequence to
$\seq{y}$ in $\seq{u} + m\SetZ^n$ is $\seq{x}(\seq{b}) = \seq{y} -
\seq{z}(\seq{b})$, where $\seq{z}(\seq{b}) = (\seq{y} -
\seq{u}(\seq{b})) \bmodIn$ is the quantization error and
$\mathcal{I} = [-\frac{m}{2},\frac{m}{2})$.  The quantizer has then to
minimize $\norm{\seq{z}(\seq{b})}$ over all $\seq{b}$'s, or
equivalently, to maximize
\begin{equation}
  \label{eq:qy}
  \qseqy(\seq{b}) = \e^{-t \norm{\seqs{z}(\seqs{b})}^2} = \prod_{j=1}^n \e^{-t(y_j-u_j(\seqs{b}))\bmodI^2}
\end{equation}
for some constant $t>0$.  The chosen $\seq{b}$ is denoted
$\seq{b}_{\seqs{y}}$, the corresponding quantization error is
$\seq{z}_{\seqs{y}} = \seq{z}(\seq{b}_{\seqs{y}})$, and the resulting
MSE \eqref{eq:mse} then becomes\footnote{For large $n$,
  \eqref{eq:mse-periodic} is mostly just an average over strongly
  typical $\seq{y}$ with respect to the uniform distribution on
  $[0,m)$, i.e.\ those whose elements are approximately uniformly
  distributed over $[0,m)$, and the rest of this paper considers such
  $\seq{y}$ only.  In shaping and DPC applications, $\seq{y}$ can be a
  modulated signal that does not follow the uniform distribution, and
  in such cases it may be necessary to ``dither'' $\seq{y}$ before
  quantization by adding to it a random sequence uniformly distributed
  in $[0,m)^n$ and known by the dequantizer, in order to obtain the
  expected MSE performance.}
\begin{equation}
  \label{eq:mse-periodic}
  \sigma^2 = \frac{1}{m^n} \cdot \frac{1}{n} \int_{[0,m)^n} \norm{\seq{z}_{\seqs{y}}}^2 \, d\seq{y}
  = \frac{1}{n} \int_{[0,1)^n} \avga{\norm{\seq{z}_{\seqs{\yt} + \seqs{a}}}^2} \, d\seq{\yt},
\end{equation}
where $\avga{\cdot}$ denotes averaging over $\seq{a} \in \{0,1,\dotsc,m-1\}^n$.

\subsection{Lower Bound of Quantization Error}
Given $\Lambda$, for each source sequence $\seq{y} \in [0,m)^n$, let
\begin{equation}
  \label{eq:Qy}
  \Qseqy = \sum_{\seqs{b} \in \{0,1\}^{nR}} \qseqy(\seq{b}).
\end{equation}
Since $\qseqy(\seq{b}_{\seqs{y}}) \le \Qseqy$, we can lower-bound the
mean-square quantization error $\frac{1}{n}
\norm{\seq{z}_{\seqs{y}}}^2$ as
\begin{equation}
    \frac{1}{n} \norm{\seq{z}_{\seqs{y}}}^2
  = -\frac{1}{nt} \ln \qseqy(\seq{b}_{\seqs{y}})
\ge -\frac{1}{nt} \ln \Qseqy.
\end{equation}
Now let $\seq{y} = \seq{\yt} + \seq{a}$ with $\seq{\yt} \in
[0,1)^n$ and $\seq{a} \in \{0,1,\dotsc,m-1\}^n$, and average over
$\seq{a}$, then from Jensen's inequality
\begin{equation}
  \label{eq:lb-avga}
  \frac{1}{n} \avga{\norm{\seq{z}_{\seqs{\yt} + \seqs{a}}}^2}
\ge -\frac{1}{nt} \avga{\ln \Qseqyta}
\ge -\frac{1}{nt} \ln \avga{\Qseqyta},
\end{equation}
where $\avga{\Qseqyta}$ can easily be found to be
\begin{equation}
  \label{eq:Qyt}
  \avga{\Qseqyta} = \frac{2^{nR}}{m^n} \prod_{j=1}^n Q_{\yt_j}, \quad \text{with}\ Q_{\yt} = \sum_{a=0}^{m-1} \e^ {-t (\yt + a) \bmodI^2 }.
\end{equation}
$\sigma^2$ in \eqref{eq:mse-periodic} can be lower-bounded by
integrating \eqref{eq:lb-avga} over $\seq{\yt}$.  For
asymptotically large $n$, we only need to consider (strongly) typical
$\seq{\yt}$ with respect to the uniform distribution on $[0,1)$,
i.e.\ whose $n$ elements are nearly uniformly distributed over $[0,1)$.
We thus have
\begin{align}
  \label{eq:sigma2min_1}
  \sigma^2 &\ge -\frac{1}{nt} \int_{[0,1)^n} \ln \avga{\Qseqyta} \, d\seq{\yt} \\
  \label{eq:sigma2min}
  &\asympteq \frac{1}{t} \left( \ln m - R\ln 2 - \int_0^1 \ln Q_{\yt} \, d\yt \right).
\end{align}
This bound holds for any $t>0$ and is found to be tightest for
$t$ satisfying $H_t = \log m - R$ (this $t$ is hence denoted
$t_0(R)$), when it becomes $\sigma^2 \ge P_t$.  $H_t$ and $P_t$ are
defined as
\begin{align}
  \label{eq:Ht}
  H_t &= -\int_{\mathcal{I}} \pz(z) \log\pz(z) \,dz, \\
  \label{eq:Pt}
  P_t &= \int_{\mathcal{I}} z^2 \pz(z) \,dz, \\
  \label{eq:pz}
  \pz(z) &= \frac{\e^{-t z^2}}{Q_{\yt}}, \quad z\in\mathcal{I}, \quad \yt = z \bmod [0,1).
\end{align}

\subsection{Achievable Quantization Error with Random Coding}
\label{sec:random-coding}
For asymptotically large $n$, we will see that the aforementioned
lower bound is actually achievable by random coding, that is, with the
$2^{nR}$ codewords in $\mathcal{U}$ independently and uniformly
sampled from $\{0,1,\dotsc,m-1\}^n$ (allowing for duplicates) and
using the nearest-neighbor quantizer.

Again we assume $\seq{y} \in [0,m)^n$, and since the MSE $\frac{1}{n}
\norm{\seq{z}_{\seqs{y}}}^2$ is bounded for any $\seq{y}$, we can
consider only typical $\seq{y}$'s with respect to the uniform
distribution on $[0,m)$.  Define
\begin{equation}
  \label{eq:SetUy}
  \mathcal{U}_{\seqs{y}} = \left\{ \seq{u} \in \{0,\dotsc,m-1\}^n \Bigm|
    \norm{(\seq{y}-\seq{u}) \bmodIn}^2 \le nP_t \right\}
\end{equation}
as the set of possible codewords that are ``sufficiently close'' to
$\seq{y}$, and we can compute $\frac{1}{n} \log
\cardinal{\mathcal{U}_{\seqs{y}}}$ with large deviation theory.  If it
is larger than $\log m - R$, with asymptotically high probability
$\mathcal{U} \cap \mathcal{U}_{\seqs{y}} \neq \emptyset$, thus some
$\seq{x} \in \mathcal{U}+ m\SetZ^n$ can be found whose MSE toward
$\seq{y}$ is no more than $P_t$.  Since this is true for most typical
$\seq{y}$, the average MSE $\sigma^2$ cannot exceed $P_t$ by more than
a vanishingly small value.

To compute $\frac{1}{n} \log\cardinal{\mathcal{U}_{\seqs{y}}}$ for a
typical $\seq{y}$, we define the type $p_y(u)$ of a sequence $\seq{u}$
as the fraction of each $u \in \{0,1,\dotsc,m-1\}$ at the positions in
$\seq{u}$ whose corresponding elements in $\seq{y}$ are approximately
$y$.  Denoting the number of sequences $\seq{u}$ with this type as
$N[p_y(u)]$, we have
\begin{equation}
  \label{eq:ldtN}
  \frac{1}{n} \log N[p_y(u)] \asympteq \frac{1}{m} \int_0^m H_y(u) \,dy,
\end{equation}
where $H_y(u)$ is the entropy
\begin{equation}
  H_y(u) = -\sum_{u=0}^{m-1} p_y(u) \log p_y(u),
\end{equation}
and $\seq{u} \in \mathcal{U}_{\seqs{y}}$ becomes the constraint
\begin{equation}
  \label{eq:ldtP}
  \frac{1}{m} \int_0^m \left( \sum_{u=0}^{m-1} (y-u)\bmodI^2 p_y(u) \right) \,dy \le P_t.
\end{equation}
According to large deviation theory, $\frac{1}{n}
\log\cardinal{\mathcal{U}_{\seqs{y}}}$ is asymptotically the maximum
of \eqref{eq:ldtN} under the constraints \eqref{eq:ldtP} and
\begin{equation}
  p_y(u) \ge 0, \quad \sum_{u=0}^{m-1} p_y(u) = 1, \quad y\in [0,m).
\end{equation}
This is a convex functional optimization problem over $p_y(u)$ (a
function of both $y$ and $u$), which can be easily solved with
Lagrange multipliers.  The maximizing $p_y(u)$ is found to be
\begin{equation}
  \label{eq:opt-pyu}
  p_y(u) = \frac{\e^{-t (y-u)\bmodI^2 }}{Q_{\yt}}, \quad \yt = y\bmod [0,1),
\end{equation}
and the resulting
\begin{equation}
  \frac{1}{n} \log\cardinal{\mathcal{U}_{\seqs{y}}} \asympteq H_t.
\end{equation}
By the argument above, as long as $H_t > \log m - R$, i.e.\ $t <
t_0(R)$, random coding can achieve $\sigma^2 \le P_t$ for
asymptotically large $n$.

\subsection{Marginal Distribution of Quantization Error}
From the $p_y(u)$ result in \eqref{eq:opt-pyu}, the marginal
distribution of an individual $z_j = (y_j - u_j) \bmodI$ under random
coding can also be obtained as
\begin{align}
  \pz(z) &= \frac{1}{m} \int_0^m \left( \sum_{u=0}^{m-1} p_y(u) \delta(z - (y-u)\bmodI) \right) \,dy \\
  &= \frac{1}{m} \int_0^m \left( \sum_{u=0}^{m-1} \frac{\e^{-t z^2}}{Q_{\yt}} \delta(z - (y-u)\bmodI) \right) \,dy \\
  &= \frac{\e^{-t z^2}}{Q_{\yt}},\quad z\in\mathcal{I},
\end{align}
where $\delta(\cdot)$ is the Dirac delta function and $\yt = z\bmod
[0,1) = y\bmod [0,1)$.  This is simply the $\pz(z)$ in \eqref{eq:pz},
and $H_t$ in \eqref{eq:Ht} and $P_t$ in \eqref{eq:Pt} are respectively
the entropy and average power of this distribution.

\subsection{The Random-Coding Loss}
\label{sec:random-coding-loss}
We have shown that a random quantization codebook with the
nearest-neighbor quantizer is asymptotically optimal among rate-$R$
quantization codes of the form $\Lambda = \mathcal{U} + m\SetZ^n$.
Therefore, its shaping loss represents the performance limit of such
codes, and can be viewed as the cost incurred by the period-$m$
structure.

\begin{figure}[!t]
  \centering
  \subfigure[binary code ($m=2$)]{\includegraphics{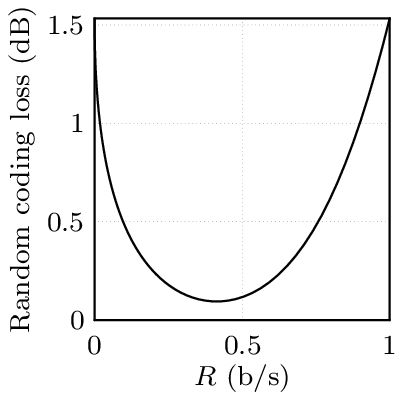}}%
  \hspace{2mm}%
  \subfigure[4-ary code ($m=4$)]{\includegraphics{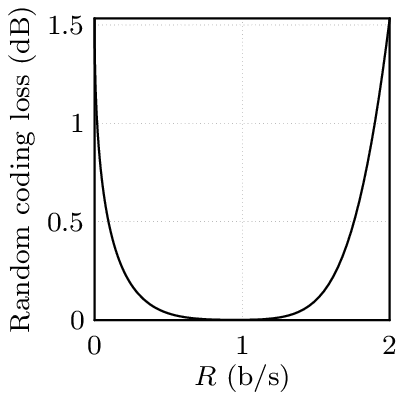}}%
  \caption{Random-coding losses of binary and 4-ary quantization
    codes.  For binary quantization codes, the minimum loss is
    approximately \unit[0.0945]{dB} at $t=3.7$ and
    $R=\unit[0.4130]{b/s}$.  For 4-ary codes, the minimum loss is only
    \unit[0.0010]{dB} at approximately $t=2$ and
    $R=\unit[0.9531]{b/s}$.}%
  \label{fig:power-entropy}%
\end{figure}

For asymptotically large $n$, the random $m$-ary quantizer has average
MSE $\sigma^2 = P_t$ with $t = t_0(R)$ and density $\rho =
2^{nR}/m^n$, so the achieved $G(\Lambda) = \sigma^2 \rho^{2/n} = P_t
(2^R/m)^2$.  The shaping loss $10\log_{10}(G(\Lambda)/G^*)$ can then
be expressed as $10\log_{10} (P_t / P^*_t)$, where $P^*_t =
\frac{1}{2\pi\e} (m/2^R)^2$ is the power of a Gaussian with entropy
$H_t = \log m - R$.  We called it the \emph{random-coding loss}, and
it is plotted in Fig.~\ref{fig:power-entropy} for $m=2$ and $m=4$.
For large $m$ and moderate $R$, $Q_{\yt}$ in \eqref{eq:Qyt} approaches
a constant, $\pz(z)$ is close to a Gaussian distribution, thus $P_t
\approx P_t^*$ and the random-coding loss is close to zero.

\section{The Binary LDGM Quantizer}
\label{sec:binary}
As random quantization codes with the nearest-neighbor quantizer are
obviously impractical to implement, it is natural to look into
sparse-graph codes as practical candidates for achieving near-zero
shaping losses.  In \cite{it-quant-codes-graphs}, it has been shown
that LDPC codes are unsuitable for BEQ but LDGM codes work well,
therefore we will also use LDGM codes in MSE quantization.  We
consider the simplest $m=2$ case first, and in
Section~\ref{sec:nonbinary} we will look into codes with larger $m$
that are not as limited by the random-coding loss.

\begin{figure}[!t]
  \centering
  \subfigure[original form]{\label{fig:binary-fg}\raisebox{7mm}{\includegraphics{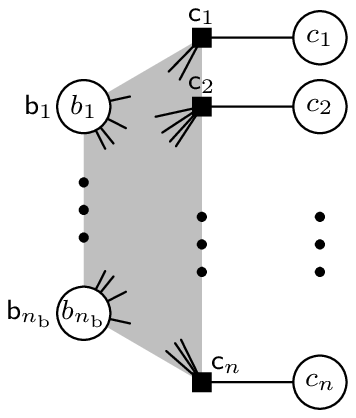}}}%
  \hspace{3mm}%
  \subfigure[perturbed form]{\label{fig:binary-fg-a}\includegraphics{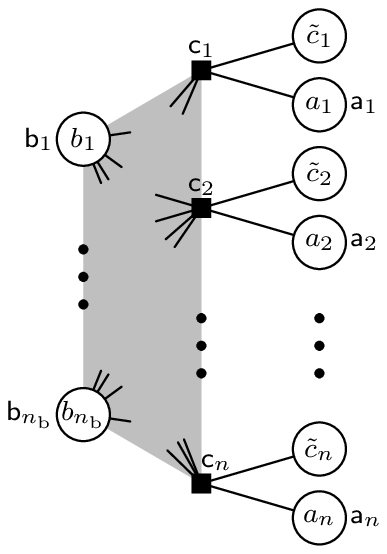}}%
  \caption{The factor graph of the binary LDGM quantizer.  Circles are
    variable nodes and black squares are factor nodes.  The gray area
    contains random \node{b}-to-\node{c} edges, and each edge from
    $\node{b}_i$ to $\node{c}_j$ corresponds to $g_{ij}=1$ in the
    generator matrix $G$.  We mostly use the original form (a) with
    the priors $\priprobx{c}{j}{c} = \pz((y_j - c) \bmodI)$.  The
    equivalent ``perturbed'' form (b) is used in
    Section~\ref{sec:de-mse-intro}, where $\seq{\yt} = \seq{y} \bmod
    [0,1)^n$, $\seq{a} = (\seq{y} - \seq{\yt}) \bmod 2 \in \{0,1\}^n$, and
    $\seq{\ut} = \seq{\ct} = (\seq{c} - \seq{a}) \bmod 2$.  With
    $\seq{y}$ fixed, each prior $\priprob{a}{j}$ is a hard decision
    $\constmsg{a_j}$.  Since $\seq{z} = (\seq{y} - \seq{c}) \bmodI =
    (\seq{\yt} - \seq{\ct}) \bmodI$, the prior on $\ct_j$ is
    $\priprobx{\ct}{j}{\ct} = \pz((\yt_j - \ct)\bmodI)$.}%
\end{figure}

We thus consider $\Lambda = \mathcal{U} + 2\SetZ^n$ with $\mathcal{U}$
being the codeword set of an LDGM code, i.e.\ each $\seq{u} \in
\mathcal{U}$ is of the form $\seq{u} = \seq{c} = \seq{b} \mat{G}$,
where $\seq{b} \in \{0,1\}^{\nb}$, $\nb=nR$ and the low-density
generator matrix $\mat{G}=(g_{ij})_{\nb\times n}$ is randomly
generated from some degree distribution that will be optimized below.
Given such a code, $\qseqy(\seq{b})$ in \eqref{eq:qy} can be
represented by the factor graph \cite{factor-graphs-sum-product} in
Fig.~\ref{fig:binary-fg}.\footnote{\label{fn:bp-defs}In the factor
  graph, symbols such as $\node{b}_i$ and $\node{c}_j$ denote variable
  and factor nodes, while $b_i$ and $c_j$ are the variables
  themselves.  $\neighbor{bc}{\cdot j}=\neighbor{cb}{j\cdot}$ denote
  the set of indices $i$ for which there is an edge connecting
  $\node{b}_i$ and $\node{c}_j$.  In belief propagation,
  $\priprob{b}{i}$ is the priors on variable $b_i$, $\extprob{b}{i}$
  is the computed extrinsic probabilities for $b_i$, $\msg{bc}{ij}$
  denotes a message from node $\node{b}_i$ to $\node{c}_j$, and so on.
  The priors, posteriors and messages are all probability
  distributions \cite{factor-graphs-sum-product}, in this case over
  $\{0,1\}$, and here we represent them by probability tuples (rather
  than $L$-values, which are equivalent).  For example,
  $\priprob{b}{i}$ is viewed as a tuple
  $(\priprobx{b}{i}{0},\priprobx{b}{i}{1})$ satisfying
  $\priprobx{b}{i}{0}+\priprobx{b}{i}{1}=1$ (the normalization is done
  implicitly), which corresponds to $L$-value $\ln(\priprobx{b}{i}{0}
  / \priprobx{b}{i}{1})$.  ``$\odot$'' and ``$\oplus$'' refer to the
  variable-node and check-node operations in LDPC literature, i.e.\
  $(\vpp{\mu}_0,\vpp{\mu}_1) \odot (\mu'_0,\mu'_1) =
  (\vpp{\mu}_0\mu'_0, \vpp{\mu}_1\mu'_1)$ (implicitly normalized) and
  $(\vpp{\mu}_0,\vpp{\mu}_1) \oplus (\mu'_0,\mu'_1) =
  (\vpp{\mu}_0\mu'_0+\vpp{\mu}_1\mu'_1,\vpp{\mu}_0\mu'_1+\vpp{\mu}_1\mu'_0)$.
  $\constmsg{0}=(1,0)$, $\constmsg{1}=(0,1)$ and
  $\constmsg{*}=(\frac{1}{2},\frac{1}{2})$ are respectively the
  ``sure-0'', ``sure-1'' and ``unknown'' messages.  $H(\mu)=-\mu_0\log
  \mu_0-\mu_1\log \mu_1$ is the entropy function for $\mu =
  (\mu_0,\mu_1)$.}  The \node{c}-nodes (shorthand for the factor nodes
$\node{c}_j$, $j=1,\dotsc,n$) represent the relationship $\seq{c} =
\seq{b} \mat{G}$, whereas each factor $\e^{-t (y_j - c_j) \bmodI^2}$
in \eqref{eq:qy} is included in the prior $\priprob{c}{j}$ on variable
$c_j$ as
\begin{equation}
  \label{eq:priprobs}
  \priprobx{c}{j}{c} = \frac{1}{Q_{\yt_j}} \e^{-t (y_j - c)\bmodI^2}
  = \pz((y_j - c) \bmodI),
\end{equation}
where $Q_{\yt_j}$ (with $\yt_j = y_j \bmod [0,1)$) serves as the
normalization factor.

The \emph{belief propagation} algorithm (also known as the sum-product
algorithm) can then be run on this factor graph.  Unlike the case of
LDPC decoding, here BP does not usually converge by itself.%
\footnote{Intuitively speaking, when doing LDPC decoding with SNR
  higher than threshold, the transmitted codeword is usually much
  closer to the received sequence (and thus much more likely) than any
  other codeword, allowing BP to converge it.  In the case of
  quantization with LDGM codes, there are usually a large number of
  similarly close codewords to the source sequence, and BP cannot by
  itself make a decision among them.}  Instead, we rely on BP to
generate ``extrinsic probabilities'' $\extprob{b}{i}$ for each $b_i$
after a number of iterations, with which hard decisions are made on
some $b_i$'s (called \emph{decimation} following
\cite{lossy-src-enc-msgpass-dec-ldgm}).  Subsequent BP iterations use
these hard decisions as priors $\priprob{b}{i}$, and the resulting
updated $\extprob{b}{i}$'s are used for more decimations.  This
iterative process continues until a definite $\seq{b}$ is obtained
that hopefully has a large $\qseqy(\seq{b})$ and thus a small
quantization error.  This quantization algorithm is shown in
Fig.~\ref{alg:binary}, with a BP part and a decimation part in each
iteration.  As is intuitively reasonable, each time we decimate the
``most certain'' bit $b_{i^*}$, with
\begin{equation}
  \label{eq:dec-choice-det}
  i^* = \argmax_{i\in\mathcal{E}} \max_{b\in\{0,1\}} \extprobx{b}{i}{b},
\end{equation}
and it is decimated to its most likely value
\begin{equation}
  \label{eq:dec-value-det}
  b^* = \argmax_{b\in\{0,1\}} \extprobx{b}{i^*}{b}.
\end{equation}
This is called the \emph{greedy decimator} (GD).  Alternatively, for
the convenience of analysis we will also look at the \emph{typical
  decimator}, implementable but with worse performance in practice, in
which the bit index $i^*$ to decimate is chosen randomly in
$\mathcal{E}$ (the set of yet undecimated bits) with equal
probabilities, and its decimated value $b^*$ is $b \in \{0,1\}$ with
probability $\extprobx{b}{i^*}{b}$.

\begin{figure}[!t]
  \centering\footnotesize
  \begin{algorithmic}
    \STATE $\priprobx{c}{j}{c} \assign \pz((y_j-c)\bmodI)$,
    $j=1,\dotsc,n$, $c=0,1$ \COMMENT{$\mathcal{I}=[-1,1)$}
    \STATE $\msg{bc}{ij} \assign \constmsg{*}$, $i=1,\dotsc,\nb$, $j=1,\dotsc,n$
    \STATE $\priprob{b}{i} \assign \constmsg{*}$, $i=1,\dotsc,\nb$
    \STATE $\mathcal{E} \assign \{1,2,\dotsc,\nb\}$ \COMMENT{the set of bits not yet decimated}
    \STATE $\deltamax \assign 0$, $\Info{bc} \assign 0$
    \REPEAT[belief propagation iteration]
      \FOR[BP computation at $\node{c}_j$]{$j=1$ to $n$}
        \STATE $\msg{cb}{ji} \assign \priprob{c}{j} \oplus
        \left(\oplus_{i'\in\neighbor{bc}{\cdot j} \excluding{i}}
          \msg{bc}{i'j}\right)$, $i\in\neighbor{cb}{j\cdot}$
      \ENDFOR
      \FOR[BP computation at $\node{b}_i$]{$i=1$ to $\nb$}
        \STATE $\msg{bc}{ij} \assign \priprob{b}{i} \odot
        \left(\odot_{j'\in\neighbor{cb}{\cdot i} \excluding{j}}
          \msg{cb}{j'i}\right)$, $j\in\neighbor{bc}{i\cdot}$
        \STATE $\extprob{b}{i} \assign \odot_{j'\in\neighbor{cb}{\cdot i}} \msg{cb}{j'i}$
      \ENDFOR
      \STATE $\nextInfo{bc} \assign 1 - (\nb\db)^{-1} \sum_{i,j} H(\msg{bc}{ij})$ \COMMENT{estimate new $\Info{bc}$}
      \STATE $\delta \assign 0$ \COMMENT{amount of decimation so far in this iteration}
      \STATE Set $\minprog$ according to the desired pace (e.g.\ to \eqref{eq:pace-approx})
      \IF[little progress, do decimation]{$\nextInfo{bc} < \Info{bc} + \minprog$}
        \REPEAT
          \STATE $i^* \assign \argmax_{i\in\mathcal{E}} \max_b \extprobx{b}{i}{b}$
          \COMMENT{$b_{i^*}$ is the most certain bit\ldots}
          \STATE $b^* \assign \argmax_{b\in\{0,1\}} \extprobx{b}{i^*}{b}$
          \COMMENT{\ldots whose likely value is $b^*$}
          \STATE $\delta \assign \delta + (-\log \extprobx{b}{i^*}{b^*})$
          \vspace{0.5mm}
          \STATE $\nextInfo{bc} \assign \nextInfo{bc} +
          (\nb\db)^{-1} \sum_{j\in\neighbor{bc}{i^*\cdot}} H(\msg{bc}{i^*j})$
          \STATE $\priprob{b}{i^*} \assign \constmsg{b^*}$,
          $\msg{bc}{i^*j} \assign \constmsg{b^*}$, $j\in\neighbor{bc}{i^*\cdot}$ \COMMENT{decimate $b_i$ to $b^*$}
          \STATE $\mathcal{E} \assign \mathcal{E} \excluding{i^*}$
        \UNTIL{$\delta > \deltamax$ or $\nextInfo{bc} \ge \Info{bc} + \minprog$ or $\mathcal{E}=\emptyset$}
      \ENDIF
      \STATE $\deltamax \assign \max(0.8 \deltamax, 1.25 \delta)$
      \STATE $\Info{bc} \assign \nextInfo{bc}$
      \vspace{0.5mm}
    \UNTIL{$\mathcal{E}=\emptyset$}
    \STATE $b_i \assign 0$ (resp. $1$) if $\priprob{b}{i}=\constmsg{0}$ (or $\constmsg{1}$), $i=1,\dotsc,\nb$
    \STATE $\seq{c} \assign \seq{b} \mat{G}$, $\seq{u} \assign \seq{c}$
    \STATE $z_j = (y_j - c_j) \bmodI$, $x_j = y_j - z_j$, $j=1,\dotsc,n$
  \end{algorithmic}
  \caption{The binary quantization algorithm.  The throttled version
    is shown above, while the unthrottled version is without the
    $\delta > \deltamax$ condition in the \textbf{until} statement.
    The choice of $i^*$ and $b^*$ corresponds to the greedy
    decimator.}
  \label{alg:binary}
\end{figure}

The number of bits to decimate is controlled through the estimated
mutual information $\Info{bc}$ in \node{b}-to-\node{c} messages (i.e.\
the $\msg{bc}{ij}$'s), which is made to increase by about $\minprog$
in each iteration.  This amount of increase $\minprog$, possibly a
function of the current $\Info{bc}$ and hence called the \emph{pace of
  decimation}, makes the algorithm terminate within $L_0$ iterations
if followed exactly, though the actual iteration count $L$ can be
somewhat different.  Uniform pacing is used in
\cite{ldgm-vq-globecom07}, i.e.\ $\minprog$ is a constant $1/L_0$.  In
this paper, the pacing is optimized in Section~\ref{sec:opt-pacing} to
obtain somewhat better MSE performance.  Increasing $L_0$ also improves
MSE performance, but more iterations would be necessary.

The decimation algorithm can either be \emph{unthrottled} or
\emph{throttled}.  The unthrottled version used in most of our
simulations simply decimates until the increase of $\Info{bc}$ in the
iteration reaches $\minprog$.  In the throttled version introduced in
\cite{ldgm-vq-globecom07}, the amount of decimation per iteration is
instead controlled by $\deltamax$, which is smoothly adapted, as shown
in Fig.~\ref{alg:binary}, to make $\Info{bc}$ increase eventually at
the desired pace.

More will be said on the decimation algorithm in
Section~\ref{sec:decimation}, but we will first discuss the
optimization of LDGM's degree distribution and the choice of $t$ in
Sections~\ref{sec:deg-opt-erasure} and \ref{sec:deg-opt-mse}\@.

\section{Degree Distribution Optimization for Binary Erasure Quantization}
\label{sec:deg-opt-erasure}
Like LDPC codes, LDGM quantization codes require optimized degree
distributions for good MSE performance.  The performance of LDGM
quantizers has been analyzed previously in
\cite{analysis-ldgm-loss-compression} for binary sources, but this
analysis, based on codeword-counting arguments, is applicable only to
nearest-neighbor quantization and not very useful for the above BP
quantizer.  In \cite{lossy-src-enc-msgpass-dec-ldgm}'s treatment of
LDGM quantization of binary sources, degree distributions of good LDPC
codes in \cite{design-cap-approaching-ldpc} are used directly,
inspired by the duality between source and channel coding in the
erasure case \cite{it-quant-codes-graphs}.  In our previous work
\cite{ldgm-vq-globecom07}, LDGM degree distributions are instead
designed by directly fitting the EXIT curves under the \emph{erasure
  approximation} (EA), also known as the BEC (binary erasure channel)
approximation \cite{exit-model-ec-prop}.  Both methods perform well,
but they are heuristic in their analysis of decimation, and may thus
be suboptimal.

In this and the next section, we will give a detailed analysis on
degree distribution optimization of BP-based LDGM quantizers that
properly takes decimation into account, which should allow better MSE
performance to be attained.  Under the erasure approximation, we are
in effect designing an LDGM quantization code for the simpler
\emph{binary erasure quantization} problem and using it in MSE
quantization.\footnote{In this paper we only consider codes chosen
  randomly, through random edge assignment, from the LDGM code
  ensemble with a given degree distribution, therefore only the degree
  distribution is subjected to optimization, and we will not
  distinguish between codes and degree distributions.} Therefore, we
will first focus on BEQ in this section, and in
Section~\ref{sec:deg-opt-mse} the methods given here will be extended
to MSE quantization, with or without the erasure approximation.

\subsection{Binary Erasure Quantization}
The binary erasure quantization problem can be formulated as follows
\cite{it-quant-codes-graphs}.  The source sequence has the form
$\seq{y} \in \{0,1,*\}^n$, where ``$*$'' denotes erased positions and
occurs with probability $\epsilon$.  A binary code $\mathcal{U}$
consisting of $2^{nR}$ codewords $\seq{u} = \seq{u}(\seq{b}) \in
\{0,1\}^n$, each labeled by $\seq{b} \in \{0,1\}^{nR}$, is then
designed according to $\epsilon$ and the rate $R$.  For each
$\seq{y}$, the quantizer should find a codeword $\seq{u} \in
\mathcal{U}$ such that $y_j = u_j$ or $y_j = *$ for all
$j=1,\dotsc,n$, i.e.\ $\seq{u}$ agrees with $\seq{y}$ on all
non-erased positions.  The number of non-erased positions in a given
$\seq{y}$ is denoted by $\nne$, which is approximately $n(1-\epsilon)$
for large $n$.  Ideally $\nb = nR$ can be as small as this
$n(1-\epsilon)$, i.e.\ $R=1-\epsilon$, but in practice higher rates
are necessary.

Similar to \eqref{eq:qy}, $\qseqy(\seq{b})$ can be defined as
\begin{equation}
  \label{eq:qy-erasure}
  \qseqy(\seq{b}) = \prod_{j=1}^n q_{y_j}(u_j(\seq{b})), \quad
  q_{y_j}(u_j) =
  \begin{cases}
    1 & y_j = u_j \text{ or } * , \\
    0 & \text{otherwise},
  \end{cases}
\end{equation}
and the quantizer can equivalently find, for a given $\seq{y}$, some $\seq{b}$
such that $\qseqy(\seq{b}) > 0$ (which then equals 1).

When $\mathcal{U}$ is the codeword set of an LDGM code and $\seq{u} =
\seq{c} = \seq{b} \mat{G}$ as in Section~\ref{sec:binary},
$\qseqy(\seq{b})$ can be described by the factor graph in
Fig.~\ref{fig:binary-fg} as well, where each $\priprobx{c}{j}{c}$ is a
normalized version of $q_{y_j}(c)$, i.e.\ $\priprob{c}{j}$ is
$\constmsg{0}$, $\constmsg{1}$ or $\constmsg{*}$ if $y_j$ is
respectively $0$, $1$, $*$.  Apart from this difference in
$\priprob{c}{j}$, the algorithm in Fig.~\ref{alg:binary} with the
typical decimator can be used here for the purpose of analysis, though
the recovery algorithm in Section~\ref{sec:impact-dec-erasure} will be
necessary for good performance in practice.

The BEQ problem may alternatively be viewed as a set of linear
equations
\begin{equation}
  \label{eq:beq-eqs}
  \seq{b} \Gne = \yne
\end{equation}
over the binary field $\GF(2) = \{0,1\}$, where $\Gne$ and $\yne$ are the $\nne$ columns
of $\mat{G}$ and $\seq{y}$ that correspond to non-erased positions of
$\seq{y}$.  Denoting by $\nr$ the rank of $\Gne$, \eqref{eq:beq-eqs}
then has $2^{\nb-\nr}$ solutions for $2^{\nr}$ of the $2^{\nne}$ possible
$\yne$'s, and for other $\yne$'s there is no solution at all.

We first assume that \eqref{eq:beq-eqs} has a set $\mathcal{B}$ of
$2^{\nb-\nr}$ solutions, then $\pseqy(\seq{b}) = 2^{-(\nb-\nr)}
\qseqy(\seq{b})$ is a probability distribution for $\seq{b}$ that is
uniform over $\mathcal{B}$\@.  Using this $\pseqy(\seq{b})$, similar to
the BP-derived extrinsics $\extprob{b}{i}$, the true extrinsic
probabilities $\trueextprob{b}{i}$ of $b_i$ can now be defined as
\begin{equation}
  \label{eq:trueextprob-b-beq}
  \trueextprobx{b}{i}{b} = \pseqy(b_i = b \condmid \seq{b}_{\mathcal{F} \excluding{i}} = \seq{b}^*_{\mathcal{F} \excluding{i}}),
  \quad i = 1, \dotsc, \nb,
\end{equation}
which depends on the set $\mathcal{F}$ of decimated bits and their
decimated values $\seq{b}^*_{\mathcal{F}}$.  Note that
$\trueextprob{b}{i}$ can only be $\constmsg{0}$, $\constmsg{1}$, or
$\constmsg{*}$: it is $\constmsg{b}$ if all solutions with
$\seq{b}_{\mathcal{F} \excluding{i}} = \seq{b}^*_{\mathcal{F}
  \excluding{i}}$ have $b_i = b \in \{0,1\}$, and otherwise there must
be the same number of solutions with $b_i = 0$ and with $b_i = 1$,
making $\trueextprob{b}{i} = \constmsg{*}$.

Without loss of generality, the typical decimator can be assumed to
decimate in the order of $b_1, b_2, \dotsc, b_{\nb}$.  Decomposing
$\pseqy(\seq{b})$ into
\begin{equation}
  \label{eq:qy-decomp-beq}
  \pseqy(\seq{b}) = \pseqy(b_1) \pseqy(b_2 \condmid b_1) \dotsm \pseqy(b_{\nb} \condmid \seq{b}_1^{\nb-1}),
\end{equation}
each factor $\pseqy(b_i \condmid \seq{b}_1^{i-1})$ is then the
$\trueextprob{b}{i}$ after the decimation of $\seq{b}_1^{i-1}$ into
$\seq{b}_1^{i-1,*}$.  We therefore construct the fictitious \emph{true
  typical decimator} (TTD), which is just like the TD except that
decimation of $b_i$ is done according to $\trueextprob{b}{i}$ rather
than $\extprob{b}{i}$.  Moreover, the TTD shares the source of
randomness with the TD, so decimation is still done in the order of
$b_1, \dotsc, b_{\nb}$, and each $b_i$ is decimated to the same value
except to account for the difference between $\extprob{b}{i}$ and
$\trueextprob{b}{i}$.\footnote{For example, the TD and the TTD can use
  the same i.i.d.\ random sequence $\tau_1, \tau_2, \dotsc,
  \tau_{\nb}$ in decimation with each $\tau_i$ uniformly distributed
  in $[0,1)$, and each $b_i$ is decimated to 0 in the TD if $\tau_i <
  \extprobx{b}{i}{0}$ and in the TTD if $\tau_i <
  \trueextprobx{b}{i}{0}$, and to 1 otherwise.  In this way, the
  decimation results are always the same if $\extprob{b}{i} =
  \trueextprob{b}{i}$, and are rarely different if $\extprob{b}{i}$
  and $\trueextprob{b}{i}$ are close.} The TTD in effect samples a
$\seq{b}^*$ according to the probability distribution
$\pseqy(\seq{b})$, so it must yield a random solution $\seq{b}^* \in
\mathcal{B}$\@.  If, for every $i=1,\dotsc,\nb$, the TD at the time of
$b_i$'s decimation has $\extprob{b}{i} = \trueextprob{b}{i}$, then it
will run synchronously with the TTD and yield the same solution in
$\mathcal{B}$\@.  Otherwise, e.g.\ if $\extprob{b}{i} = \constmsg{*}$
and $\trueextprob{b}{i} = \constmsg{0}$ for some $i$, then the TD
might decimate $b_i$ to 1, which will eventually result in a
contradiction.  Therefore, our first requirement for TD to find a
solution to \eqref{eq:beq-eqs} is that BP must compute the correct
extrinsic probabilities after enough iterations, which is hence called
the \emph{extrinsic probability condition}.

How, then, to ensure the existence of solutions to \eqref{eq:beq-eqs}
for any $\yne$?  We may define $\Qseqy$ with \eqref{eq:Qy} which, for
each $\yne$, gives the number of solutions to \eqref{eq:beq-eqs} and
is $2^{\nb-\nr}$ for $2^{\nr}$ $\yne$'s and zero for the rest.
$\Qseqy$, if normalized by $2^{-\nb}$, is again a uniform distribution
over these $2^{\nr}$ $\yne$'s.  We then require $\nr = \nne$, making
$\Qseqy$ a uniform distribution over all $2^{\nne}$ possible $\yne$'s,
so that the BEQ problem have $2^{\nb-\nne}$ solutions for any
$\yne$.  This is the other condition for BEQ to be always solvable by
the TD, hence called the \emph{equi-partition condition}.

For $n \to\infty$, the two conditions above are now suitable for
analysis with density evolution methods, which in the BEQ case
can be accurately done with EXIT charts, as will be discussed in the
following subsections.

\subsection{Fixed Points and EXIT Curves}
\label{sec:exit-erasure}
\begin{figure*}[!t]
  \centering
  \subfigure[EBP curves of $(4,2)$ regular LDGM code]{\label{fig:ebp42}\includegraphics{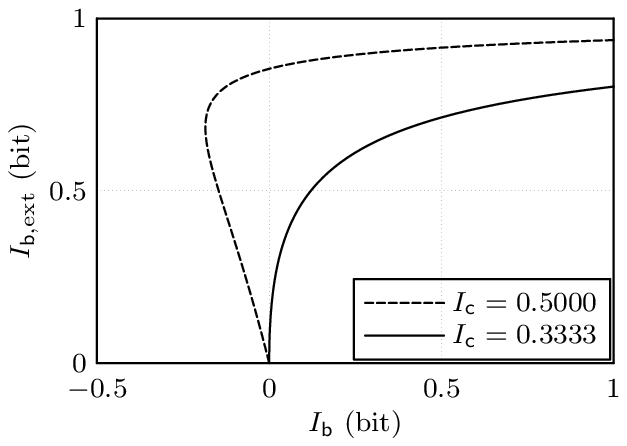}}%
  \hspace{1mm}%
  \subfigure[EBP curves of $(5,3)$ regular LDGM code]{\label{fig:ebp53}\includegraphics{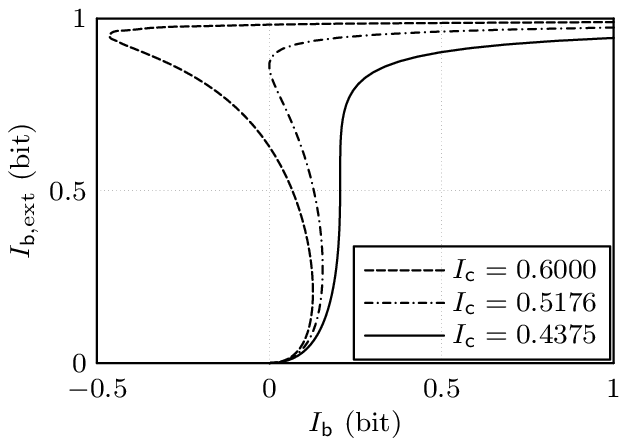}}%
  \hspace{1mm}%
  \subfigure[Comparison of EBP, BP and MAP]{\label{fig:ebp-map}\includegraphics{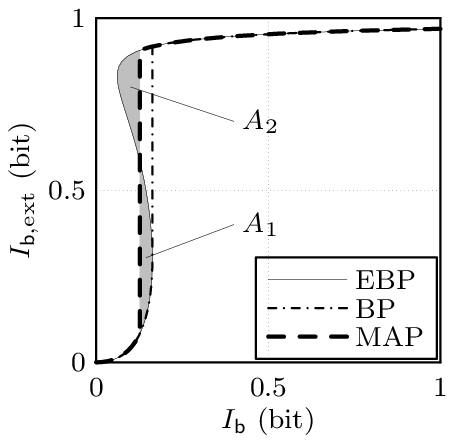}}%
  \caption{The EBP curves of some $(\db,\dc)$ regular LDGM codes, in
    which all \node{b}-nodes have right-degree $\db$ and all
    \node{c}-nodes have left-degree $\dc$.  (a) The $(4,2)$ regular
    code has rate $R=0.5$ and monotonicity threshold $\Icthr = 1/3$.
    For $\Icthr < \Info{c} \le R$, part of the EBP curve lies in the
    $\Info{b}<0$ half-plane, although $\Info{b}$ is monotonically
    increasing once it becomes positive.  This implies a violation of
    the equi-partition condition.  For $\Info{c} < \Icthr$, the
    monotonicity conditions are satisfied.  (b) The $(5,3)$ regular
    code has rate $R=0.6$ and monotonicity threshold $\Icthr = 7/16 =
    0.4375$.  When $\Info{c}$ is reduced to 0.5176, the EBP curve no
    longer extends into the $\Info{b}<0$ half-plane, but it is still
    not monotonic until $\Info{c}$ is further reduced to $\Icthr$.
    (c) A comparison of the EBP, BP and MAP curves of the $(5,3)$
    regular code at $\Info{c} = 0.5$, assuming that the results in
    \cite{maxwell-constr} remain true.  The area $A_1$ to the right of
    the MAP curve represents the $b_i$'s whose $\trueextprob{b}{i} =
    \constmsg{b_i^*}$ but $\extprob{b}{i} = \constmsg{*}$ and thus
    violate the extrinsic probability condition.  That is, the values
    of these bits are determined by previous decimation results but
    not available from BP at the time; they are apparently ``guesses''
    until they are ``confirmed'' by an equal number of equations
    encountered later represented by $A_2$.  Here $A_2 = A_1$, which
    intuitively means that all confirmations constrain earlier guesses
    rather than $\yne$, so the equi-partition condition is satisfied.
    This is not the case for e.g.\ the $(4,2)$ regular code at
    $\Info{c} = 0.5$ in (a): there the MAP and the BP curves overlap
    with the EBP curve in the $\Info{b}\ge 0$ half-plane but does not
    extend to the left, and the area between the EBP curve and the
    $\Info{b} = 0$ axis represent ``confirmations'' that, having no
    earlier guesses, must be satisfied by $\yne$, therefore the
    equi-partition condition is not satisfied.}%
  \label{fig:ebp}%
\end{figure*}

We use \node{b}-regular, \node{c}-irregular LDGM codes for
quantization as suggested by the LDGM-LDPC duality in
\cite{it-quant-codes-graphs}.  Let $\db$ be the right-degree of all
\node{b}-nodes, and denote $\wc{d}$ as the fraction of \node{c}-nodes
with left-degree $d$ and $\vc{d}=d \wc{d}/(R \db)$ as the
corresponding fraction of edges.

Assuming that the BEQ problem does have solutions for the given
$\seq{y}$, with the one found by TTD denoted $\seq{b}^*$ and
$\seq{u}^* = \seq{c}^* = \seq{b}^* \mat{G}$.  Assuming additionally
that our quantizer based on TD has decimated a fraction $\Info{b}$ of
the \node{b}-nodes and has so far maintained synchronization with the
TTD in decimation decisions, $\seq{b}^*$ is then consistent with the
current priors and can serve as the \emph{reference codeword}: all
$\msg{bc}{ij}$ and $\msg{cb}{ji}$'s, with $b_i$ decimated or not, must
be either $\constmsg{b_i^*}$ or $\constmsg{*}$ and never contradict
the reference codeword.  Denoting by e.g.\ $\Info{bc}$ the average
mutual information (MI) in the $\msg{bc}{ij}$'s from the previous
iteration about their respective reference values $b_i^*$, which in
this case is simply the fraction of $\msg{bc}{ij}$ that equals
$\constmsg{b_i^*}$,\footnote{In this paper, all such MIs and EXIT
  curves are also averaged over the LDGM code ensemble with the given
  degree distribution.  Assuming that relevant concentration results
  hold, for $n\to\infty$ we can also talk about the convergence
  behavior of a specific code using these ensemble-averaged MIs.} and
using the usual fact that the factor graph becomes locally tree-like
with high probability as $n\to\infty$, we can find the EXIT curve
relating the input $\Info{bc}$ for the \node{c}-nodes and their output
$\Info{cb}$, hence called the \node{c}-curve, to be
\begin{equation}
  \label{eq:exit-binary-c}
  \Info{cb} = \Info{c} \sum_d \vc{d} \Info{bc}^{d-1},
\end{equation}
where $\Info{c}$ is the MI of the $\priprob{c}{j}$'s, in this case
$1-\epsilon$.  The \node{b}-curve relating $\Info{cb}$ and the output
$\Info{bc}$ from the \node{b}-nodes (denoted by $\nextInfo{bc}$ as it
refers to the next iteration) is likewise
\begin{equation}
  \label{eq:exit-binary-b}
  \nextInfo{bc} = 1-(1-\Info{b})(1-\Info{cb})^{\db-1}.
\end{equation}

To analyze the extrinsic probability condition, it is necessary to
look into the behavior of BP's fixed points, which are characterized by
the EBP EXIT curve first proposed in
\cite{maxwell-constr} for LDPC decoding over BEC.  The EBP curve
relates the \textit{a priori} MI $\Info{b}$ at fixed points (i.e.\ the
$\Info{b}$ making $\nextInfo{bc} = \Info{bc}$),
\begin{equation}
  \label{eq:Ib}
  \Info{b} = 1-(1-\Info{bc}) / (1-\Info{cb})^{\db-1},
\end{equation}
and the extrinsic MI in the $\extprob{b}{i}$'s, i.e.\ the fraction of
$\extprob{b}{i}$ that are $\constmsg{b_i^*}$ rather than $\constmsg{*}$,
\begin{equation}
  \label{eq:Ibext}
  \Ibext = 1-(1-\Info{cb})^{\db},
\end{equation}
as $\Info{bc}$ goes from 0 to 1 and $\Info{cb}$ given by
\eqref{eq:exit-binary-c}.  Fig.~\ref{fig:ebp} shows the EBP curves of
some codes for example.  Note that $\Ibext$ is always non-negative and
monotonically increasing with $\Info{bc}$, but $\Info{b}$ in
\eqref{eq:Ib} is not necessarily so.

Every crossing the EBP curve makes with a constant-$\Info{b}$ vertical
line corresponds to a fixed point of BP at this $\Info{b}$, and when
the number of iterations $L\to\infty$, it is clear that BP will follow
the minimum-$\Ibext$ fixed point as $\Info{b}$ goes from 0 to 1,
forming the BP EXIT curve in \cite{maxwell-constr}.  The MAP (maximum
\textit{a posteriori} probability) EXIT curve in
\cite[Definition~2]{maxwell-constr} is simply the relationship between
the fraction $\Info{b}$ of decimated bits and the average true
extrinsic MI in the $\trueextprob{b}{i}$'s, as is evident from
\cite[Theorem~2]{maxwell-constr}, where the random vector $\seq{b}$
(currently taking value $\seq{b}^*$) is the $X$ in
\cite{maxwell-constr}, the \node{b}-priors $\priprob{b}{i}$ are the
BEC output $Y$, and the \node{c}-priors $\priprob{c}{j}$ (or
$\seq{y}$) are the additional observation $\Omega$.

\begin{figure}[!t]
  \centering
  \includegraphics{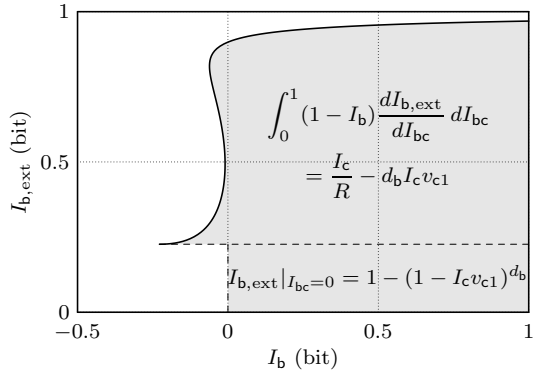}%
  \caption{The area under the EBP curve (the thick solid curve) when
    $\vc{1}>0$.  In such cases the EBP curve does not start from
    $(0,0)$, and we define the area below it as the total area of the
    two gray regions, whose respective areas are shown in the figure.
    Note that the lower area $1-(1-\Info{c}\vc{1})^{\db}$ is smaller
    than $\db\Info{c}\vc{1}$, so the total area is smaller than
    $\Info{c}/R$, but is very close to it in the codes we will
    encounter since $\db\Info{c}\vc{1}$ is at most 0.03 or so.}%
  \label{fig:ebp-area}%
\end{figure}

Interestingly, our BEQ problem is now very similar to the LDPC
decoding problem on BEC considered in \cite{maxwell-constr}, as both
involve a system of linear equations over $\GF(2)$ that has at least
one solution ($\seq{b}^*$ for LDGM-based BEQ and the transmitted
codeword for LDPC-over-BEC) consistent with all previous
guesses.\footnote{The only difference is that the number of equations
  in BEQ, $\nne$, is random whereas in LDPC decoding over BEC it is
  always the number of check nodes.  This should not be essential
  though.}  In particular, the area above the MAP curve is $H(\seq{b}
\condmid \seq{y})/\nb$ \cite[Theorem~1]{maxwell-constr}, with
$H(\seq{b} \condmid \seq{y})$ being the entropy of the aforementioned
$\pseqy(\seq{b})$; under the equi-partition condition
\eqref{eq:beq-eqs} should have $2^{\nb - \nne} \asympteq
2^{\nb(1-\Info{c}/R)}$ solutions, so this area is $1-\Info{c}/R$, and
the area below the MAP curve is $\Info{c}/R$, while if the
equi-partition condition is violated \eqref{eq:beq-eqs} will have more
solutions for the current $\seq{y}$ (and none for many other
$\seq{y}$'s), and the MAP curve will have a smaller area below it.  On
the other hand, the area below the EBP curve can be computed directly
from \eqref{eq:Ib} and \eqref{eq:Ibext}; this area is also
$\Info{c}/R$ if $\vc{1}=0$, and when $\vc{1}>0$ it is defined as the
total gray area in Fig.~\ref{fig:ebp-area}, which is smaller than but
close to $\Info{c}/R$.

If the results in \cite{maxwell-constr} on the relationship between
MAP, BP and EBP curves remain true, these three curves should be given
by Fig.~\ref{fig:ebp-map}.  Heuristic arguments below the figure
suggest that the extrinsic probability and equi-partition conditions
above for the TD to solve the BEQ problem are satisfied, with a
vanishing fraction of exceptions as $n\to\infty$, if and only if the
EBP curve satisfies the following \emph{monotonicity
  conditions}:\footnote{Note that this has nothing to do with
  monotonicity with respect to a class of channels, which appears
  often in LDPC literature \cite{cap-ldpc-msgpassing-dec}.}
\begin{align}
  \label{eq:mono-cond1}
  \Info{b} |_{x=0} &\ge 0, \\
  \label{eq:mono-cond2}
  \frac{d\Info{b}}{dx} &\ge 0, \quad x\in [0,1],
\end{align}
where $\Info{b}$ is viewed as a function \eqref{eq:Ib} of
$x=\Info{bc}$.  We now prove this using similar methods to
\cite{maxwell-constr}.

\emph{Necessity.}  The extrinsic probability condition means that
$\extprob{b}{i} = \trueextprob{b}{i}$ for all but a vanishing fraction
of $i \in \{1,\dotsc,\nb\}$ at any $\Info{b}$ after enough iterations,
which implies that the two have at least the same average MI, i.e.\
the BP curve coincides with the MAP curve, the area below which is in
turn $\Info{c}/R$ under the equi-partition condition.  Since the BP
curve follows the minimum fixed points on the EBP curve, and the area
under the latter is at most $\Info{c}/R$, the two curves must coincide
as well, which immediately leads to \eqref{eq:mono-cond1} and
\eqref{eq:mono-cond2}.

\emph{Sufficiency.}  Under \eqref{eq:mono-cond1} and
\eqref{eq:mono-cond2}, the BP curve obviously coincides with the EBP
curve, and since \eqref{eq:mono-cond1} implies $\vc{1} = 0$, the area
below them is $\Info{c}/R$.  BP can never give any information not
implied by $\seq{y}$ and previous decimation results, i.e.\ for any
$i$ we have either $\extprob{b}{i} = \trueextprob{b}{i}$ or
$\extprob{b}{i} = \constmsg{*}$, so the MAP curve cannot lie below the
BP curve and the area below it is at least $\Info{c}/R$.  We have also
shown that the area below the MAP curve is at \emph{most}
$\Info{c}/R$, therefore equality must hold and the equi-partition
condition is satisfied.  Now that the MAP and BP curves also coincide,
for any $\Info{b}$ the $\extprob{b}{i}$'s will have the nearly the
same average MI as the $\trueextprob{b}{i}$'s (with the difference
vanishing after many iterations when $n\to\infty$), and since any
$\extprob{b}{i} \ne \trueextprob{b}{i}$ implies $\extprob{b}{i} =
\constmsg{*}$ and $\trueextprob{b}{i} = \constmsg{b_i^*}$ and thus a
difference in MI, it can only occur for a vanishingly small fraction
of $i$'s.  Therefore the extrinsic probability condition also
holds.\qed

We will see below that the monotonicity conditions are more easily
satisfied for smaller $\Info{c}$, so for a given code, we can define
the maximum $\Info{c}$ that satisfies them as the \emph{monotonicity
  threshold}, denoted by $\Icthr$.  This is the maximum $(1-\epsilon)$
for which the BEQ problem can, in an asymptotic sense, be solved by
the TD.  The same performance is expected for the greedy decimator,
since in BEQ it is basically identical to TD.

It should be noted that the monotonicity conditions are sufficient for
the extrinsic probability and equi-partition conditions only in the
sense that the \emph{fraction} of violations approaches zero as the
block size $n$ and the iteration count $L$ go to infinity.
Therefore, in practice some contradictions will occur in the TD, and
some equations in \eqref{eq:beq-eqs} will be unsatisfied.  In
Section~\ref{sec:impact-dec-erasure}, we will propose a method to deal
with such contradictions, such that the number of unsatisfied
equations remains a vanishing fraction of $n$.

\subsection{Optimization of the Monotonicity Threshold}
\label{sec:mono-opt-beq}
We can now optimize the degree distribution so that $\Icthr$ is
maximized and approaches its ideal value $R$.

From \eqref{eq:Ib} and \eqref{eq:exit-binary-c}, it is easy to show
that the condition \eqref{eq:mono-cond1} is equivalent to $\vc{1}=0$,
i.e.\ there are no degree-1 \node{c}-nodes.  As for the second
condition \eqref{eq:mono-cond2}, differentiating \eqref{eq:Ib} with
respect to $x=\Info{bc}$ gives (hence we denote $y=1-\Info{cb}$)
\begin{equation}
  \label{eq:Ib_x}
  \frac{d\Info{b}}{dx} = y^{-\db} \left( y - \Info{c} \cdot (\db-1) (1-x) \sum_d (d-1)\vc{d} x^{d-2} \right).
\end{equation}
Making \eqref{eq:Ib_x} nonnegative, we get
\begin{equation}
  \label{eq:s-constr-ml}
  \Info{c} \le \frac{1}{s(x)}, \quad x \in [0,1]
\end{equation}
where
\begin{equation}
  \label{eq:s}
  s(x) = \sum_d \vc{d} x^{d-1} + (\db - 1)(1-x) \sum_d (d-1) \vc{d} x^{d-2}.
\end{equation}
Therefore, the monotonicity threshold is
\begin{equation}
  \label{eq:Icthr}
  \Icthr = \left( \max_{x \in [0,1]} s(x) \right)^{-1},
\end{equation}
and it can be maximized by solving the following optimization problem
over $\smax = 1/\Icthr$ and $\vc{d}$, $d = 2, 3, \dotsc$:
\begin{equation}
  \label{eq:binary-opt}
  \begin{split}
    \text{minimize}\ & \smax \\
    \text{subject to}\ & s(x) \le \smax, \quad \forall x \in [0,1], \\
    & \sum_d \vc{d} = 1,\quad \sum_d \frac{\vc{d}}{d}=\frac{1}{R\db},\\
    & \vc{d}\ge 0, \quad \forall d.
  \end{split}
\end{equation}
In practice, the $s(x) \le \smax$ constraint is applied to a number of
discrete $x$'s (1000 values uniformly spaced over $[0,1]$ seem to
suffice), and the set of \node{c}-degrees is chosen to be the
exponential-like sequence
\begin{equation}
  \label{eq:degree-set}
  \mathcal{D} = \{ d_k \mid k = 1,2,\dotsc,\cardinal{\mathcal{D}},\
  d_1 = 2,\ d_{k+1} = \lceil \beta \cdot d_k \rceil \},
\end{equation}
where we set $\beta = 1.1$, and $\cardinal{\mathcal{D}}$ is made large
enough not to affect the final result.  Since $s(x)$ is linear
in $\vc{d}$, \eqref{eq:binary-opt} then becomes a linear programming
problem that is easily solved using usual numerical methods.

In Table~\ref{tab:impact-db-beq} we list the optimal $\Icthr$ achieved
at different values of $\db$ as well as the resulting maximum
\node{c}-degree $\dcmax$.  We see that $\Icthr$ approaches its ideal
value $R$ exponentially fast with the increase of $\db$, but the
necessary $\dcmax$ also increases exponentially.  Due to the problem's
simplicity, it is probably not difficult to prove this.

\begin{table}[!t] 
  \centering
  \caption{Impact of $\db$ in BEQ ($R=\unit[0.4461]{b/s}$)}
  \label{tab:impact-db-beq}
  \begin{tabular}{cccccccc}
    \toprule
    $\db$ & 6 & 7 & 8 & 9 & 10 & 11 \\ 
    \midrule
    $\Icthr$ & 0.4110 & 0.4294 & 0.4376 & 0.4416 & 0.4437 & 0.4448 \\
    $\dcmax$ & 6 & 10 & 19 & 37 & 70 & 127 \\
    \bottomrule
  \end{tabular}
\end{table}

\section{Degree Distribution Optimization for MSE Quantization}
\label{sec:deg-opt-mse}
It is well known that long LDPC channel codes can be effectively
analyzed and designed using density evolution methods, not only over
BEC but also over general binary-input symmetric channels
\cite{design-cap-approaching-ldpc}.  Such methods are also useful for
LDGM quantization codes, but their application is not as
straightforward as the LDPC case due to the stateful nature of
decimation, its use of extrinsic probabilities (which is available in
DE only for the final iteration, at the root node of the tree-like
neighborhood), and the lack of a ``natural'' reference codeword in
quantization as is available in channel decoding.

In Section~\ref{sec:deg-opt-erasure}, we have solved these problems in
the BEQ case by introducing the TTD: the result of TTD is used as the
reference codeword, with which decimation can be modeled by the priors
$\priprob{b}{i}$ with a single parameter $\Info{b}$, and the extrinsic
probabilities at each decimation step can be analyzed separately.  In
this section, we will extend this TTD-based method to MSE quantization
so that code optimization can likewise be carried out with DE.  When
the erasure approximation is used in DE, we obtain the same optimized
degree distributions for BEQ, but we can also avoid EA and do a more
accurate optimization using the quantized DE method a la
\cite{design-cap-approaching-ldpc, on-design-ldpc-45e-4-dB}.

\subsection{Density Evolution in MSE Quantization}
\label{sec:de-mse-intro}
Without loss of generality, suppose the source sequence $\seq{y} \in
[0,m)^n$, which can, as in Section~\ref{sec:basics}, be decomposed
into $\seq{y} = \seq{\yt} + \seq{a}$, where $\seq{\yt} \in [0,1)^n$ is
assumed to be typical with respect to the uniform distribution over
$[0,1)$, and $\seq{a} \in \{0,1,\dotsc,m-1\}^n$.  For a fixed
$\seq{\yt}$, we may define, similar to \eqref{eq:qy},
\begin{equation}
  \label{eq:qba}
  q(\seq{b};\seq{a}) = q_{\seqs{\yt}+\seqs{a}}(\seq{b})
  = e^{-t \norm{(\seqs{\yt} + \seqs{a} - \seqs{u}(\seqs{b}))\bmodIn}^2},
\end{equation}
which can be regarded as a probability distribution over $\seq{b}$ and
$\seq{a}$ after normalization.  With $\Qseqyta$ defined in
\eqref{eq:Qy}, this distribution can be decomposed into
\begin{equation}
  \label{eq:qba-decomp}
  q(\seq{b};\seq{a}) = \Qsum \cdot p(\seq{b};\seq{a}) = \Qsum \cdot P(\seq{a}) \pseqa(\seq{b}),
\end{equation}
where
\begin{equation}
  P(\seq{a}) = \frac{Q_{\seqs{\yt}+\seqs{a}}}{\Qsum}, \quad \pseqa(\seq{b}) = \frac{q(\seq{b};\seq{a})}{Q_{\seqs{\yt}+\seqs{a}}}
\end{equation}
are respectively probability distributions over $\seq{a}$ and over $\seq{b}$ conditioned on $\seq{a}$, and
\begin{align}
  \Qsum &= \sum_{\seqs{a}} Q_{\seqs{\yt}+\seqs{a}} = m^n \avga{\Qseqyta}
  \label{eq:Qsum0}
  = 2^{\nb} \prod_{j=1}^n Q_{\yt_j} \\
  \label{eq:Qsum}
  &\asympteq 2^{\nb} \exp \left( n \int_0^1 \ln Q_{\yt} \, d\yt \right)
\end{align}
using \eqref{eq:Qyt} and the typicality of $\seq{\yt}$.

The quantization of $\seq{y} = \seq{\yt} + \seq{a}$ is equivalent to
finding a $\seq{b}$ for a given $\seq{a}$ that (approximately) maximizes
$q(\seq{b};\seq{a})$.  Again, we consider the typical decimator since
the greedy decimator is difficult to analyze, and the order of
decimation is assumed to be $b_1, b_2, \dotsc, b_{\nb}$ without loss
of generality.  With the true extrinsic probabilities
$\trueextprob{b}{i}$ of $b_i$ defined like
\eqref{eq:trueextprob-b-beq} according to $\pseqa(\seq{b})$, the decomposition
\begin{equation}
  \label{eq:pa-decomp}
  \pseqa(\seq{b}) = \pseqa(b_1) \pseqa(b_2 \condmid b_1) \dotsm \pseqa(b_{\nb} \condmid \seq{b}_{1}^{\nb-1})
\end{equation}
again has each factor $\pseqa(b_i = b \condmid \seq{b}_1^{i-1} =
\seq{b}_1^{i-1,*})$ equaling the $\trueextprobx{b}{i}{b}$ when
previous $\seq{b}_1^{i-1}$ has been decimated into
$\seq{b}_1^{i-1,*}$.  The TTD is then the decimator similar to TD but
using $\trueextprob{b}{i}$ instead of $\extprob{b}{i}$, so it yields
decimation result $\seq{b}^*$ with probability $\pseqa(\seq{b}^*)$,
and the TD attempts to synchronize with it.

In addition, $\seq{a}$ can be viewed as the product of a \emph{source
  generator} before quantization but after $\seq{\yt}$ is determined.
This can be shown more clearly on the equivalent factor graph
Fig.~\ref{fig:binary-fg-a}.  All priors on $a_j$ and $b_i$,
$\priprob{a}{j}$ and $\priprob{b}{i}$, being initially $\constmsg{*}$,
the source generator first determines $a_1,\dotsc,a_n$ by setting
$\priprob{a}{j}$ to hard decisions, and the quantizer then determines
$b_1,\dotsc,b_{\nb}$.  In the source generation process, BP can be run
to yield the extrinsics $\extprob{a}{j}$, and the true extrinsic
probabilities $\trueextprob{a}{j}$ can likewise be defined with
$P(\seq{a})$.  Similar to the TTD, we define the \emph{true typical
  source generator} (TTSG) as one generating each $\seq{a}$ with
probability $P(\seq{a})$.  Since
\begin{equation}
  \label{eq:P-decomp}
  P(\seq{a}) = P(a_1) P(a_2 \condmid a_1) \dotsm P(a_n \condmid \seq{a}_1^{n-1}),
\end{equation}
and each factor $P(a_j = a \condmid \seq{a}_1^{j-1})$ is the
$\trueextprobx{a}{j}{a}$ when $\seq{a}_1^{j-1}$ has been determined, the
TTSG simply sets each $a_j = a$ with probability
$\trueextprobx{a}{j}{a}$.  In reality, all $2^n$ possible values of
$\seq{a}$ are equally likely to occur, so we can safely assume that
$\seq{a}$ comes from the TTSG if and only if $P(\seq{a})$ is a uniform
distribution, that is, each $\trueextprob{a}{j}$ must be $\constmsg{*}$
when $\seq{a}_1^{j-1}$ has been determined.

When both the TTSG and the TTD are used, each possible
$(\seq{b},\seq{a})$ is generated with probability
$p(\seq{b};\seq{a})$.  Define $\seq{\ut} = (\seq{u}(\seq{b}) -
\seq{a}) \bmod m$, each $\seq{\ut}$ then corresponds to $2^{\nb}$
$(\seq{b},\seq{a})$'s, all of which having the same
\begin{equation}
  p(\seq{b};\seq{a}) = \frac{1}{\Qsum} e^{-t\norm{(\seqs{\yt} - \seqs{\ut})\bmodIn}^2},
  \label{eq:pba}
\end{equation}
and the total probability of generating $\seq{\ut}$ becomes
\begin{align}
  p(\seq{\ut}) &= 2^{\nb} p(\seq{b};\seq{a}) 
  = \prod_{j=1}^n \frac{1}{Q_{\yt_j}} e^{-t (\yt_j - \ut_j)\bmodI^2} \\
  \label{eq:prob-ut}
  &= \prod_{j=1}^n \pz((\yt_j - \ut_j)\bmodI) = \prod_{j=1}^n \pz(z_j)
\end{align}
from \eqref{eq:Qsum0} and \eqref{eq:pz}, noting that $\seq{z} =
(\seq{y} - \seq{u})\bmodIn = (\seq{\yt} - \seq{\ut})\bmodIn$.
Eq.~\eqref{eq:prob-ut} shows that $\ut_j$ can be viewed as i.i.d.\
samples conditioned on $\yt_j$ with probability density
$p(\ut\condmid\yt) = \pz((\yt-\ut)\bmodI)$, so for $n\to\infty$
$\seq{\ut}$ will be strongly typical according to this conditional
distribution with high probability, and the quantization error
$\seq{z}$ is likewise strongly typical with respect to $\pz(z)$, so
the resulting MSE is $P_t$.

To achieve this $P_t$ with the TD, again we have
\begin{itemize}
\item \emph{extrinsic probability condition}: $\extprob{b}{i}$ must
  be close to $\trueextprob{b}{i}$ when decimating each $b_i$, so that the
  TD can synchronize with the TTD;
\item \emph{equi-partition condition}: $P(\seq{a})$ must be a uniform
  distribution so that the use of TTSG here matches reality and does
  not pick ``easy'' source sequences with large $P(\seq{a})$ too
  often.
\end{itemize}
It may be interesting to note the relationship between the two
conditions and the two inequalities in \eqref{eq:lb-avga}.

Similar to the BEQ case, we assume that $\seq{y}$ is generated by the
TTSG and use TTD's final result $\seq{b}^*$ and the corresponding
$\seq{u}^* = \seq{c}^* = \seq{b}^* \mat{G}$ as the \emph{reference
  codeword}, then each $\priprob{b}{i}$, $\extprob{b}{i}$,
$\msg{bc}{ij}$ and $\msg{cb}{ji}$ have reference value $b_i^*$ and
each $\priprob{c}{j}$ has reference value $c_j^*$, and DE can be
carried out with respect to these reference values to analyze the
above two conditions.  The density of $\priprob{c}{j}$ (actually that
of $\priprobx{c}{j}{c_j^*}$) can be obtained from \eqref{eq:priprobs}
using the strong typicality of $\seq{z} = (\seq{y} -
\seq{u}^*) \bmodIn$ with respect to $\pz(z)$.  Furthermore,
assuming that the TD had been synchronized with the TTD in all
previous decimation decisions, $\priprob{b}{i}$ is then
$\constmsg{b_i^*}$ at the decimated positions (whose fraction is
denoted $\Info{b}$ as before) and $\constmsg{*}$ elsewhere.  We thus
have all the necessary information for DE.

In BEQ, we have found \eqref{eq:mono-cond1} and \eqref{eq:mono-cond2}
to be sufficient and necessary for the equi-partition and extrinsic
probability conditions to be satisfied with a vanishing fraction of
exceptions.  According to the definition of the EBP curve,
\eqref{eq:mono-cond2} and \eqref{eq:mono-cond1} correspond to two
properties of the code and $\Info{c}$ in DE:
\begin{itemize}
\item Starting from any $\Info{b} \in [0,1]$, DE converges to a unique
  fixed point regardless of the initial message density, provided that
  this initial density is intuitively ``consistent'', i.e.\ free of
  contradictions and not over- or
  under-confident;\footnote{\label{fn:symmetry}For binary quantization
    codes, this consistency can be defined rigorously as the symmetry
    condition of a message density in
    \cite[Sec.~III-D]{design-cap-approaching-ldpc}.  In BEQ, symmetry
    with respect to $\seq{b}^*$ of e.g.\ the density of $\msg{bc}{ij}$
    means that each $\msg{bc}{ij}$ is either $\constmsg{b_i^*}$ or
    $\constmsg{*}$ but never the opposite ``sure'' value (which would
    indicate a contradiction).  In MSE quantization, it means that,
    with $\msg{bc}{ij}$ being a randomly chosen \node{b}-to-\node{c}
    message, the probability density of $\msgx{bc}{ij}{b_i^*}$ at $p$
    and at $1-p$ have ratio $p : (1-p)$ for any $p \in [0,1]$.  All
    priors have symmetric densities when using binary codes, and the
    symmetry of the initial message density will thus be maintained
    throughout the DE process.  The symmetry condition is not
    necessarily true in non-binary cases, so we keep using the term
    ``consistency'' for generality.}
\item The fixed point at $\Info{b} = 0$ is at $\Ibext = 0$,
  corresponding to both $\msg{bc}{ij}$'s and $\msg{cb}{ji}$'s being
  all-$\constmsg{*}$.
\end{itemize}
We conjecture that these properties, which are again called the
\emph{monotonicity conditions}, are sufficient and necessary for MSE
quantization as well.

Proving this equivalence rigorously appears
difficult.\footnote{\label{fn:map-exit}The MAP EXIT curve can use
  basically the same definition \cite[Definition~2]{maxwell-constr};
  the area theorem \cite[Theorem~1]{maxwell-constr} still holds
  because only the $\Omega$ there is different, while the $Y$ there,
  corresponding to the $\priprob{b}{i}$'s, can still be viewed as BEC
  outputs.  The area below the MAP curve is therefore $1 - H(\seq{b}
  \condmid \seq{y})/\nb$, where $H(\seq{b} \condmid \seq{y})$ is the
  entropy of the distribution $\pseqa(\seq{b})$, and this area is
  again $\Info{c}/R$ under the equi-partition condition using
  \eqref{eq:pba}.  The EBP curve can also be obtained through DE,
  although its unstable branches may require tricks similar to
  \cite[Sec.~VIII]{gen-area-theorem-conseq} to find; but we no longer
  know the area below it.  More importantly, the ``erasure''
  relationship $\extprob{b}{i} = \trueextprob{b}{i}$ or
  $\extprob{b}{i} = \constmsg{*}$ in BEQ is no longer true, so it is
  difficult to relate the average MIs to the closeness of individual
  $\extprob{b}{i}$ and $\trueextprob{b}{i}$'s, which was essential in
  our BEQ analysis.}  We can, however, provide the following heuristic
argument.  For any number of iterations $l$, when $n$ is sufficiently
large, a randomly selected node $\node{b}_i$ will likely have a
tree-like neighborhood in the factor graph within depth $2l$.  If DE
has a unique fixed point, for sufficiently large $l$ the message
density after $l$ iterations no longer depends much on the initial
message density from the un-tree-like part of the factor graph, so the
resulting $\extprob{b}{i}$'s from BP, which is accurate for a
tree-like factor graph, should be mostly accurate here.\footnote{The
  un-tree-like part of the factor graph is apparently difficult to
  deal with rigorously.  A related proof is
  \cite[Sec.~X]{gen-area-theorem-conseq} on the accuracy of individual
  BP-extrinsic probabilities (represented by conditional means) when
  the BP and MAP generalized EXIT (GEXIT) curves match, which is based
  on the concavity of the GEXIT kernel relating conditional means and
  the ``generalized entropy'' used by GEXIT.  However, given the
  factors in the un-tree-like part of the factor graph, it is not
  clear why we have $\mu_i^{(l)}(Y) = E[X_i \condmid Y_{\sim
    i}^{(l)}]$ in \cite[Lemma~15]{gen-area-theorem-conseq}.} 
As for the equi-partition condition, when the fixed-point at
$\Info{b}=0$ does not correspond to all-$\constmsg{*}$ messages, in
Fig.~\ref{fig:binary-fg-a} the $\trueextprob{a}{j} \approx
\extprob{a}{j}$ will not be all-$\constmsg{*}$ when the TTSG
determines the last elements of $\seq{a}$, so $P(\seq{a})$ will not be
a uniform distribution.

Experiments show that these monotonicity conditions are more easily
satisfied when $t$ is small, but the resulting MSE $P_t$ will be
larger.  We thus define the \emph{monotonicity threshold} $\tthr$ of a
code as the maximum $t$ that satisfies these conditions.

As in BEQ, the above conditions are only sufficient in an asymptotic
sense.  In practice, even if $t \le \tthr$, the TD will desynchronize
with the TTD due to the finite block length $n$ and iteration count
$L$, and a recovery algorithm from ``incorrect'' decimations is
necessary to achieve acceptable performance with TD, though the greedy
decimator usually performs adequately without recovery.  This will be
discussed in detail in Section~\ref{sec:impact-dec-mse}\@.

Unlike BEQ, in which the monotonicity conditions mean the difference
between being able and unable to find a solution (allowing for a
vanishing fraction of unsatisfied equations), in MSE quantization the
non-satisfaction of these conditions simply causes the asymptotic MSE
to be higher than $P_t$, which is dependent on $t$ anyway.  We will
set $t=\tthr$, so that we have an MSE $P_{\tthr}$ that is
asymptotically (as the block length $n$ and the iteration count $L$ go
to infinity) achievable and analytically tractable, and we can then
design the degree distribution to maximize $\tthr$ and make it
approach its ideal value $t_0(R)$, which corresponds to random-coding
performance in Section~\ref{sec:random-coding}\@.  However, further
optimization on the choice of $t$ is possible.

\subsection{The Erasure Approximation}
\label{sec:mse-erasure-approx}
Similar to BEQ, the average MIs $\Info{b}$, $\Ibext$, $\Info{bc}$,
$\Info{cb}$ and $\Info{c}$ can now be defined for the densities of
respectively $\priprob{b}{i}$, $\extprob{b}{i}$, $\msg{bc}{ij}$,
$\msg{cb}{ji}$ and $\priprob{c}{j}$, e.g.\ $\Info{bc}$ is the average
$1-H(\msg{bc}{ij})$ with $H(\cdot)$ defined in
footnote~\ref{fn:bp-defs}.  When the message densities satisfy the
symmetry condition in footnote~\ref{fn:symmetry}, this is actually the
average mutual information between the messages and their respective
reference values.

In particular, from \eqref{eq:priprobs} we can eventually obtain
$\Info{c}$ as
\begin{equation}
  \label{eq:Ic}
  \Info{c} = \log 2 - H_t = 1 - H_t,
\end{equation}
with $H_t$ defined in \eqref{eq:Ht}.  This relationship allows us to
define the monotonicity threshold alternatively in terms of
$\Info{c}$, as $\Icthr = 1 - H_{\tthr}$, or $\tthr = t_0(\Info{c})$.

When all densities are \emph{erasure-like}, i.e.\ every message, as in
BEQ, is either $\constmsg{*}$ or $\constmsg{b}$ where $b$ is the
message's reference value, \eqref{eq:exit-binary-c} and
\eqref{eq:exit-binary-b} obviously hold.  In general, $\Info{cb}$ is
not uniquely determined by $\Info{bc}$ and $\Info{c}$, nor is
$\Info{bc}$ by $\Info{cb}$ and $\Info{b}$, but
\eqref{eq:exit-binary-c} and \eqref{eq:exit-binary-b} are still
approximately true
\cite{bounds-info-combining,extremes-info-combining}, and the erasure
approximation assumes them to be exact.  The fixed points of DE are
then characterized by the same EBP curve \eqref{eq:Ib} and
\eqref{eq:Ibext}, and according to the conditions above, the
monotonicity threshold $\Icthr$ is the same as that given by
\eqref{eq:Icthr}.  In other words, \emph{the optimized degree
  distribution that maximizes the monotonicity threshold for MSE
  quantization under the EA is the same as that for BEQ}.  Of course,
the true $\Icthr$ of this EA-optimized code will differ from that in
\eqref{eq:Icthr}.

\subsection{Quantized Density Evolution}
Besides the erasure approximation method, the analysis given above
also enables density evolution to be carried out directly on quantized
messages, which allows for arbitrarily good precision.  Our DE scheme
is similar to that in \cite{on-design-ldpc-45e-4-dB}.  Without loss of
generality, we can assume that $\seq{b}^*$ and thus $\seq{u}^*$ and
$\seq{c}^*$ are all-zero, in which case $\seq{z} = (\seq{y}) \bmodIn$
should be strongly typical with respect to $\pz(z)$, and the density
of $\priprob{c}{j}$'s can accordingly be computed with
\eqref{eq:priprobs}.  The messages are represented by uniformly
quantized $L$-values, plus two values representing $\constmsg{0}$ and
$\constmsg{1}$.  The \node{b}-node operations, which simply add up the
$L$-values, become convolutions on densities that can be computed with
fast Fourier transform (FFT), while \node{c}-node operations are
decomposed into that between two messages and computed by table
lookup.\footnote{In LDPC optimization there are only one or two
  distinct check-degrees, but in LDGM quantization codes many more
  different \node{c}-degrees may exist, therefore it may seem tempting
  to represent the densities by instead the ``dual'' $L$-values,
  $\conjL = -\sgn(L)\ln\tanh(\abs{L}/2)$ (see e.g.\
  \cite[Sec.~III-B]{design-cap-approaching-ldpc}), so that the
  check-operations can be computed faster with convolutions.
  Unfortunately, uniformly quantized $\conjL$ is not able to represent
  high-confidence messages (those with a large $\abs{L}$) with
  sufficient accuracy for this approach to work.}

To verify the monotonicity conditions at a
certain $t$, two DE processes are then performed, one starting from
all-$\constmsg{*}$ $\msg{bc}{ij}$ density with $\Info{b}$ gradually
increasing from 0 to 1 (recall that $\priprob{b}{i}$'s density is
always erasure-like), and the other starting from all-$\constmsg{0}$
with $\Info{b}$ gradually decreasing from 1 to 0.  For the uniqueness
of fixed points required by the extrinsic probability condition, it
appears sufficient to check that the above two processes converge to
the same fixed point at the same $\Info{b}$ within the accuracy of
quantized DE, and the equi-partition condition can be checked by
observing whether the latter process converges to all-$\constmsg{*}$
messages when $\Info{b}$ reaches zero.  The monotonicity threshold
$\tthr$ (corresponding to an $\Icthr$) is then the maximum $t$ that
satisfies these conditions.


\subsection{The EXIT Curves for MSE Quantization}
\label{sec:exit-de}
In principle, it is possible to use directly the quantized DE method
to find the monotonicity threshold of a given code, with which the
code's degree distribution can be optimized with e.g.\ local search
methods or differential evolution \cite{design-cap-approaching-ldpc}.
However, this is computationally intensive and unintuitive.

The inaccuracy of EA is mainly due to the erasure-like densities used
for computing the EXIT curves \eqref{eq:exit-binary-c} and
\eqref{eq:exit-binary-b} being very different from the actual message
densities encountered in DE.  If the EXIT curves are computed using
instead the densities encountered in DE of some \emph{base code} under
a \emph{base $t$}, then they are obviously accurate for that code and
$t$.  Moreover, \emph{locally}, i.e.\ for codes with similar degree
distributions and for similar values of $t$, the densities encountered
in DE are usually similar, therefore it is reasonable to expect the
error in EXIT caused by EA to be approximately the same.  If we model
this error by a ``correction factor'' $r(x)$, optimization of the
monotonicity threshold can then be carried out with EXIT curves just
like the BEQ case, simplifying it immensely.

Specifically, given a base code and a base $t$, we model its EXIT
curves with three functions $f(\cdot)$, $g(\cdot)$ and $h(\cdot)$,
such that the average MIs in DE satisfy, under that $t$,
\begin{align}
  \label{eq:exit-binary-c-de}
  \Info{cb} &= \Info{c} \cdot f(\Info{bc}), \\
  \label{eq:exit-binary-b-de}
  \nextInfo{bc} &= 1-(1-\Info{b}) \cdot g(1-\Info{cb}), \\
  \label{eq:Ibext-de}
  \Ibext &= 1-h(1-\Info{cb}).
\end{align}
Note that the erasure approximation corresponds to
\begin{align}
  \label{eq:exit-f-beq}
  f(x) &= \sum_d \vc{d} x^{d-1}, \\
  \label{eq:exit-g-beq}
  g(y) &= y^{\db - 1}, \\
  \label{eq:exit-h-beq}
  h(y) &= y^{\db}.
\end{align}

$f$, $g$ and $h$ are obtained from quantized DE results.  We start
with e.g.\ the base code optimized with EA, and the base $t$ is chosen
near its $\tthr$.  DE is then performed, starting from
all-$\constmsg{*}$ $\msg{bc}{ij}$ density, with $\Info{b}$ increasing
from 0 to 1 slowly enough that the message densities are always close
to fixed points.  The average MI is computed for each density
encountered, and we thus obtain a number of data points that can be
interpolated to form $f$, $g$ and $h$.  The derivatives $f'(x)$,
$g'(y)/g(y) = d(\ln g(y))/dy$ and $h'(x)$ used in the optimization
below are then computed with finite differences.

\begin{figure}[!t]
  \centering
  \subfigure{\includegraphics{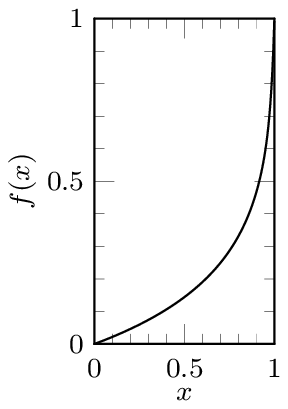}}\hspace{1mm}%
  \subfigure{\includegraphics{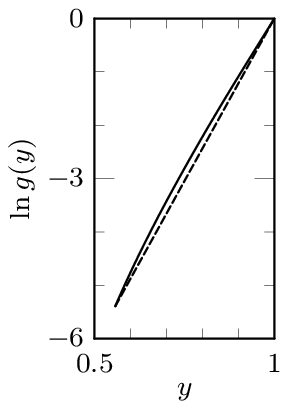}}\hspace{1mm}%
  \subfigure{\includegraphics{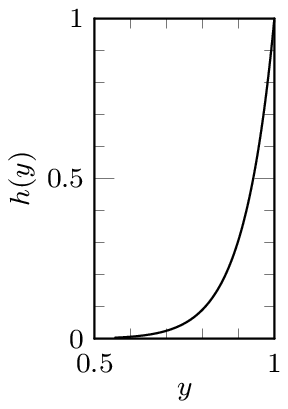}}%
  \caption{The $f(\cdot)$, $\ln g(\cdot)$ and $h(\cdot)$ curves of an
    optimized LDGM quantization code with $R=\unit[0.4461]{b/s}$ and
    $\db=12$ at $t=3.97$ ($\Info{c} = 0.4429$).  Each curve is
    obtained from quantized density evolution results by connecting
    one data point from each iteration.  The dashed straight line in
    the $\ln g(\cdot)$ plot is meant to show its approximate
    concavity.}%
  \label{fig:exit-de}%
\end{figure}

Under EA, we observe from 
\eqref{eq:exit-f-beq}--\eqref{eq:exit-h-beq} that
\begin{itemize}
\item $f(\cdot)$ and $h(\cdot)$ are increasing and convex, so
  $f'(\cdot)$ and $h'(\cdot)$ are nonnegative and increasing;
\item $\ln g(\cdot)$ is increasing and concave, so its derivative
  $g'(\cdot)/g(\cdot)$ is nonnegative and decreasing.
\end{itemize}
In our numerical experiments (e.g.\ Fig.~\ref{fig:exit-de}), we find
that these observations remain approximately true for quantized DE
results except for a slight non-concavity of $\ln g(y)$ for $y$ close
to 1.  This will be useful in the optimization below.

\subsection{Optimization of the Monotonicity Threshold}
\label{sec:mono-opt-de}
Similar to the erasure case, the EBP curve can be obtained if we
equate $\nextInfo{bc}$ in \eqref{eq:exit-binary-b-de} and $\Info{bc}$
in \eqref{eq:exit-binary-c-de} and plot the relationship between
\begin{equation}
  \label{eq:Ib-de}
  \Info{b} = 1 - \frac{1-x}{g(y)}
\end{equation}
(where $x=\Info{bc}$ and $y=1-\Info{cb}$) and $\Ibext$.  The
monotonicity conditions for the base code then again become
\eqref{eq:mono-cond1} and \eqref{eq:mono-cond2}.  The condition
\eqref{eq:mono-cond1} means that BP does not progress at all when
$\Info{b} = 0$ starting from all-$\constmsg{*}$ \node{b}-to-\node{c}
messages, which still implies $\vc{1}=0$, i.e.\ no degree-1
\node{c}-nodes.  As for \eqref{eq:mono-cond2}, since
\begin{equation}
  \label{eq:Ib_x-de}
  \frac{d\Info{b}}{dx} = \frac{g(y) - \Info{c} \cdot (1-x) f'(x) g'(y)}{(g(y))^2},
\end{equation}
the condition is equivalent to (noting that $g'(y)\ge 0$)
\begin{equation}
  \label{eq:mono-cond-de}
  \frac{g(y)}{g'(y)} \ge \Info{c} \cdot (1-x) f'(x).
\end{equation}
According to our observations above, $g(y)/g'(y)$ is nonnegative and
mostly increasing with respect to $y$ and thus decreasing with respect
to $\Info{c}$, while the right side of \eqref{eq:mono-cond-de} is
nonnegative and increasing with respect to $\Info{c}$.  Therefore, for
each $x \in [0,1]$, \eqref{eq:mono-cond-de} is usually satisfied by
all $\Info{c}$ up to a maximum $\IcDE(x) = 1/\sDE(x)$ which can be
found with e.g.\ the bisection method, and the base code's
monotonicity threshold is thus
\begin{equation}
  \label{eq:Icthr-de}
  \Icthr = \left( \max_{x \in [0,1]} \sDE(x) \right)^{-1},
\end{equation}
which has a similar form to \eqref{eq:Icthr}.

\begin{figure}[!t]
  \centering
  \includegraphics{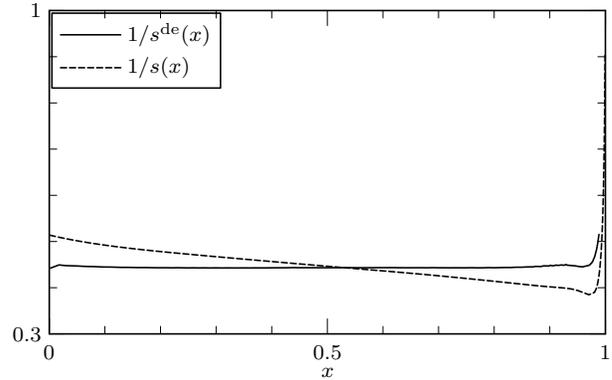}%
  \caption{The $1/s(x)$ and $1/\sDE(x)$ curves for the optimized
    $R=\unit[0.4461]{b/s}$, $\db=12$ LDGM quantization code.  As this
    base code is already well optimized, its $1/\sDE(x)$ is almost a
    flat line except for $x$ close to 1, and its minimum
    $\unit[0.4427]{b/s}$ is $\Icthr$ by \eqref{eq:Icthr-de}, which is
    quite close to $R$.  If this code had instead been optimized under
    EA, $1/s(x)$ would be almost flat but $1/\sDE(x)$ would not be,
    and $\Icthr$ in \eqref{eq:Icthr-de} would be smaller.  Note that
    $\sDE(x)$ and $r(x)$ cannot be computed for $x$ very close to 1,
    as \eqref{eq:mono-cond-de} is then unsatisfied only for $\Info{c}$
    so large that $y$ lies outside the range of available DE data.
    However, since $\sDE(x)$ is expected to be large for $x$ close to
    1, the constraints $r(x)s(x) \le \smax$ in
    \eqref{eq:binary-opt-de} are not usually tight for such $x$ and
    can simply be removed.}%
  \label{fig:exit-de-s}%
\end{figure}

A comparison of $s(x)$ and $\sDE(x)$ is shown in
Fig.~\ref{fig:exit-de-s}.  We can then define the ``correction
factor'' of the base code due to EA as
\begin{equation}
  \label{eq:r-de}
  r(x) = \frac{\sDE(x)}{s(x)},\quad x\in [0,1].
\end{equation}
This $r(x)$ does turn out to be relatively code-independent.
Therefore, for any code with a similar degree distribution to the base
code, its $\Icthr$ can be approximately obtained from
\eqref{eq:Icthr-de} with $\sDE(x) = r(x)s(x)$ and $s(x)$ in
\eqref{eq:s}.  Denoting $\smax = 1/\Icthr$, the optimization of
$\Icthr$ now becomes
\begin{equation}
  \label{eq:binary-opt-de}
  \begin{split}
    \text{minimize}\ & \smax \\
    \text{subject to}\ & r(x)s(x) \le \smax, \quad \forall x \in [0,1], \\
    & \sum_d \vc{d} = 1,\quad \sum_d \frac{\vc{d}}{d}=\frac{1}{R\db},\\
    & \vc{d}\ge 0, \quad \forall d \in \mathcal{D},
  \end{split}
\end{equation}
which is a linear programming problem similar to \eqref{eq:binary-opt}
that can be solved in the same manner.  The solution of
\eqref{eq:binary-opt-de}, presumably better than the original base
code, can be used as the base code for another iteration of the
optimization process in order to obtain a more accurate $r(x)$.  2--3
iterations of this process usually give sufficient accuracy.

\subsection{Relationship to Previous Methods}
It is now instructive to analyze the code optimization
approaches previously proposed in \cite{lossy-src-enc-msgpass-dec-ldgm} and
\cite{ldgm-vq-globecom07}.

In \cite{lossy-src-enc-msgpass-dec-ldgm}, the duals of optimized LDPC
codes are used in the LDGM quantizer for binary symmetric sources.
Under EA, this duality is in fact exact \cite{it-quant-codes-graphs}.
Specifically, if the variable-nodes and check-nodes in the LDPC
decoder are denoted respectively as \node{q}-nodes and \node{p}-nodes,
the erasure-approximated EXIT curves can be given using similar
notation by
\begin{gather}
  \label{eq:exit-binary-q}
  \Info{qp} = 1 - (1 - \Info{q}) \sum_d \vq{d} (1-\Info{pq})^{d-1}, \\
  \label{eq:exit-binary-p-Iqext}
  \nextInfo{pq} = \Info{qp}^{\degp-1}.
\end{gather}
They become identical to \eqref{eq:exit-binary-c} and
\eqref{eq:exit-binary-b} when we replace each \node{q} with \node{c},
\node{p} with \node{b}, each MI $I$ with $1-I$, and let $\Info{b} =
0$.  At the threshold of the LDPC code, the only fixed point is at
$\Info{qp} = \Info{pq} = 1$, which translates to the LDGM code's EBP
curve crossing $\Info{b} = 0$ at $\Info{bc} = \Info{cb} = \Ibext = 0$
only.  The method in \cite{lossy-src-enc-msgpass-dec-ldgm} thus, in
effect, maximizes the maximum $t$ and $\Info{c}$ at which the EBP
curve satisfies this condition, without additionally requiring
$\Info{b}$ to monotonically increase along the curve (see the
$\Info{c}=0.5176$ case in Fig.~\ref{fig:ebp53}).  Also, this duality
is not exact in non-erasure cases
\cite[Fig.~3]{extremes-info-combining}, though such dual
approximations are common in LDPC literature
\cite{design-ldpc-mod-det}.

In \cite{ldgm-vq-globecom07}, curve-fitting is carried out between the
erasure-approximated EXIT curves \eqref{eq:exit-binary-c} and
\eqref{eq:exit-binary-b} at $\Info{b}=0$ and $\Info{c}=R$ (i.e.\
$t=t_0(R)$).  This is roughly equivalent to making $\Info{b}$ as close
to zero as possible along the EBP curve at $\Info{c} = R$.

The three EBP curves in Fig.~\ref{fig:ebp53} illustrate the difference
among the three optimization criteria.  Clearly, the methods in
\cite{lossy-src-enc-msgpass-dec-ldgm} and \cite{ldgm-vq-globecom07}
do not maximize the monotonicity threshold, which has been shown
above to be a reliable indicator of MSE quantizers' performance.
Nevertheless, for reasonably large $\db$ all three criteria tend to
make the EBP curve close to the $\Info{b}=0$ axis except where $\Ibext
\approx 1$, thus the difference among the resulting degree
distributions is not large.  This explains the good performance
obtained in these previous works.

\section{Decimation}
\label{sec:decimation}
Decimation, i.e.\ guessing the values of some $b_i$'s and fixing them
to hard decisions, is an essential component of our LDGM-based
quantization algorithm.  Apart from the aforementioned
\cite{lossy-src-enc-msgpass-dec-ldgm} and \cite{binary-quant-bp-dec-ldgm}, ideas similar to decimation
have also appeared in \cite{on-dec-ldpc-bec} and \cite{maxwell-constr}
in the context of LDPC decoding over BEC.  In \cite{on-dec-ldpc-bec},
guessing is used when a stopping set is encountered, and backtracking
within a limited depth allows guesses leading to contradictions to be
recovered from.  In \cite{maxwell-constr}, the use of guessing with
full backtracking (the Maxwell decoder) leads to the relationship
between the MAP, BP and EBP EXIT curves mentioned in
Section~\ref{sec:exit-erasure}.  The area argument in
Fig.~\ref{fig:ebp-map} suggests that amount of guessing needed by the
Maxwell decoder is dependent on the non-monotonicity of the EBP curve
and is also proportional to the block length $n$.  In practice, the
backtracking depth is limited by its exponential complexity, so
backtracking is not expected to provide much gain for large $n$ and
will not be considered here.

Without backtracking, there will unavoidably be ``wrong'' decimation
decisions, which in the above analysis means that the TD decimates
some $b_i$ to a different value from the TTD due to a difference
between $\extprob{b}{i}$ and $\trueextprob{b}{i}$.  This difference
can be caused by non-satisfaction of the monotonicity conditions, the
finiteness of block length $n$, or most importantly, because the
limited iteration count $L$ has not allowed BP to converge.  In this
section, we will attempt to get a rough idea of the impact of such
incorrect decimation, how to recover from them, and how to minimize
this impact within a given number of iterations.

\subsection{Controlling the Decimation Process}
\label{sec:dec-control}
Within a limited number of iterations $L$, the determination of how
much decimation to do in each iteration, possibly based on the current
progress of convergence, is obviously important in minimizing the
amount of ``incorrect'' decimations.  In
\cite{lossy-src-enc-msgpass-dec-ldgm}, bits that are more ``certain''
than some threshold are decimated every few iterations.  In
\cite{binary-quant-bp-dec-ldgm}, upper and lower limits on the number
of bits to decimate at each time are introduced in addition.  An early
version of our quantization algorithm, instead, decimates a number of
bits whenever the quantizer gets ``stuck'' for a number of iterations.
The downside of these decimation strategies is their reliance on
manual adjustment of various thresholds, which can be cumbersome in
code optimization, as different codes may require different thresholds
for acceptable performance.  Instead, our unthrottled decimation
strategy controls the amount of decimation by forcing $\Info{bc}$ to
increase by $\minprog$ per iteration, with $\minprog$ possibly
dependent on the current $\Info{bc}$.\footnote{The bit granularity of
  the amount of decimation as well as random variations in the
  $\Info{bc}$ estimate can cause the actual iteration count $L$ to
  differ from the intended $L_0$.  If, instead of making $\Info{bc}$
  increase \emph{by} a certain amount depending on its current value,
  we make it increase \emph{to} some value according to the elapsed
  number of iterations, then $L$ will be more predictable, which is
  desirable in practice.  However, our current unthrottled and
  throttled strategies are yet unable to control the decimation
  process well enough in this case, resulting in a worse tradeoff
  between iteration count $L$ and the achieved MSE, therefore this
  will not be adopted here.} Although this pace can also be optimized
according to the code, as will be done in
Section~\ref{sec:opt-pacing}, a uniform pace of $\minprog = 1/L_0$
already performs well, making the strategy very convenient to use.

The throttled decimation strategy shown in Fig.~\ref{alg:binary} was
introduced in \cite{ldgm-vq-globecom07}.  It is based on the
observation that the $\Info{bc}$ estimated in the algorithm is noisy
and tends to progress somewhat erratically, sometimes even decreasing,
which in the unthrottled algorithm causes unintended variation in the
amount of decimation in each iteration.  To reduce this variation, the
throttled algorithm introduces $\deltamax$, which can roughly be
viewed as the amount of decimation per iteration.  $\deltamax$ is
slowly adjusted according to the actual pace of convergence, and upon
reaching the steady state $\Info{bc}$ should be increasing at the
desired pace.

In practice, at a given $L_0$, throttling does improve MSE performance
but also increases the actual iteration count $L$.  In terms of the
$L$-versus-MSE tradeoff, the unthrottled algorithm is better for small
$L$, when the iterations necessary for $\deltamax$ to reach its
steady-state value represent a significant overhead, but for $L_0$
greater than about $10^3$ the throttled algorithm perform better,
therefore both will be used in our simulations.  A detailed analysis
and optimization of the throttling strategy is an interesting problem
of optimal control, and may be worthy of further study.

\subsection{Impact of Imperfect Decimation in BEQ}
\label{sec:impact-dec-erasure}
We begin analyzing the performance impact of non-ideal decimation by
looking at the simpler BEQ problem, viewed as a set of linear
equations \eqref{eq:beq-eqs} over variables $b_1,\dotsc,b_{\nb}$.
With finite block size $n$ and iteration count $L$, BP cannot be
expected to find an exact solution, so our aim is to minimize the
number of unsatisfied equations.

Incorrect decimations are indicated by contradictions in BP, e.g.\
$\constmsg{0} \odot \constmsg{1}$.  If we proceed with BP after
contradictions by simply setting $\constmsg{0} \odot \constmsg{1} =
\constmsg{*}$, a large fraction of unsatisfied equations will
result.\footnote{A more elaborate treatment of contradictions in BEQ
  can be given as follows.  Instead of setting $\priprob{c}{j}$ to
  hard decisions $\constmsg{0}$ and $\constmsg{1}$ when the source
  symbol $y_j = 0$ and $1$, it is ``softened'' to probability tuples
  $(1-\delta,\delta)$ and $(\delta,1-\delta)$, respectively, where
  $\delta > 0$ is an infinitesimal constant.  Now let $L_0 =
  \log((1-\delta)/\delta)$, and each message $\mu = (\mu_0,\mu_1)$ can
  then be represented by the scaled $L$-value $l(\mu) = (1/L_0)
  \log(\mu_0 / \mu_1)$.  For $\delta \to 0$ and with $l=l(\mu)$,
  $l'=l(\mu')$, the definitions of ``$\odot$'' and ``$\oplus$'' imply
  that $l(\mu \odot \mu') = l + l'$ and $l(\mu \oplus \mu') =
  \max(l+l',0) - \max(l,l')$, thus belief propagation can be run using
  this scaled $L$-value representation.  This results in a slightly
  lower, but still large, fraction of unsatisfied equations.}
Intuitively, as the contradictory messages propagate, they essentially
set a variable $b_i$ to $0$ in some equations and to $1$ elsewhere and
determine the values of other variables with these contradictory
values, and the confusion thus spreads.

To avoid this problem, each known variable should be made to possess a
consistent value in all equations.  A class of ``serial'' algorithms
of the following form have this property.  Initially all variables are
unknown, and in each step the quantizer may either \emph{guess} the
value of one unknown variable, or \emph{discover} the value of one
unknown variable with an equation in which all variables but that one
are known.\footnote{The choice is left to the individual algorithms
  within the class.}  This process repeats until all variables become
known.  Suppose $\nguess$ guesses are made, then the remaining
$\nb-\nguess$ variables are each determined by one unique equation.
These $\nb-\nguess$ equations are always satisfied, while the
remaining
\begin{equation}
  \label{eq:nignore}
  \nignore = \nne - (\nb - \nguess)  
\end{equation}
equations have been ignored in the process and half of them are
expected to be unsatisfied.

For the original ``parallel'' BP algorithm,\footnote{Of course, BEQ
  itself is more efficiently solved by a serial algorithm, but only a
  ``parallel'' BP algorithm can be extended to MSE quantization.} a
``recovery'' step from contradictions can be introduced into each BP
iteration, which changes some \node{c}-priors $\priprob{c}{j}$ (in
effect making BP use a different source sequence) to fix the
contradiction.  Specifically,
\begin{itemize}
\item If all incoming $\msg{bc}{ij}$'s to some $\node{c}_j$ are
  ``known'' ($\constmsg{0}$ or $\constmsg{1}$), and $\priprob{c}{j}$
  is ``known'' and disagrees with them, flip $\priprob{c}{j}$
  ($\constmsg{0}$ to $\constmsg{1}$ and vice versa) such that they
  agree, and compute the outgoing $\msg{cb}{ji}$'s accordingly.
\item If the incoming $\msg{cb}{ji}$'s to some $\node{b}_i$ include
  both $\constmsg{0}$ and $\constmsg{1}$,
  \begin{itemize}
  \item randomly pick one ``known'' $\msg{cb}{ji}$ and denote its
    value by $\constmsg{b}$ with $b\in\{0,1\}$;
  \item for each $j \in \neighbor{cb}{\cdot i}$ that $\msg{cb}{ji} \ne \constmsg{*}$ and
    $\msg{cb}{ji} \ne \constmsg{b}$, flip $\priprob{c}{j}$ and
    recompute all messages from $\node{c}_j$;
  \item compute the outgoing $\msg{bc}{ij}$'s from $\node{b}_i$
    according to the new incoming messages.
  \end{itemize}
\end{itemize}
With this recovery step, the parallel BP algorithm works like one of
the aforementioned class of serial algorithms.  In each iteration,
\begin{itemize}
\item BP at \node{c}-nodes assigns tentative values to
  previously unknown variables $b_i$ according to equations, and all
  equations that are already unsatisfied are ignored due to the first
  rule above.
\item BP at \node{b}-nodes with the second rule above picks one
  assignment among possibly many for each newly known variable.  This
  assignment becomes one ``discovery'' step in the serial algorithm,
  while all other assignments are ignored.
\item Each decimation of a $b_i$ with $\extprob{b}{i} = \constmsg{*}$
  constitutes a ``guess'' step in the serial algorithm.
\end{itemize}
Therefore it does view every variable consistently, and
\eqref{eq:nignore} is applicable.  Clearly, incorrect decimations now
only cause more flips in the recovery step, but they do not affect the
fraction of ``known'' variables and messages in each iteration, which
can then be computed assuming that all decimations have been correct.
For asymptotically large $n$ and the typical decimator, this is given
by the evolution of MIs according to the EXIT curves
\eqref{eq:exit-binary-c}, \eqref{eq:exit-binary-b} and
\eqref{eq:Ibext}.

\begin{figure}[!t]
  \centering
  \subfigure[the EBP and the actual curves]{\label{fig:ebp-actual-Ai}\includegraphics{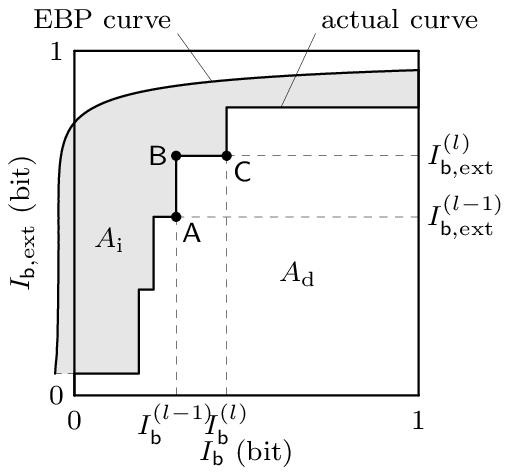}}%
  \nolinebreak\hspace{-5mm}%
  \subfigure[flowchart of one iteration]{\label{fig:ebp-actual-flow}\includegraphics{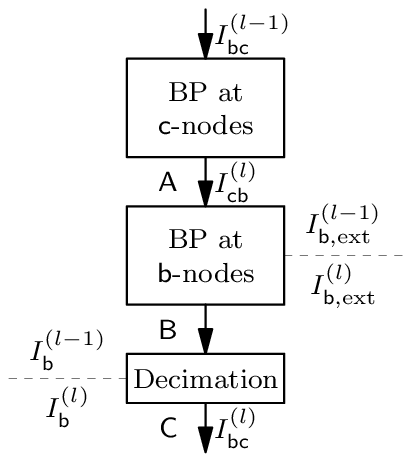}}%
  \caption{A comparison of the EBP and the actual \IbvsIbext\ curves.
    Here $\vc{1}>0$ so the EBP curve does not start from $(0,0)$.  The
    gray area between the two curves is the delta-area $\Aignore$.
    The flowchart on the right shows the trajectory followed by the
    quantizer on the actual curve in a single iteration.}%
%
  \label{fig:ebp-actual}%
\end{figure}

The path followed by $(\Info{b},\Ibext)$ during the actual
quantization process has thus a staircase shape as shown in
Fig.~\ref{fig:ebp-actual}, and it is hence called the \emph{actual
  curve}.  Since $\Info{b}$ indicates the fraction of decimated bits,
and in each iteration $\Ibext$ is the fraction among newly decimated
$b_i$'s that have $\extprob{b}{i} = \constmsg{0}$ or $\constmsg{1}$,
the area above the actual curve $\Aguess = 1-\Adet$ is the overall
fraction of guesses $\nguess/\nb$.  We have found the area below the
EBP curve to be $\Ane = \Info{c} / R = n\Info{c} / \nb$ (approximate
when $\vc{1}>0$), so from \eqref{eq:nignore} the \emph{delta-area}
$\Aignore = \Ane - \Adet$ between the two curves is asymptotically
$\nignore / \nb$, and it can thus serve as a measure of the number of
unsatisfied equations.  As the number of iterations goes to infinity,
the actual curve approaches the BP curve, and the delta-area goes to
zero if and only if the monotonicity conditions \eqref{eq:mono-cond1}
and \eqref{eq:mono-cond2} are satisfied.

\subsection{Impact of Imperfect Decimation: MSE Quantization}
\label{sec:impact-dec-mse}
For MSE quantization, simulation results show that the typical
decimator by itself again has poor performance.  The reason is similar
to the BEQ case: imperfect decimation causes message densities to be
no longer consistent, in effect containing ``soft'' contradictions
that slow down future convergence if not recovered from.  The greedy
decimator in Fig.~\ref{alg:binary}, however, does achieve satisfactory
performance in this case, presumably because it tends to choose
better-than-typical codewords and the resulting gain can usually
compensate for the effect of imperfect decimation.

It is still of interest to make the more analytically tractable TD
perform acceptably by recovering properly from incorrect decimations.
The method is similar to the recovery at \node{c}-nodes in the BEQ
case: if the prior $\priprob{c}{j}$ at some $\node{c}_j$ is inconsistent
with the incoming messages $\msg{bc}{ij}$, as summarized by the
extrinsic probability
\begin{equation}
  \extprob{c}{j} = \bigoplus_{i'\in\neighbor{bc}{\cdot j}} \msg{bc}{i'j},
  \label{eq:extprob-c}
\end{equation}
then $\priprob{c}{j}$ is adjusted to fix the inconsistency, by using a
slightly different $\yest_j$ (which is recomputed in every iteration)
instead of $y_j$ in \eqref{eq:priprobs}.

We first analyze the relationship that $y_j$ (or $\priprob{c}{j}$) and
$\extprob{c}{j}$ should have if all decimations are correct, i.e.\ the
equi-partition condition is satisfied and our TD is perfectly
synchronized with a TTD.  Assuming $\seq{y} \in \mathcal{I}^n =
[-1,1)^n$ without loss of generality, and using TTD's final result
$\seq{b}^*$ and the corresponding $\seq{u}^*=\seq{c}^*=\seq{b}^*
\mat{G}$ as the reference codeword, we can define $\seq{z} = (\seq{y}
- \seq{u}^*) \bmodIn$ and $\seq{p}$ with $p_j = \extprobx{c}{j}{c_j^*}$, which should then
asymptotically satisfy the following typicality properties: with $j$
being random,
\begin{itemize}
\item $z_j$ has pdf $\pz(z)$, because of
  $\seq{z}$'s strong typicality shown in Section~\ref{sec:de-mse-intro};
\item $p_j$'s pdf at $p$ and $(1-p)$ have ratio $p:(1-p)$ for any
  $p \in [0,1]$, due to the symmetry condition
  (footnote~\ref{fn:symmetry}) also satisfied by the density of
  $\extprob{c}{j}$;
\item $z_j$ and $p_j$ are independent, since $p_j$ comes from the
  extrinsic $\extprob{c}{j}$, which only depends on information other
  than $y_j$ in $\node{c}_j$'s tree-like neighborhood in
  the factor graph.
\end{itemize}
In the actual quantizer $\seq{b}^*$ is unknown, so instead of $p_j$
only $q_j = \extprobx{c}{j}{1}$ is observable.  From the above
property of $\seq{p}$, among those $j$'s with $q_j$ near some $q \in
[0,1]$, a fraction $q$ should have $c_j^* = 1$ and the rest have
$c_j^* = 0$, therefore the density of $y_j$ at these positions should
be
\begin{equation}
  \label{eq:pyq}
  \pyq{q}(y) = (1-q) \cdot \pz(y) + q \cdot \pz((y-1)\bmodI).
\end{equation}
This relationship \eqref{eq:pyq} can be checked by comparing the
actual cumulative distribution function (CDF) of $y_j$ at the
positions where $q_j \approx q$, denoted by $\Fyqest{q}(y)$, to the
CDF $\Fyq{q}(y)$ corresponding to $\pyq{q}(y)$.  When they are
different, our recovery algorithm attempts to find a $\seq{\yest}$
close to $\seq{y}$ such that the corresponding CDF of $\yest_j$
matches $\Fyq{q}(y)$, thus allowing BP to continue as if decimation
had been perfect.

Denote $F_0(y) = \Fyq{1/2}(y)$ as the CDF of the uniform
distribution over $\mathcal{I}$, and $G(y) = \Fyq{1}(y) -
\Fyq{0}(y)$, then
\begin{equation}
  \label{eq:Fyq-G10}
  \Fyq{q}(y) = F_0(y) + (q-1/2) G(y). 
\end{equation}
To help estimate $\Fyqest{q}(y)$, it is similarly approximated as
\begin{equation}
  \label{eq:Fyq-G1}
  \Fyqest{q}(y) = F_0(y) + (q-1/2) \Gest(y),
\end{equation}
so that only $\Gest(\cdot)$ has to be estimated.  For any $y \in \mathcal{I}$, $\Fyqest{q}(y)$ is the average of $\oneif[y_j \le y]$ over
positions $j$ with $q_j \approx q$,\footnote{$\oneif[y_j \le y]$ is
  defined as 1 if $y_j \le y$, 0 otherwise.} therefore
$\Gest(y)$ can be unbiasedly estimated by
\begin{equation}
  \label{eq:G1-est}
  \Gest(y) = \frac{\sum_{j=1}^n (q_j - 1/2) (\oneif[y_j \le y] - F_0(y))}{\sum_{j=1}^n (q_j - 1/2)^2}.
\end{equation}
Having obtained $\Gest(\cdot)$ and thus $\Fyqest{q}(\cdot)$, we can let
\begin{equation}
  \label{eq:y-adjust}
  \yest_j = \Fyq{q_j}^{-1} \left( \Fyqest{q_j} (y_j) \right), \quad j=1,\dotsc,n,
\end{equation}
then $\yest_j$ should have the desired CDF $\Fyq{q}(\cdot)$ at positions $j$ with $q_j \approx q$.

In practice, $\Gest(y)$ is computed for a few discrete values of $y$
that divide $\mathcal{I}$ into intervals.  By first summing the
corresponding $(q_j-\frac{1}{2})$ for $y_j$'s falling in each interval,
\eqref{eq:G1-est} for these $y$'s can be computed in $\asymptO(n)$
time.  Initially this estimated $\Gest(y)$ will be rather noisy and
may need to be adjusted such that all CDFs remain monotonic and within
range.  The transform \eqref{eq:y-adjust} is then evaluated at these
$y$'s and a few discrete values of $q$, after which each $\yest_j$ is
computed by bilinear interpolation.  The symmetry of $\pyq{q}(y)$ and
$\pyqest{q}(y)$ (corresponding to $\Fyqest{q}(y)$) around $y=0$ can be
further exploited to simplify this process.  This recovery procedure
is carried out at the beginning of each iteration (or possibly once
every few iterations), after which the $\priprob{c}{j}$'s are
recomputed with \eqref{eq:priprobs} using $\yest_j$ for $y_j$.

When TD is used with recovery, the message densities can be kept
approximately consistent after imperfect decimation, allowing the
average MIs to evolve according to the EXIT curves
\eqref{eq:exit-binary-c-de}, \eqref{eq:exit-binary-b-de} and
\eqref{eq:Ibext-de}, and the actual curve as well as the areas $\Adet$
and $\Aignore$ can thus be similarly defined.  We do not know of any
definite relationship between the delta-area $\Aignore$ and the MSE,
as the amount of movement between $\seq{\yest}$ and $\seq{y}$ in
recovery is hard to analyze.  Nevertheless, simulation results suggest
that the MSE can be roughly estimated by
\begin{equation}
  \label{eq:mse-Ai-emp}
  \hat{\sigma}^2 = \left( 1-\frac{\Aignore}{\Info{c}/R} \right) \cdot P_t + \frac{\Aignore}{\Info{c}/R} \cdot P_0,
\end{equation}
where $P_0 = P_t |_{t=0}$ is the zero-rate MSE and is $\frac{1}{3}$ in
the binary case.  Intuitively speaking, each $y_j$ can be viewed as a
soft constraint on $\seq{b}$ that amounts to $\Info{c}$ hard
constraints, and the $n\Info{c}$ hard constraints in total are
represented by the area $n\Info{c}/\nb = \Info{c}/R$, which in our
simulations appears to be the area below the EBP curve just like the
BEQ case.\footnote{At least, when the monotonicity conditions are
  satisfied, we expect the EBP curve to coincide with the MAP curve,
  the area below which is indeed $\Info{c}/R$ as shown in
  footnote~\ref{fn:map-exit}.} The area below the actual curve, $\Adet
= \Info{c}/R - \Aignore$, represents satisfied constraints having MSE
$P_t$, while the delta-area $\Aignore$ represents ignored constraints,
corresponding to quantization error uniformly distributed in
$\mathcal{I}$ with MSE $P_0$, therefore we obtain an explanation for
\eqref{eq:mse-Ai-emp}.  Even though \eqref{eq:mse-Ai-emp} is not
exact, it does give a reasonably accurate relationship between
$\Aignore$ and the MSE, and the minimization of $\Aignore$ will thus
be our objective in the optimization of the pace of decimation below.

\subsection{Optimal Pacing of Decimation}
\label{sec:opt-pacing}
We can observe from Fig.~\ref{fig:ebp-actual} that a large number of
iterations is needed to make the actual curve fit closely to the EBP
curve and achieve a small delta-area, which is necessary for good MSE
performance.  Under a fixed number of iterations, this tradeoff can be
improved somewhat by optimizing the pace of decimation, as will be
discussed in this subsection.  This iteration count will be denoted by
$L$ in the analysis here; it corresponds to $L_0$ in the quantization
algorithm, which may take a slightly different number of iterations to
converge.

Denote the MIs at each iteration $l$ by e.g.\ $\Infoiter{bc}{l}$.  If
the deviation of the actual curve from the EBP curve is sufficiently
small such that the DE results
\eqref{eq:exit-binary-c-de}--\eqref{eq:Ibext-de} remain valid, we then
have, for each $l=1,\dotsc,L$,
\begin{align}
  \label{eq:exit-binary-c-pace}
  \Infoiter{cb}{l} &= \Info{c} \cdot f(\Infoiter{bc}{l-1}), \\
  \label{eq:exit-binary-b-pace}
  \Infoiter{bc}{l} &= 1-(1-\Infoiter{b}{l}) \cdot g(1-\Infoiter{cb}{l}), \\
  \label{eq:Ibext-pace}
  \Ibextiter{l} &= 1-h(1-\Infoiter{cb}{l}).
\end{align}
All these MIs can be viewed as functions of
$\Infoiter{bc}{1},\dotsc,\Infoiter{bc}{L-1} \in [0,1]$, subjected to
boundary conditions
\begin{equation}
  \label{eq:pace-boundary}
  \Infoiter{bc}{0} = 0, \quad \Infoiter{bc}{L} = 1,
\end{equation}
and monotonicity constraint (since there can only be more decimated bits after more iterations)
\begin{equation}
  \label{eq:pace-mono}
  0 \le \Infoiter{b}{1} \le \dotsb \le
  \Infoiter{b}{L-1} \le \Infoiter{b}{L} = 1.
\end{equation}
The area below the actual curve is then
\begin{equation}
  \label{eq:Adet}
  \Adet = \sum_{l=0}^{L-1} (1-\Infoiter{b}{l}) (\Ibextiter{l+1} - \Ibextiter{l}).
\end{equation}
where we have set $\Infoiter{b}{0} = \Ibextiter{0} = 0$ for
convenience.  The uniform pacing used in \cite{ldgm-vq-globecom07}
corresponds to $\Infoiter{bc}{l} = l/L$, and we now optimize
$\Infoiter{bc}{1},\dotsc,\Infoiter{bc}{L-1}$ to minimize the
delta-area $\Aignore$, or equivalently, to maximize $\Adet$ in
\eqref{eq:Adet}.

Usually $\Info{c} \le \Icthr$ (or only slightly larger), in which case
the monotonicity constraint \eqref{eq:pace-mono} is frequently
redundant.  Ignoring this constraint, the maximization of $\Adet$ can
then be efficiently solved by dynamic programming.  Specifically, for
each $\Infoiter{bc}{l-1} = x \in [0,1]$, define
\begin{equation}
  \Adetiter{l}(x) = \max_{\Infoiter{bc}{l},\dotsc,\Infoiter{bc}{L-1}}
  \sum_{l'=l}^{L-1} (1-\Infoiter{b}{l'}) (\Ibextiter{l'+1} - \Ibextiter{l'}),
\end{equation}
and it satisfies the recursive formula
\begin{equation}
  \label{eq:pace-dp-recurse}
  \Adetiter{l}(x) = \max_{\Infoiter{bc}{l}}
  \left[ (1-\Infoiter{b}{l}) (\Ibextiter{l+1} - \Ibextiter{l}) + \Adetiter{l+1}(\Infoiter{bc}{l}) \right]
\end{equation}
with $\Adetiter{L} \equiv 0$.  The maximum of $\Adet$ is then
$\Adetiter{1}(0)$ plus the constant term
\begin{equation}
  (1-\Infoiter{b}{0}) (\Ibextiter{1} - \Ibextiter{0}) = 1-h(1-\Info{c} f(0)).
\end{equation}
After discretizing $x$, the recursion \eqref{eq:pace-dp-recurse} can
be evaluated numerically, obtaining the optimal
$\Infoiter{bc}{1},\dotsc,\Infoiter{bc}{L-1}$.

If the solution thus obtained violates \eqref{eq:pace-mono}, that is,
this constraint turns out to be tight, a good but suboptimal solution
can be found by imposing the constraint ``greedily'' during the
recursion \eqref{eq:pace-dp-recurse}: when computing the previous
$\Adetiter{l+1}(x)$, the $\Infoiter{b}{l+1}$ corresponding to the
optimal $\Infoiter{bc}{l+1}$ is recorded along with the maximum for
each $x$, and then the maximization with respect to $\Infoiter{bc}{l}$
is done under the constraint $\Infoiter{b}{l} \le \Infoiter{b}{l+1}$.

When $L$ is large, the above optimization can be simplified, which also
enables us to analyze the asymptotic performance as $L\to\infty$.  For
each $l$, the $\Info{b}$ corresponding to $\Ibextiter{l}$ on the EBP
curve, $\Ibfixiter{l}$, is determined by
\begin{equation}
  \label{eq:exit-binary-b-pace-ebp}
  \Infoiter{bc}{l-1} = 1-(1-\Ibfixiter{l}) \cdot g(1-\Infoiter{cb}{l}).
\end{equation}
Comparing \eqref{eq:exit-binary-b-pace} and
\eqref{eq:exit-binary-b-pace-ebp}, $\deltaIbiter{l} = \Infoiter{b}{l}
- \Ibfixiter{l}$ should satisfy
\begin{equation}
  \label{eq:pace-vs-deltaIb-disc}
  \Infoiter{bc}{l} - \Infoiter{bc}{l-1} = \deltaIbiter{l} g(1-\Infoiter{cb}{l}).
\end{equation}
For large $L$, $l$ can be viewed as a continuous-valued variable and
$x = \Infoiter{bc}{l-1}$ is an increasing function of it, with
$dx/dl \approx \Infoiter{bc}{l} - \Infoiter{bc}{l-1}$.
$\deltaIbiter{l}$ \textit{et al} can then be viewed as functions of
$x$ rather than of $l$, and defining $y = 1-\Infoiter{cb}{l} =
1-\Info{c}f(x)$ as before, \eqref{eq:pace-vs-deltaIb-disc} becomes
\begin{equation}
  \label{eq:pace-vs-deltaIb}
  \frac{dx}{dl} = \deltaIb(x) \cdot g(y).
\end{equation}
The number of iterations is then
\begin{equation}
  \label{eq:pace-L}
  L = \int_0^1 \frac{dl}{dx}\, dx = \int_0^1 \frac{dx}{\deltaIb(x) \cdot g(y)},
\end{equation}
and since $\Ibextiter{l} = 1-h(y) = 1-h(1-\Info{c}f(x))$, $\Aignore$ becomes
\begin{align}
  \Aignore 
  &= \int_0^1 \deltaIb(x) \frac{d\Ibext}{dx} \,dx \\
  \label{eq:pace-Ai}
  &= \int_0^1 \deltaIb(x) \cdot \Info{c} \cdot f'(x) \cdot h'(y) \,dx.
\end{align}
The constraint \eqref{eq:pace-mono} basically requires $\Info{b}(x) =
\Info{b}^*(x) + \deltaIb(x)$ to be non-negative and increasing with
$x$.  Note that this reduces to \eqref{eq:mono-cond1} and
\eqref{eq:mono-cond2} when $L\to\infty$ and thus $\deltaIb(x)\to 0$.

Again, in practice \eqref{eq:pace-mono} is usually not tight and can
be ignored at first, and the minimization of \eqref{eq:pace-Ai} (a
functional of $\deltaIb(x)$) under constraint \eqref{eq:pace-L} can
then be solved with Lagrange multipliers.  Setting
\begin{equation}
  \frac{\delta[\Aignore + \lambda^{-1} L]}{\delta[\deltaIb(x)]} 
= \Info{c} f'(x) h'(y) - \frac{\lambda^{-1}}{ (\deltaIb(x))^2 g(y) } = 0,
\end{equation}
we find the optimal $\deltaIb(x)$ 
\begin{equation}
  \label{eq:opt-deltaIb}
  \deltaIb(x) = \left(\lambda \Info{c} \cdot f'(x) \cdot g(y) \cdot h'(y) \right)^{-1/2},
\end{equation}
and \eqref{eq:pace-vs-deltaIb} then gives the desired increase of $\Info{bc}$ per iteration.

Substitute \eqref{eq:opt-deltaIb} into \eqref{eq:pace-L} and
\eqref{eq:pace-Ai} and we get
\begin{equation}
  \label{eq:LAi}
  L \Aignore = \left( \int_0^1 \sqrt{\frac{\Info{c} \cdot f'(x) h'(y)}{g(y)}}\,dx \right)^2.
\end{equation}
Therefore, \emph{$L$ and $\Aignore$ are inversely proportional when
  $L$ is large and \eqref{eq:pace-mono} is not tight}, which is an
interesting result on the loss-complexity tradeoff of LDGM
quantization codes.  The right-hand side of \eqref{eq:LAi} can be
numerically evaluated and is generally slightly smaller than 4.  For
example, it is 3.365 for the optimized $\db=12$ code used in the
simulations below, and under the erasure approximation and
\eqref{eq:close-fit-approx-pace} below we get $4(\db-1)/\db$, which
approaches 4 for large $\db$.  Indeed, when $L$ is large,
$\deltaIb(x)$ is basically scaled by different constants to achieve
different tradeoffs between $L$ and $\Aignore$, so from
\eqref{eq:pace-L} and \eqref{eq:pace-Ai} we see that this inverse
proportional relationship is also true for other paces.  For example,
from \eqref{eq:pace-vs-deltaIb}, uniform pacing corresponds to
$\deltaIb(x) = 1/Lg(y)$, which results in
\begin{equation}
  \label{eq:LAi-upace}
  L \Aignore = \int_0^1 \frac{\Info{c} \cdot f'(x) h'(y)}{g(y)} \,dx.
\end{equation}
For the same optimized $\db=12$ code, \eqref{eq:LAi-upace} evaluates
to 4.701, therefore for large $L$ the optimized pacing of decimation is
expected to require approximately $3.365/4.701 = 72\%$ as many
iterations as uniform pacing to achieve the same MSE performance.

In practice, $\Aignore$ is not very sensitive to $\deltaIb(x)$, so
\eqref{eq:opt-deltaIb} can be further approximated.  We can observe
that the EBP curves of good codes have $\Ibfix \approx 0$ for all $x$
but those very close to 1, which means $x \approx 1-g(y)$.  Taking
derivatives, we have
\begin{equation}
  \label{eq:close-fit-approx-pace}
  \Info{c}\cdot f'(x) \cdot g'(y) \approx 1,
\end{equation}
and \eqref{eq:opt-deltaIb} and \eqref{eq:pace-vs-deltaIb} then become
\begin{equation}
  \frac{dx}{dl} = \sqrt{\frac{g(y)\cdot g'(y)}{\lambda \cdot h'(y)}}.
\end{equation}
If the erasure approximations \eqref{eq:exit-g-beq} and
\eqref{eq:exit-h-beq} are used in addition, we get a simple
formula dependent only on $\db$:
\begin{align}
  \frac{dx}{dl} &= \sqrt{\frac{\db - 1}{\lambda \db}} y^{(\db-2)/2} \\
  \label{eq:pace-approx}
  &\approx \frac{2(\db-1)}{L\db} (1-x)^{\frac{\db-2}{2(\db-1)}},
\end{align}
where we have used $x\approx 1-g(y)$ and \eqref{eq:pace-L} in
\eqref{eq:pace-approx}.  Eq.~\eqref{eq:pace-approx} is still
near-optimal: its $L\Aignore$ for the optimized $\db=12$ code is
3.443, only slightly larger than the optimal 3.365.

In the actual decimation algorithm, we adopt such a pace by setting
$L$ to $L_0$ and $\minprog$ to this $dx/dl$, with $x$ being the
$\Info{bc}$ estimated in the algorithm.

\subsection{Pacing-Aware Code Optimization}
\label{sec:pacing-aware-code-opt}
Our code design method in Sections \ref{sec:deg-opt-erasure} and
\ref{sec:deg-opt-mse} has focused on maximizing the monotonicity
threshold $\Icthr$, and with $t$ chosen such that $\Info{c} = \Icthr$,
this minimizes the resulting MSE $P_t$ as the delta-area approaches
zero with $L\to\infty$ and $n\to\infty$.  We have mentioned at the end
of Section~\ref{sec:de-mse-intro} that this is not necessarily
optimal; ideally $t$ and the degree distribution should be jointly
optimized, and when $L$ is finite, the pace of decimation should be
included in the joint optimization as well.  Doing this optimization
precisely would be prohibitively complicated with limited benefit, so
below we will look at a simple heuristic adjustment on the degree
distribution optimization process for finite $L$ that nevertheless
results in some performance gain.

According to our analysis above, for large $L$, if the optimized pace
of decimation given by \eqref{eq:opt-deltaIb} and
\eqref{eq:pace-vs-deltaIb} does not violate the monotonicity
constraint \eqref{eq:pace-mono}, then the resulting $\Aignore$ is
inversely proportional to $L$, and the product $L\Aignore$ given by
\eqref{eq:LAi} is not very dependent on the code.  When optimizing the
code's degree distribution for a fixed $L$, we can therefore
approximately view $\Aignore$ as a constant, and \eqref{eq:mse-Ai-emp}
suggests that the optimization should maximize the maximum $\Info{c}$
satisfying \eqref{eq:pace-mono}, hence denoted $\IcthrL$.  As $L$ goes
to infinity, \eqref{eq:pace-mono} reduces to the code's monotonicity
conditions \eqref{eq:mono-cond1} and \eqref{eq:mono-cond2}, and this
optimization method reduces to that in
Section~\ref{sec:mono-opt-de}\@.

The optimized pace of decimation is approximated by the
code-independent \eqref{eq:pace-approx}, which can be integrated to
yield
\begin{equation}
  \label{eq:xl-pace-approx}
  x(l) = 1-(1-l/L)^{2(\db-1)/\db}.
\end{equation}
We also define $l(x)$ as the inverse function of $x(l)$, and $p^+(x) =
l^{-1}(l(x)+1)$ as the mapping from $\Infoiter{bc}{l-1}$ to
$\Infoiter{bc}{l}$.  Now let $\Info{bc} = x$ and $\nextInfo{bc} =
p^+(x)$ in the EXIT curves \eqref{eq:exit-binary-c-de} and
\eqref{eq:exit-binary-b-de}, and we obtain the $\Info{b}$ needed for
this pace of decimation:
\begin{equation}
  \Info{b} = 1-\frac{1-p^+(x)}{g(y)} = 1-\frac{1-p^+(x)}{g(1-\Info{c} f(x))}.
\end{equation}
The condition \eqref{eq:pace-mono} means that $\Info{b} |_{x=0} \ge 0$
and $d\Info{b}/dx \ge 0$.  Since $f(0) = \vc{1}$ (when $\Info{bc} =
0$, the \node{c}-to-\node{b} messages from degree-1 \node{c}-nodes
have average MI $\Info{c}$ while all other \node{c}-nodes output
all-$\constmsg{*}$, so $\Info{cb} = \Info{c} \vc{1}$), the former is
equivalent to
\begin{equation}
  \label{eq:v1-pace}
  \vc{1} \le \frac{1-g^{-1}(1-p^+(0))}{\Info{c}}.
\end{equation}
On the other hand, $d\Info{b}/dx \ge 0$ is equivalent to
\begin{equation}
  \frac{g(y)}{g'(y)} \ge \Info{c} \cdot \frac{f'(x)\cdot (1-p^+(x))}{p^{+\prime}(x)},
\end{equation}
which is similar to \eqref{eq:mono-cond-de} except with $(1-x)$
replaced by $q(x) := (1-p^+(x)) / p^{+\prime}(x)$.  Under the erasure
approximation, where $g(y) / g'(y) = y / (\db - 1)$ by
\eqref{eq:exit-g-beq}, it is thus sufficient to change the $s(x)$ in
\eqref{eq:binary-opt} into
\begin{equation}
  \label{eq:sL}
  \sL(x) = \sum_d \vc{d} x^{d-1} + (\db-1) q(x) \sum_d (d-1)\vc{d} x^{d-2},
\end{equation}
and replace $\vc{1}=0$ with the linear constraint
\begin{equation}
  \label{eq:v1-pace-constr}
  \vc{1} \le \smax \cdot (1 - g^{-1}(1-p^+(0)))
\end{equation}
corresponding to \eqref{eq:v1-pace}.  When not using the EA, the
counterpart of $\sDE(x)$, $\sDEL(x)$, can be defined in a similar
manner to Section~\ref{sec:mono-opt-de}, and $r(x)$ becomes $\rL(x) =
\sDEL(x) / \sL(x)$.  The maximization of $\IcthrL$ is then a
linear programming problem similar to \eqref{eq:binary-opt-de}, except
with $r(x)s(x)$ replaced by $\rL(x)\sL(x)$ and $\vc{1}$ constrained by
\eqref{eq:v1-pace-constr}.


\section{Non-binary LDGM Quantizers}
\label{sec:nonbinary}
The binary LDGM quantization codes designed in the last few sections
could, as we shall see in Section~\ref{sec:numerical}, achieve shaping
losses that are very close to the random-coding loss.  However, the
random-coding loss of binary codes is at least \unit[0.0945]{dB}; this
limitation has been observed in \cite{near-cap-dpc-tcq-ira} in view of
the performance advantage of 4-ary TCQ compared to the binary
convolutional codes used for shaping in \cite{close-to-cap-dpc}, and
it is more evident in LDGM quantization codes.  From
Fig.~\ref{fig:power-entropy}, it is clear that non-binary codes, i.e.\
those with a larger $m$, are necessary.

In channel coding, two types of approaches exist in dealing with
non-binary modulation schemes such as 4-PAM/16-QAM: one may use a
binary channel code and modulate multiple coded bits onto each channel
symbol, as in bit-interleaved coded modulation (BICM) with iterative
detection \cite{bicm,bicm-id}; alternatively, a non-binary channel
code such as trellis-coded modulation (TCM)
\cite{chan-coding-multi-level-phase} or a non-binary LDPC code can be
used, such that one coded symbol is mapped directly to a channel
symbol.  Similar methods can be applied to MSE quantization.  TCQ, for
example, has a 4-ary trellis structure just like TCM.  The use of LDGM
codes over Galois field $\GF(2^K)$ for quantization, as proposed in
\cite{quant-sig-sp-gen-fg-codes}, also fits in this category.
However, \cite{quant-sig-sp-gen-fg-codes} does not consider decimation
issues and degree distribution optimization much, and these problems
are more complex for such non-binary LDGM codes.  In MSE quantization,
where the mapping between $\GF(2^K)$ and the modulo-$2^K$ structure of
the reproduction alphabet is not natural anyway, such complexity seems
unjustified.  Therefore, we have instead adopted a BICM-like approach
in \cite{ldgm-vq-globecom07}, where the LDGM code itself is still
binary and every two coded bits in a codeword are Gray-mapped to a
4-ary reproduction symbol, and we have found that this approach allows
near-ideal codes to be designed under the erasure approximation with
relative ease.

In this section, we will propose an improved version of the scheme in
\cite{ldgm-vq-globecom07}, which also has near-ideal MSE performance
but allows even simpler code optimization, and is applicable to
general $2^K$-ary, not just 4-ary, cases.  Most of the optimization
methods proposed in the previous sections will then be extended to
this scheme.

\subsection{Quantizer Structure}
\begin{figure}[!t]
  \centering
  \includegraphics{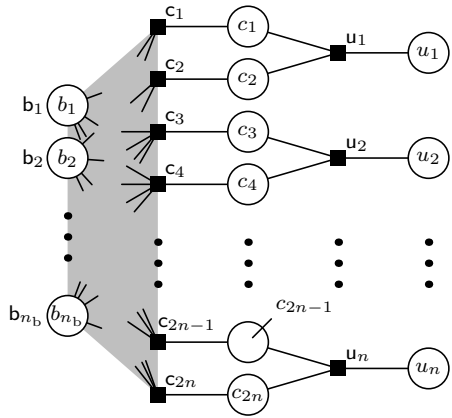}%
  \caption{The factor graph of the $2^K$-ary LDGM quantizer when $K=2$.
    Note that all \node{c}-nodes connecting to the same \node{u}-node
    have the same left-degree.  The factor graph also has a perturbed
    form akin to Fig.~\ref{fig:binary-fg-a}, with a $2^K$-ary variable
    node $\node{a}_j$ connecting to each $\node{u}_j$.}%
  \label{fig:mary-fg}%
\end{figure}

The $m$-ary LDGM quantizer with $m=2^K$ uses the codebook $\Lambda =
\mathcal{U} + m\SetZ^n$, where each codeword $\seq{u} \in \mathcal{U}$
is obtained by Gray-mapping every $K$ consecutive bits in a binary
LDGM codeword $\seq{c}$ of length $\nc = Kn$ into an $m$-ary symbol in
$\seq{u}$.  Denoting the generator matrix of the \emph{binary}
$(\nc,\nb)$ LDGM code by $\mat{G}$, its $\nb = nR$ information bits by
$\seq{b}$, the Gray mapping function by $\phi(\cdot)$ (e.g.\
$\phi(00)=0$, $\phi(10)=1$, $\phi(11)=2$, $\phi(01)=3$ for
$K=2$),\footnote{The optimization methods below appear to be usable
  for other mappings $\phi(\cdot)$ as well.  Indeed, $\phi(\cdot)$ can
  even conceivably be a vector-valued mapping for $\seq{y}$ being a
  sequence of \emph{vectors}, which results in a form of vector
  precoding \cite{vector-perturbation-mod-precoding1}, though various
  details remain to be worked out.} and denoting $j_k = K(j-1)+k$, we
have
\begin{gather}
  \label{eq:mary-dequant-c}
  \seq{c} = \seq{b} \mat{G}, \quad \seq{\ct}_j := (c_{j_1}, c_{j_2}, \dotsc, c_{j_K}), \\
  \label{eq:mary-dequant-u}
  \quad u_j = \phi(\seq{\ct}_j), \quad j=1,2,\dotsc,n.
\end{gather}
The corresponding factor graph for $\qseqy(\seq{b})$ is shown in
Fig.~\ref{fig:mary-fg}, where the \node{c}-nodes represent
\eqref{eq:mary-dequant-c} and the \node{u}-nodes represent
\eqref{eq:mary-dequant-u}.  Each factor $\e^{-t (y_j - u_j) \bmodI^2}$
in \eqref{eq:qy}, with $\mathcal{I} = [-m/2,m/2)$, is included in the
priors $\priprob{u}{j}$, which now has $m$ components since $u_j$ is
$m$-ary:
\begin{equation}
  \label{eq:priprobs-ml}
  \priprobx{u}{j}{u} = \frac{1}{Q_{\yt_j}} \e^{-t (y_j - u) \bmodI^2}
  = \pz((y_j - u) \bmodI).
\end{equation}
The quantization algorithm in Fig.~\ref{alg:mary} then follows from
the BP rules on this factor graph.

\begin{figure}[!t]
  \centering\footnotesize
  \begin{algorithmic}
    \STATE \COMMENT{compute the $2^K$-ary priors $\priprob{u}{j}$; $\mathcal{I} = [-2^{K-1}, 2^{K-1})$}
    \STATE $\priprobx{u}{j}{u} \assign \pz((y_j-u)\bmodI)$, $j=1,\dotsc,n$, $u=0,\dotsc,2^K-1$
    \vspace{0.5mm}%
    \STATE $\msg{bc}{ij_k} \assign \constmsg{*}$, $\msg{cu}{j_kj} \assign \constmsg{*}$, $i=1,\dotsc,\nb$, $j=1,\dotsc,n$, $k=1,\dotsc,K$
    \STATE $\priprob{b}{i} \assign \constmsg{*}$, $i=1,\dotsc,\nb$
    \STATE $\mathcal{E} \assign \{1,2,\dotsc,\nb\}$ \COMMENT{the set of bits not yet decimated}
    \STATE $\deltamax \assign 0$, $\Info{bc} \assign 0$
    \REPEAT[belief propagation iteration]
      \FOR[BP computation at $\node{u}_j$]{$j=1$ to $n$}
        \STATE Compute $\msg{uc}{jj_k}$ with \eqref{eq:msg_uc} for $k=1,\dotsc,K$
      \ENDFOR
      \FOR[BP computation at $\node{c}_{j_k}$]{$s=j_k=1$ to $\nc$}
        \STATE $\msg{cb}{si} \assign \msg{uc}{js} \oplus
          \left(\oplus_{i'\in\neighbor{bc}{\cdot s} \excluding{i}} \msg{bc}{i's}\right)$, $i\in\neighbor{cb}{s\cdot}$
        \STATE $\msg{cu}{sj} \assign \oplus_{i'\in\neighbor{bc}{\cdot s}} \msg{bc}{i's}$
      \ENDFOR
      \FOR[BP computation at $\node{b}_i$]{$i=1$ to $\nb$}
        \STATE $\msg{bc}{is} \assign \priprob{b}{i}
          \odot \left(\odot_{s' \in\neighbor{cb}{\cdot i} \excluding{s}} \msg{cb}{s' i} \right)$,
          $s \in \neighbor{bc}{i\cdot}$
        \STATE $\extprob{b}{i} \assign \odot_{s'\in\neighbor{cb}{\cdot i}} \msg{cb}{s'i}$
      \ENDFOR
      \STATE Estimate $\nextInfo{bc}$ and do decimation as in the binary case
    \UNTIL{$\mathcal{E}=\emptyset$}
    \STATE $b_i \assign 0$ (resp. $1$) if $\priprob{b}{i}=\constmsg{0}$ (or $\constmsg{1}$), $i=1,\dotsc,\nb$
    \STATE Compute $\seq{c}$ and $\seq{u}$ from $\seq{b}$ with \eqref{eq:mary-dequant-c} and \eqref{eq:mary-dequant-u}
    \STATE $z_j = (y_j - u_j) \bmodI$, $x_j = y_j - z_j$, $j=1,\dotsc,n$
  \end{algorithmic}
  \caption{The $2^K$-ary quantization algorithm.  The decimation part
    is almost the same as the one in Fig.~\ref{alg:binary}, so it is
    not reproduced here.}
  \label{alg:mary}
\end{figure}

\subsection{Code Optimization: Erasure Approximation}
\label{sec:mono-opt-ml-erasure}
The LDGM code here is still \node{b}-regular and \node{c}-irregular,
with all \node{b}-nodes having right-degree $\db$.  To simplify
analysis, we make all \node{c}-nodes connecting to the same
\node{u}-node have the same left-degree, which is called the
\emph{\node{c}-degree} of the \node{u}-node.  We denote by $w_d$ the
fraction of \node{u}-nodes with \node{c}-degree $d$, and by $v_d = K d
w_d / (R \db)$ the corresponding fraction of edges.

Using essentially the same argument as in Section~\ref{sec:de-mse-intro},
under the monotonicity conditions a reference codeword denoted by
$\seq{u}^*$, $\seq{c}^*$ and $\seq{b}^*$ can be found with the TTD,
and the corresponding quantization error $\seq{z}^* =
(\seq{y}-\seq{u}^*) \bmodIn$ is strongly typical with respect to
$\pz(z)$.

As in the binary case, we begin with the simpler erasure
approximation, which can serve as a starting point for more accurate
methods.  Similar to Section~\ref{sec:mse-erasure-approx}, EA assumes
that e.g.\ $\Info{bc}$ is determined solely by $\Info{cb}$ and can be
computed by assuming the density of \node{c}-to-\node{b} messages to
be erasure-like with respect to the reference codeword.  Clearly, with
a fraction $\Info{b}$ of decimated \node{b}-nodes, the output
$\nextInfo{bc}$ and $\Ibext$ from \node{b}-nodes are still given by
\eqref{eq:exit-binary-b} and \eqref{eq:Ibext}.  Below we compute the
\node{c}-curve relating the output $\Info{cb}$ from \node{c}-nodes to
their input $\Info{bc}$.

Consider a \node{u}-node $\node{u}_j$ with \node{c}-degree $d$.  Due
to EA, each incoming \node{c}-to-\node{u}
message $\msg{cu}{j_kj}$ must be either $\constmsg{c^*_{j_k}}$ or
$\constmsg{*}$, with the former occurring with probability
$(\Info{bc})^d$.  Each outgoing message is given by
\begin{equation}
  \label{eq:msg_uc}
  \msgx{uc}{jj_k}{c} = \sum_{\seqs{\ct}:\ct_k=c} \priprobx{u}{j}{\phi(\seq{\ct})} \prod_{k'\ne k} \msgx{cu}{j_{k'}j}{\ct_{k'}},
  \quad c=0,1,
\end{equation}
which depends on the set
\begin{equation}
  \label{eq:S}
  \mathcal{S} = \{ k' \in \{1,\dotsc,K\} \excluding{k} \mid \msg{cu}{j_{k'}j} = \constmsg{c^*_{j_{k'}}} \}
\end{equation}
of used incoming messages that are ``known''.  It is now useful to
define auxiliary random variables $\uck$, $\seq{\cck}$ and $\yck$,
such that $\uck=\phi(\seq{\cck})$ is $0,1,\dotsc,m-1$ with equal
probability and $\yck \in [0,m)$ has conditional pdf $p(\yck \condmid
\uck) = \pz((\yck - \uck)\bmodI)$.  $p(\yck) = \sum_{\uck} p(\uck)
p(\yck\condmid\uck)$ is then a uniform distribution over $[0,m)$ and
$p(\uck \condmid \yck) = \pz((\yck - \uck)\bmodI)$, so
\eqref{eq:priprobs-ml} becomes simply
\begin{equation}
  \priprobx{u}{j}{u} = p(\uck = u \condmid \yck = y_j), \quad u = 0,1,\dotsc,m-1,
\end{equation}
and \eqref{eq:msg_uc} becomes the conditional
distribution (omitting $c$-independent factors)\footnote{$\seq{\cck}_{\mathcal{S}} =
  \seq{c}^*_{j_{\mathcal{S}}}$ is abbreviation for $\cck_k =
  c^*_{j_k}$, $\forall k \in \mathcal{S}$\@.}
\begin{align}
  \label{eq:msg_uc-alt}
  \msgx{uc}{jj_k}{c} &= \sum_{\seq{\ct}: \ct_k = c, \seq{\ct}_{\mathcal{S}} = \seq{c}^*_{j_{\mathcal{S}}}} p(\uck = \phi(\seq{\ct}) \condmid \yck = y_j) \\
  \label{eq:msg_uc-alt1}
  &= p(\cck_k = c, \seq{\cck}_{\mathcal{S}} = \seq{c}^*_{j_{\mathcal{S}}} \condmid \yck = y_j) \\
  \label{eq:msg_uc-alt2}
  &= p(\cck_k = c \condmid \seq{\cck}_{\mathcal{S}} = \seq{c}^*_{j_{\mathcal{S}}}, \yck = y_j).
\end{align}
To obtain the average MI $\Info{cb}$, we first average
$H(\msg{uc}{jj_k}) = H(\cck_k \condmid \seq{\cck}_{\mathcal{S}} =
\seq{c}^*_{j_{\mathcal{S}}}, \yck=y_j)$ over $j$ for a given $k$ and
$\mathcal{S}$\@.  For $n\to\infty$, using the typicality of $\seq{z}^*$
with respect to $\pz(z)$, this yields the average conditional entropy
\begin{equation}
  \label{eq:Hc_kS}
  \Hc(k,\mathcal{S}) = H(\cck_k \condmid \seq{\cck}_{\mathcal{S}}, \yck),
\end{equation}
which can be computed using the above probability distributions of
$\seq{\cck}$ and $\yck$.  Among \node{u}-to-\node{c} messages from
\node{u}-nodes with \node{c}-degree $d$, $k = 1,\dotsc,K$ with equal
frequency and each $\mathcal{S}$ with $\cardinal{\mathcal{S}} = k'$
occurs with probability $\Info{bc}^{dk'} \cdot
(1-\Info{bc}^d)^{K-1-k'}$, therefore if we define, for $k' =
0,\dotsc,K-1$,\footnote{This $\Info{c}$ satisfies $K\Info{c} = K -
  H(\seq{\cck} \condmid \yck) = K - H_t$ due to \eqref{eq:Hc_kS}.}
\begin{gather}
  \label{eq:Hc}
  \Hcx{k'} = \frac{1}{K}\binom{K-1}{k'}^{-1} \sum_{k=1}^K
  \sum_{\substack{\mathcal{S} \subseteq \{1,\dotsc,K\} \excluding{k} \\ \cardinal{\mathcal{S}} = k'}}
  \Hc(k,\mathcal{S}), \\
  \label{eq:Ic-ml}
  \Infox{c}{k'} = 1 - \Hcx{k'},\quad \Info{c} = \frac{1}{K}\sum_{k'=0}^{K-1} \Infox{c}{k'},
\end{gather}
the average MI of these messages is then
\begin{equation}
  \label{eq:Iuc_d}
  \Infox{uc}{d} 
  = \sum_{k'=0}^{K-1} \binom{K-1}{k'} \cdot \Infox{c}{k'} \cdot \Info{bc}^{dk'} \cdot (1-\Info{bc}^d)^{K-1-k'}.
\end{equation}
Finally, since the \node{b}-to-\node{c} message density is assumed to
be erasure-like, a look at the local tree-like neighborhood of a
\node{c}-node reveals that
\begin{equation}
  \label{eq:exit-c-ml}
  \Info{cb} = \sum_d v_d \Infox{uc}{d} \Info{bc}^{d-1}
  = \sum_{k',d} v_d \Infox{c}{k'} \cdot \alpha_{k',d}(\Info{bc}),
\end{equation}
where
\begin{equation}
  \label{eq:alpha}
  \alpha_{k',d}(x) = \binom{K-1}{k'} x^{d(k'+1)-1} (1-x^d)^{K-(k'+1)}.
\end{equation}

Having obtained the EXIT curves \eqref{eq:exit-binary-b},
\eqref{eq:Ibext} and \eqref{eq:exit-c-ml}, the EBP curve can be
defined just like the binary case, as the relationship between the
$\Info{b}$ making $\nextInfo{bc} = \Info{bc}$ and
$\Ibext$.\footnote{The area below this erasure-approximated EBP curve,
  as defined in Fig.~\ref{fig:ebp-area}, can be found to be
  $K\Info{c}/R - \db v_1 \Infox{c}{0} + (1 - (1-v_1 \Infox{c}{0})^{\db})$,
  which equals $K\Info{c}/R$ when $v_1 = 0$ and slightly smaller
  otherwise.  Interestingly, this is the same as the binary case
  except that $\Info{c}$ becomes $K\Info{c}$ and $\vc{1}\Info{c}$
  becomes $v_1 \Infox{c}{0}$.  As in footnote~\ref{fn:map-exit}, the
  MAP EXIT curve can also be defined, and the area below it under the
  equi-partition condition is now $K\Info{c}/R$ as well.} The
monotonicity conditions are again \eqref{eq:mono-cond1} and
\eqref{eq:mono-cond2}; the former means $v_1 = 0$, and the latter,
$d\Info{b}/dx \ge 0$ ($x=\Info{bc}$), becomes
\begin{equation}
  \label{eq:sK-constr-ml}
  \sum_{k'=0}^{K-1} \Infox{c}{k'} \cdot s_{k'}(x) \le 1, \quad x \in [0,1],
\end{equation}
where
\begin{equation}
  \label{eq:sK-ml}
  s_{k'}(x) = \sum_d v_d \left( \alpha_{k',d}(x) + (1-x)(\db-1) \alpha'_{k',d}(x) \right).
\end{equation}
For a given degree distribution, the monotonicity threshold $\tthr$
(or the corresponding $\Info{c}$ denoted by $\Icthr$) is the maximum
$t$ such that \eqref{eq:sK-constr-ml} holds.  Since all
$\Infox{c}{k'}$'s are increasing functions of $t$, the degree
distribution with the largest $\tthr$ can be found via a linear search
for the maximum $t$ at which the linear constraints
\eqref{eq:sK-constr-ml} and
\begin{equation}
  \label{eq:v-constr-ml}
  \sum_d v_d = 1, \quad \sum_d \frac{v_d}{d} = \frac{K}{R\db}, \quad v_d \ge 0,
\end{equation}
on $v_d$, with $d\in\mathcal{D}$ given by \eqref{eq:degree-set}, are
feasible.  As in the binary case, we can then use $t=\tthr$ in the
quantization algorithm.

In practice we often have a good guess $t^*$ (e.g.\ $t_0(R)$ when
$\db$ is large enough) of $\tthr$, along with the corresponding
$\Infox{c}{k'}^*$ and $\Info{c}^*$.  If $t^*$ is close to $\tthr$, we
can approximately view $\Infox{c}{k'}/\Info{c}$ as
$t$-independent constants $\gamma_{k'} := \Infox{c}{k'}^* /
\Info{c}^*$, and \eqref{eq:sK-constr-ml} then becomes \eqref{eq:s-constr-ml} with $s(x)$ given by
\begin{equation}
  \label{eq:s-ml}
  s(x) = \sum_{k'=0}^{K-1} \gamma_{k'} \cdot s_{k'}(x),
\end{equation}
so the above optimization is again a linear programming problem
\eqref{eq:binary-opt}.

\subsection{Code Optimization: Density Evolution}
\label{sec:mono-opt-ml-de}
As in the binary case, it is expected that discretized density
evolution will yield better codes by avoiding the erasure
approximation.  The method used is essentially the same; the only
difficulty lies in the computation of the outgoing
\node{u}-to-\node{c} message density from \node{u}-nodes with
\node{c}-degree $d$, for which the $K-1$ incoming \node{c}-to-\node{u}
messages follows i.i.d.\ a given density.  When $K=2$, this
\node{u}-to-\node{c} density can be computed with a two-dimensional
lookup table on the quantized incoming \node{c}-to-\node{u} $L$-value
and the quantized $y_j$, much like the lookup table used at
\node{c}-nodes.

For larger $K$, this table-lookup method requires a table with $K$
dimensions, and the resulting computational complexity is likely
impractical.  We have not investigated this case in detail, as $K=2$
is already sufficient for MSE quantization, but it seems that a
Monte-Carlo approach may be effective for such density computation at
\node{u}-nodes.

The DE results can be used to obtain the EXIT curves, and the
monotonicity threshold be thus optimized, in essentially the same
manner as Sections~\ref{sec:exit-de} and \ref{sec:mono-opt-de}\@.  In
the computation of the correction factor $r(x)$, \eqref{eq:s-ml}
should be used as the reference $s(x)$.

\subsection{Pacing of Decimation}
\label{sec:pacing-ml}
Under a finite number $L$ of iterations, the approximate relationship
\eqref{eq:mse-Ai-emp} between MSE and delta-area still holds according
to simulation results (but $P_0$ is now $m^2/12$), therefore we can
still optimize the pace of decimation by minimizing the delta-area
with the same methods in Section~\ref{sec:opt-pacing}\@.  In particular,
\eqref{eq:pace-approx} is unchanged from the binary case.

The method in Section~\ref{sec:pacing-aware-code-opt} can still be used
to optimize the degree distribution under a finite number of
iterations with a given pace.  However, now $\Info{c} f(0)$, the
$\Info{cb}$ when \node{b}-to-\node{c} messages are all-$\constmsg{*}$,
should be $v_1 \Infox{c}{0}$ according to \eqref{eq:exit-c-ml}, in
which the erasure approximation is exact here.  Therefore
\eqref{eq:v1-pace} should be replaced by
\begin{equation}
  \label{eq:v1-pace-ml}
  v_1 \le \frac{1-g^{-1}(1-p^+(0))}{\Infox{c}{0}} \approx \frac{1-g^{-1}(1-p^+(0))}{\gamma_0 \Info{c}},
\end{equation}
and the corresponding linear constraint \eqref{eq:v1-pace-constr}
becomes
\begin{equation}
  \label{eq:v1-pace-constr-ml}
  v_1 \le \smax \cdot \frac{1-g^{-1}(1-p^+(0))}{\gamma_0}.
\end{equation}
Finally, $\sL(x)$ now has the same form as $s(x)$ in \eqref{eq:s-ml},
except with the $(1-x)$ factor in \eqref{eq:sK-ml} replaced by $q(x)
= (1-p^+(x)) / p^{+\prime}(x)$.

\section{Simulation Results}
\label{sec:numerical}

\begin{table*}[!t]
  \centering
  \caption{Performance of LDGM Quantization Codes at $n=10^5$}
  \label{tab:long-code-perf}
  \begin{tabular}{cccccccccccc}
    \toprule
    \mrowa{2}{$L_0$} & \mrowa{2}{$K$} & \mrowa{2}{$R$ (\unit{b/s})} & \mrowa{2}{$\db$} & \mrowa{2}{Method}
    & \mrowa{2}{$K\IcthrorL$} & \mrowa{2}{$L\Aignore$} & \mcol{4}{Losses: $10\log_{10}(\cdot / P^*_{t_0(R)})$ (dB)} & \mrowa{2}{$L$} \\
    \cmidrule(lr){8-11}
    & & & & & & & $P_{t_0(R)}$ & $P_{\tthr}$ & \eqref{eq:mse-Ai-emp} & Actual $\hat{\sigma}^2$ \\
    \midrule
    \mrow{5}{$10^2$} & \mrow{2}{1} & \mrow{2}{0.4461} & \mrow{2}{12} & DE & 0.4427 & 3.44 & \mrow{2}{0.0976} & 0.1174 & 0.3479 & 0.3241 & 100 \\ 
    & & & & DE-PO & \textit{0.4525} & 3.46 & & N/A & 0.2921 & 0.2721 & 100 \\ 
    \cmidrule(lr){2-9}
    & \mrow{3}{2} & \mrow{2}{0.9531} & \mrow{2}{11} & DE & 0.9460 & 3.44 & 0.0010 & 0.0437 & 0.6441 & 0.4949 & 100 \\ 
    & & & & DE-PO & \textit{0.9672} & 3.46 & 0.0010 & N/A & 0.5282 & 0.3962 & 100 \\ 
    & & 0.4898 & 20 & DE-PO & \textit{0.5010} & 3.47 & 0.0369 & N/A & 0.2306 & 0.2676 & 99 \\ 
    \midrule
    \mrow{2}{$10^3$} & \mrow{2}{1} & \mrow{2}{0.4461} & \mrow{2}{12} & DE & 0.4427 & 3.44 & \mrow{2}{0.0976} & 0.1174 & 0.1466 & 0.1537 & 809 \\ 
    & & & & DE-PO & \textit{0.4442} & 3.45 & & N/A & 0.1377 & 0.1443 & 815 \\ 
    \cmidrule(lr){1-12}
    \mrow{4}{$10^{3\prime}$} & \mrow{2}{1} & \mrow{2}{0.4461} & \mrow{2}{12} & DE & 0.4427 & 3.44 & \mrow{2}{0.0976} & 0.1174 & 0.1402 & 0.1426 & 1036 \\ 
    & & & & DE-PO & \textit{0.4442} & 3.45 & & N/A & 0.1318 & 0.1400 & 1023 \\ 
    \cmidrule(lr){2-9}
     & \mrow{2}{2} & 0.9531 & 11 & DE & 0.9460 & 3.44 & 0.0010 & 0.0437 & 0.1049 & 0.0876 & 1046 \\ 
    & & 0.6285 & 17 & DE-PO & \textit{0.6256} & 3.49 & 0.0130 & N/A & 0.0660 & 0.0741 & 1022 \\ 
    \midrule
    $10^4$ & 2 & 0.9531 & 11 & DE & 0.9460 & 3.44 & 0.0010 & 0.0437 & 0.0608 & 0.0565 & 3778 \\ 
    \cmidrule(lr){1-12}
    \mrow{2}{$10^{4\prime}$} & 1 & 0.4461 & 12 & DE & 0.4427 & 3.44 & 0.0976 & 0.1174 & 0.1210 & 0.1245 & 6678 \\ 
    & 2 & 0.9531 & 11 & DE & 0.9460 & 3.44 & 0.0010 & 0.0437 & 0.0514 & 0.0423 & 8356 \\ 
    \bottomrule
  \end{tabular}
\end{table*}

In this section we evaluate the MSE performance of our quantization
codes by Monte Carlo simulation.  For our $m$-ary code ($m=2,4$),
without loss of generality each source sequence $\seq{y}$ is uniformly
sampled from $[0,m]^n$, quantized to $\seq{x}$, and the MSE is then
evaluated as $\frac{1}{n} \sum_{j=1}^n \abs{y_j-x_j}^2$.  Denoting
$\hat{\sigma}^2$ as the average MSE over a number of source sequences
used in the simulation (usually 20 at $n=10^5$ and more for smaller
$n$), the shaping loss can be estimated by $10\log_{10}
(\hat{G}(\Lambda) / G^*)$, with
\begin{equation}
  \label{eq:sim-loss}
  \frac{\hat{G}(\Lambda)}{G^*} \approx \frac{\hat{\sigma}^2\rho^{2/n}}{(2\pi\e)^{-1}} = 
  \left(2^R/m\right)^2 2\pi\e \hat{\sigma}^2.
\end{equation}

We will first evaluate long-block performance ($n=10^5$) of binary
and 4-ary codes, then the impact of smaller block lengths $n$
will be investigated.  Unless otherwise noted:
\begin{itemize}
\item The degree distribution is optimized with one of the following methods:
  \begin{enumerate}
  \item \emph{DE}: maximize $\Icthr$ with quantized density evolution
    (Sections~\ref{sec:mono-opt-de}, \ref{sec:mono-opt-ml-de});
  \item \emph{EA}: maximize $\Icthr$ under the erasure approximation
    (Sections~\ref{sec:mse-erasure-approx}, \ref{sec:mono-opt-beq},
    \ref{sec:mono-opt-ml-erasure});
  \item \emph{DE-PO}: maximize $\IcthrL$ ($L=L_0$) with quantized DE
    (Sections~\ref{sec:pacing-aware-code-opt}, \ref{sec:pacing-ml});
  \item \emph{EA-PO}: maximize $\IcthrL$ with EA
    (Sections~\ref{sec:pacing-aware-code-opt}, \ref{sec:pacing-ml}).
  \end{enumerate}
\item The code is randomly generated from the degree distribution by
  random edge assignment, followed by the removal by pairs of
  duplicate edges between two nodes.
\item The $t$ used in the quantization algorithm is $t_0(K\Icthr)$ or
  $t_0(K\IcthrL)$, such that $\Info{c} = \IcthrorL$.  When the EA or
  EA-PO method is used, this $\IcthrorL$ is the erasure-approximated
  result; the true $\IcthrorL$ is lower.
\item The greedy decimation algorithm is used.
\item The pace of decimation is given by \eqref{eq:pace-approx}.
\item The decimation process is controlled to make the actual
  iteration count $L$ close to the target $L_0$, using the throttled
  algorithm if $L_0$ is marked with a prime (e.g.\ $L_0 =
  10^{3\prime}$), and the unthrottled algorithm otherwise.
\item The recovery algorithm in Section~\ref{sec:impact-dec-mse} is
  not used.
\end{itemize}

\subsection{Performance of the Greedy Decimator at $n=10^5$}
\label{sec:sim-long}
For binary codes, the random-coding loss is significant, therefore we
choose the code rate $R=\nb/n=\unit[0.4461]{b/s}$ with $t_0(R)=4$,
where the random-coding loss of \unit[0.0976]{dB} is close to minimum.

For 4-ary codes, the code rate is chosen to be
$R=\nb/n=\unit[0.9531]{b/s}$ at $t=2$ in some cases, where the
random-coding loss of \unit[0.0010]{dB} is close to minimum.  However,
for moderate iteration counts $L$ there are now a large range of rates
for which the random-coding loss is small compared to the loss due to
the delta-area, and \eqref{eq:mse-Ai-emp} suggests that the latter
loss increases when higher rates are used, since $P_0$ becomes
a larger multiple of $P_t$.  Therefore, we also experiment with
somewhat lower rates that may give better MSE performance.

On the choice of $\db$, we note that gap between $K\Icthr$ and its
ideal value $R$ decreases rapidly as $\db$ increases, but
computational complexity also increases, and the finite-$n$ loss may
worsen when the factor graph is denser.  Therefore, we choose $\db$
such that the maximum \node{c}-degree is about 50--100.

Results are shown in Table~\ref{tab:long-code-perf}\@.  $K\Icthr$ is
shown for each code optimized with the DE method (the factor $K$ makes
it easy to compare $K\Icthr$ with its ideal value $R$), and when the
DE-PO method is used $K\IcthrL$ is shown instead in italics to
indicate the choice of $t=t_0(K\IcthrL)$.\footnote{In the iterative
  optimization process in Section~\ref{sec:mono-opt-de}, the $\Icthr$
  of an optimized code can either be obtained from
  \eqref{eq:binary-opt-de} as $1/\smax$, or more accurately, by making
  it the base code, rerunning DE on it, and computing $\Icthr$ from
  \eqref{eq:Icthr-de}.  $\Icthr$ (but not $\IcthrL$) in
  Table~\ref{tab:long-code-perf} is computed with the latter method.}
The $L\Aignore$ value is obtained from \eqref{eq:pace-approx},
\eqref{eq:pace-vs-deltaIb}, \eqref{eq:pace-L} and \eqref{eq:pace-Ai};
technically it is only applicable when $L\to\infty$ but in practice
its accuracy is good even when $L=100$.  The four losses that follow
are with respect to the ideal MSE $P^*_{t_0(R)}$ defined in
Section~\ref{sec:random-coding-loss}, and they are respectively
\begin{enumerate}
\item the random-coding loss;
\item the \emph{TTD loss} corresponding to the MSE $P_{\tthr}$
  achieved by the TD, when $L\to\infty$ and it is able to synchronize
  with the TTD;
\item the loss estimate \eqref{eq:mse-Ai-emp}, in which we divide
  $L\Aignore$ above by the \emph{actual} average iteration count $L$
  to obtain $\Aignore$;
\item the actual shaping loss \eqref{eq:sim-loss} from simulation
  results.
\end{enumerate}

Several observations can be made:
\begin{itemize}
\item The shaping loss decreases as the iteration count $L$ increases,
  and can approach the random-coding loss and even be lower than the
  TTD loss (because the greedy decimator is better than the TD) when
  $L$ is large.
\item At small $L$, adjusting the degree distribution according to $L$
  with the DE-PO method can improve performance significantly.
\item At a given $L$, the loss due to the finite $L$ is larger for
  higher rates.  Therefore, for 4-ary codes it is indeed helpful to
  small-$L$ performance if a smaller $R$ than that minimizing the
  random-coding loss is used.
\item For binary codes the random-coding loss becomes dominant at
  large $L$ and limits the achievable shaping loss.
\item $L\Aignore$ is virtually code-independent.
\item The shaping loss can be well predicted by \eqref{eq:mse-Ai-emp};
  it is not entirely accurate because the formula itself is only a
  heuristic, it is given for TD-with-recovery but here used with
  GD,\footnote{As will be shown in Table~\ref{tab:td-gd}, the greedy
    decimator is much less sensitive to code optimization and to the
    choice of $t$ (or $\Info{c}$) than TD with recovery, so its
    performance tends to be better than the estimate
    \eqref{eq:mse-Ai-emp} when $K\Icthr$ is significantly lower than
    its ideal value $R$.}  and also because it ignores the difference
  between the throttled and unthrottled decimation algorithms and the
  loss due to finiteness of $n$.
\end{itemize}

Through better degree distribution optimization methods, pacing of
decimation, and choice of code rate, we have achieved in
Table~\ref{tab:long-code-perf} better MSE performance than in
\cite{ldgm-vq-globecom07} at the same complexity.  In
Table~\ref{tab:long-code-perf-ind}, we analyze the contribution of
each individual improvement to the MSE performance of 4-ary LDGM
quantization codes.  Starting with the method of
\cite{ldgm-vq-globecom07} in the first row, which uses a slightly
different code construction, EA-based optimization method and uniform
pacing, we introduce one by one the following improvements in the
subsequent five rows:
\begin{enumerate}
\item \label{imp:code} The code construction in Fig.~\ref{fig:mary-fg} optimized with
  EA;
\item \label{imp:pace} Optimized pace of decimation in \eqref{eq:pace-approx};
\item \label{imp:PO} Pacing-aware code optimization in Section~\ref{sec:pacing-aware-code-opt};
\item \label{imp:rate} The use of lower rates (\unit[0.4898]{b/s} for
  $L_0 = 10^2$ and \unit[0.6285]{b/s} for $L_0 = 10^{3\prime}$) than
  the random-coding-loss-minimizing \unit[0.9531]{b/s} rate used in
  previous rows;
\item \label{imp:de} Quantized DE that avoids the erasure approximation used in
  previous rows.
\end{enumerate}
$\db = 11$ is used in all but the first row, where the right-degree of
each \node{b}-node is $2\db = 12$ \cite{ldgm-vq-globecom07}.  The
average actual iteration counts $L$ are shown in parentheses.  Since
$L$ varies considerably when $L_0 = 10^{3\prime}$, for the purpose of
a fairer comparison, we also show in italics the adjusted shaping
losses approximately corresponding to $L=10^3$.\footnote{The
  adjustment uses the tradeoff $\unit[0.66\cdot 10^{-4}]{dB}$ per
  iteration between shaping loss and $L$.  This tradeoff factor is
  obtained by reducing $L_0$ from $1000'$ to $935'$ for the last row
  in Table~\ref{tab:long-code-perf-ind}; the resulting shaping loss
  increases by \unit[0.0040]{dB} to \unit[0.0781]{dB} and $L$
  decreases by 60 to 962, and $0.0040/60 = 0.66\cdot 10^{-4}$.}

\begin{table}[!t]
  \centering
  \caption{Effects of Various Optimizations on Shaping Loss (\dB) of 4-ary LDGM Codes}
  \label{tab:long-code-perf-ind}
  \begin{tabular}{cr@{ }lr@{ }lc}
    \toprule
    Code & \mcol{2}{$L_0 = 10^2$} & \mcol{3}{$L_0 = 10^{3\prime}$} \\
    \midrule
    \cite{ldgm-vq-globecom07}, unif.~pace & 0.5420 & (100) & 0.1022 & (953) & \emph{0.0991} \\
    EA, unif.~pace & 0.5530 & (99) & 0.1037 & (948) & \emph{0.1002} \\
    EA, opt.~pace & 0.4594 & (100) & 0.0875 & (995) & \emph{0.0872}\\
    EA-PO & 0.3641 & (100) & 0.0847 & (988) & \emph{0.0839} \\
    EA-PO, low $R$ & 0.2501 & (99) & 0.0861 & (960) & \emph{0.0834} \\
    DE-PO, low $R$ & 0.2676 & (99) & 0.0741 & (1022) & \emph{0.0756} \\
    \bottomrule
  \end{tabular}
\end{table}

We observe from Table~\ref{tab:long-code-perf-ind} that improvements
\ref{imp:pace}), \ref{imp:PO}) and \ref{imp:rate}) are all important
when $L_0=10^2$, but quantized DE (compared to EA) is only helpful
when $L_0 = 10^{3\prime}$ or larger, in which case it can decrease the
shaping loss by about \unit[0.01]{dB}.  Technically, as is evident
from Fig.~\ref{fig:exit-de-s}, the codes optimized by EA usually have
significantly suboptimal true monotonicity thresholds, but apparently
the greedy decimator, unlike the TD with recovery on which our
analysis is based, can avoid most of this loss.  We will further
investigate this below.

\subsection{Performance of the Typical Decimator}
\label{sec:sim-typ}
Having discussed the greedy decimator, now we look at the typical
decimator on which most of our theoretical analysis is based.  Good
performance from the TD requires the use of the recovery algorithm,
which we have only implemented for the binary case as shown in
Section~\ref{sec:impact-dec-mse},\footnote{A similar algorithm for the
  $2^K$-ary case is conceivable but significantly more complex, since
  the desired distribution of some $y_j$ would depend on $K$ incoming
  messages $\msg{cu}{j_kj}$, rather than just one $\extprob{c}{j}$ in
  the binary case.} therefore only binary codes are considered here.

The results are shown in Table~\ref{tab:td-gd} for the two binary
codes in Table~\ref{tab:long-code-perf} optimized with method DE-PO at
respectively $L_0=10^2$ and $L_0=10^3$.  We additionally include the
code optimized with EA-PO at the same $R$, $\db$ and $L_0$ as an
example of one with a poor monotonicity threshold: its
erasure-approximated $\IcthrL$ is 0.4469, but the true $\IcthrL$ is
much lower at 0.3836 due to the use of EA.  The shaping losses of this
code for $\Info{c}$ at 0.4469 and at 0.3836 are shown respectively in
the third and fourth row of Table~\ref{tab:td-gd}.  \emph{TD} and
\emph{GD} denote the typical and the greedy decimators, while
\emph{TD-R} and \emph{GD-R} refer to the corresponding decimators with
the recovery algorithm.  The loss estimates are obtained via
\eqref{eq:mse-Ai-emp}, with $\Aignore$ computed from DE results
without using the large-$L$ approximation, so they differ slightly
from the estimates in Table~\ref{tab:long-code-perf}.

\begin{table}[!t]
  \centering
  \caption{Shaping Loss (\dB) of Binary LDGM Codes with the Typical
    and the Greedy Decimators at $n=10^5$}
  \label{tab:td-gd}
  \begin{tabular}{cccccc}
    \toprule
    Code & Est. & TD & TD-R & GD & GD-R \\
    \midrule
    $L_0=10^2$,DE-PO & 0.2894 & 0.9128 & 0.2923 & 0.2721 & 0.2291 \\ 
    $L_0=10^3$,DE-PO & 0.1330 & 0.4678 & 0.1479 & 0.1443 & 0.1296 \\ 
    $L_0=10^3$,EA-PO & 0.2530 & 0.4592 & 0.1834 & 0.1463 & 0.1741 \\ 
    ($\Info{c}$: 0.4469, 0.3836) & 0.4871 & 0.5888 & 0.4968 & 0.1526 & 0.2649 \\ 
    \bottomrule
  \end{tabular}
\end{table}

\begin{figure}[!t]
  \centering
  \includegraphics{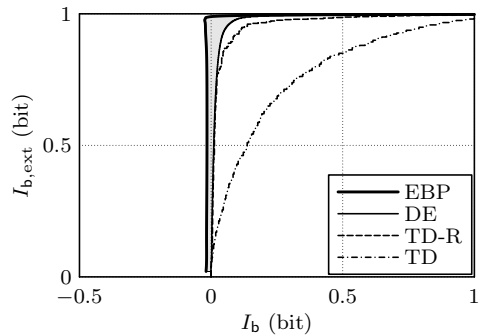}%
  \caption{Comparison of EBP and actual curves with TD and TD-R at
    $L_0 = 10^3$.  The curve labeled ``DE'' is the actual curve
    computed from DE results as used in Section~\ref{sec:opt-pacing}.
    The curves labeled ``TD'' and ``TD-R'' are the trajectories of
    $(\Info{b},\Ibext)$ followed by the actual quantizer when
    decimating a source sequence using the respective decimators,
    where $\Info{b}$ is the fraction of decimated $b_i$'s and $\Ibext$
    is the average of $1-H(\extprob{b}{i})$.}%
  \label{fig:ebp-recovery-compare}%
\end{figure}

We see that the typical decimator by itself performs rather poorly,
but with recovery its MSE performance is at least close to that
predicted by \eqref{eq:mse-Ai-emp}.  This can be observed more clearly
from Fig.~\ref{fig:ebp-recovery-compare}.  When TD is used without
recovery, imperfect decimation causes the message densities to become
far from consistent, in turn making the MI of the extrinsic
$\extprob{b}{i}$ messages far lower than the $\Ibext$ predicted by DE,
which is only accurate for consistent densities close to those
encountered in the DE process.  This, in effect, greatly increases the
delta-area and thus the MSE.  With the recovery algorithm, the
$\Ibext$ from the quantizer matches much better (though not perfectly)
with the DE result, showing that the message densities have been kept
mostly consistent.\footnote{The loss due to imperfect recovery is not
  as large as that estimated by \eqref{eq:mse-Ai-emp} though, if the
  area between the EBP curve and the TD (TD-R) curve in
  Fig.~\ref{fig:ebp-recovery-compare} is used as $\Aignore$.  The
  estimated losses are \unit[0.3925]{dB} for TD-R and
  \unit[1.4594]{dB} for TD, but the actual shaping losses are only
  respectively \unit[0.2925]{dB} and \unit[0.8797]{dB} for the source
  sequence used.  The likely reason for this discrepancy is that our
  method for estimating message MIs in
  Section~\ref{sec:mse-erasure-approx} is accurate only for symmetric
  message densities, so it does not well characterize the deviations
  of the message densities from consistency (symmetry).}

Table~\ref{tab:td-gd} also shows that, for the first two
well-optimized codes whose $\IcthrL$ are close to ideal, TD-R and GD
have similar performance, and GD-R works even better, suggesting that
the recovery algorithm (whose complexity is a moderate $O(n)$ per
iteration) is also useful in practical quantizers.  However, when
using the code optimized with EA-PO and thus having low $\IcthrL$, GD
performs decidedly better than TD-R and even GD-R; apparently, GD is
much less sensitive to the code or to the choice of $\Info{c}$.

\subsection{Finite-Length Effects}
\label{sec:sim-short}
Like LDPC codes with random edge assignment, LDGM quantization codes
require large block sizes $n$ to perform well.  As an example, we
consider the $R=\unit[0.6285]{b/s}$ 4-ary code designed with the DE-PO
method for $L_0 = 10^{3\prime}$ in Table~\ref{tab:long-code-perf}, and
its small-$n$ shaping losses at this $L_0$ are shown in
Table~\ref{tab:short-code-perf}.  For comparison, we also include in
Table~\ref{tab:short-code-perf} the shaping losses of TCQ, as well as
the sphere-covering (SC) bound
\cite{tcq-memoryless-gauss-markov} 
\begin{equation}
  \label{eq:vq-blocklen-loss}
  \frac{G(\Lambda)}{G^*} \ge \frac{e \Gamma(n/2+1)^{2/n}}{n/2+1},
\end{equation}
which is a lower bound of MSE at finite $n$, derived for exactly
spherical Voronoi regions of $\Lambda$.

\begin{table}[!t]
  \centering
  \caption{Shaping Loss (\dB) of Short 4-ary LDGM Codes at $L_0 = 10^{3\prime}$}
  \label{tab:short-code-perf}
  \begin{tabular}{rccc}
    \toprule
    $n$ & LDGM (\unit[0.6285]{b/s},DE-PO) & $2^{11}$-state TCQ & SC bound \\
    \midrule
    100 000 & 0.0741 & 0.1335 & 0.0005 \\
    30 000 & 0.0929 & 0.1339 & 0.0014 \\
    10 000 & 0.1297 & 0.1362 & 0.0036 \\
    3 000 & 0.2096 & 0.1394 & 0.0104 \\
    1 000 & 0.3225 & 0.1515 & 0.0263 \\
    300 & 0.5100 & 0.1901 & 0.0703 \\
    \bottomrule
  \end{tabular}
\end{table}

We observe from Table~\ref{tab:short-code-perf} that LDGM quantization
codes suffer significant loss when $n$ is small.  In particular, the
loss in the sphere-covering bound scales as $n^{-1}\ln n$, and TCQ's
performance loss due to small $n$ appears to scale similarly, but for
LDGM-based quantizers this small-$n$ loss decreases much more slowly
as $n$ increases.


\subsection{Comparison to TCQ}
For comparison purposes, we show the MSE performance of TCQ with long
block length $n=10^5$ in Table~\ref{tab:tcq-perf}.  The codes have the
same structure as the $\tilde{m}=1$ case in
\cite{chan-coding-multi-level-phase} and have $2^{\nu}$ states.  In
our terminology, they are thus 4-ary codes of rate
$R=\unit[(1+\nu/n)]{b/s}$ including tail bits.  To study the
performance trends of TCQ codes with more states than those found in
the literature, we optimize the generator polynomials ourselves via
random search.  The resulting shaping losses agree with the results in
\cite[Table~IV]{trellis-shaping} and
\cite[Table~I]{near-cap-dpc-tcq-ira} available for $\nu \le 11$,
suggesting that the random search method, though not exhaustive,
already gives near-optimal TCQ codes.

\begin{table}[!t]
  \centering
  \caption{Shaping Loss (\dB) of $2^{\nu}$-State TCQ at $n=10^5$}
  \label{tab:tcq-perf}
  \begin{tabular}{rlrlrlrl}
    \toprule
    $\nu$ & loss (dB) & $\nu$ & loss (dB) & $\nu$ & loss (dB) & $\nu$ & loss (dB) \\
    \midrule
    2 & 0.5371 & 6 & 0.2664 & 10 & 0.1484 & 14 & 0.0951 \\
    3 & 0.4464 & 7 & 0.2321 & 11 & 0.1335 & 15 & 0.0853 \\
    4 & 0.3781 & 8 & 0.1921 & 12 & 0.1155 & 16 & 0.0784 \\
    5 & 0.3183 & 9 & 0.1757 & 13 & 0.1033 & 17 & 0.0705 \\
    \bottomrule
  \end{tabular}    
\end{table}

The results in Table~\ref{tab:tcq-perf} confirm that TCQ can also
achieve near-zero shaping losses, but the loss decreases only
slightly faster than $1/\nu$, therefore the number of states $2^\nu$
(and thus the time and memory complexity) increases exponentially as
the loss approaches zero.  For example, the \unit[0.2676]{dB} loss of
LDGM-based quantization at $L \approx 10^2$ can be achieved by TCQ with $2^6$
to $2^7$ states, but the \unit[0.0741]{dB} loss at $L \approx 10^3$ would
require an astronomical $2^{16}$ to $2^{17}$ states to achieve with
TCQ, so the proposed LDGM-based quantizer is much better than TCQ at
achieving near-zero shaping losses when $n$ is large.\footnote{One
  may note that the LDGM code and the TCQ code have different rates
  $R$.  However, in shaping and DPC applications, the rate of the
  shaping code does not matter much as long as the desired shaping
  loss is achieved, therefore it should be fair to compare TCQ and
  LDGM at their respective ``natural'' rates.}  However, TCQ remains
advantageous for small $n$ as we have shown in
Table~\ref{tab:short-code-perf}.

\section{Complexity Analysis}
\label{sec:discussion}

\subsection{Computational Complexity}
\label{sec:complexity}
We now analyze the time complexity, per block of $n$ source symbols,
of a serial implementation of the proposed quantization algorithm.
Dequantization obviously has much lower complexity and will therefore
not be discussed.

The time complexity of the belief propagation part in the binary case
(Fig.~\ref{alg:binary}) is clearly linear in the number of edges in
the factor graph,%
\footnote{Note that the computation at each \node{b}- or \node{c}-node
  with degree $d$ requires $\asymptO(d)$ time per iteration using the
  forward-backward algorithm (similar to BCJR), not $\asymptO(d^2)$ as
  is required by the naive implementation.}  i.e.\ $\asymptO(nR\db)$
per iteration.  In the $2^K$-ary algorithm in Fig.~\ref{alg:mary}, BP
at \node{b}- and \node{c}-nodes also has this complexity, while at
each $\node{u}_j$ the $K$ $\msg{uc}{jj_k}$'s take $\asymptO(K2^K)$
time to compute with \eqref{eq:msg_uc},\footnote{Again, the
  forward-backward algorithm is responsible for the reduction in
  complexity from $\asymptO(K^2 2^K)$ to $\asymptO(K 2^K)$.} therefore
the total complexity of BP is $\asymptO(n(R\db+K 2^K))$ per iteration,
whose $K=1$ version is also applicable to the binary case.

Within the decimation part, only the greedy decimator's selection of
the most certain bits to decimate may have higher complexity.  In a
straightforward implementation of the GD in Fig.~\ref{alg:binary}, the
most certain $b_i$'s are selected one by one until either $\deltamax$
or $\minprog$ is reached.  This incremental selection problem can be
solved with partial quicksort; if $\nbx{l}$ bits end up being
decimated in iteration $l$, the selection has complexity
$\asymptO(\nb+\nbx{l}\log \nbx{l})$ in that iteration.  Since
$\nb=\sum_l \nbx{l}$, this complexity averaged over $L$ iterations is
at most $\asymptO(\nb(1+\frac{\log \nb}{L}))$ per iteration, which
usually reduces to $\asymptO(\nb)$ since generally $\log\nb \ll L$.
For even larger $\nb$, we note that the quantization algorithm is
unaffected even if the decimated bits in an iteration are selected
non-incrementally and unsorted among themselves by certainty, which
has only $\asymptO(\nb)$ time complexity per iteration using partial
quicksort, and the limits $\deltamax$ and $\minprog$ can still be
enforced by appropriate summing within each partition at the same
complexity.  This method is probably slower in practice, but it shows
that $\asymptO(\nb)$ selection complexity per iteration is possible in
principle even when $\log\nb \gg L$.

We thus conclude that our quantization algorithm has complexity
$\asymptO(n(R\db+K 2^K))$ per block per iteration, or
$\asymptO(L(R\db+K 2^K))$ per symbol summed over all iterations.  In
practice, the most certain bits to decimate can also be selected with
a priority queue or even by a full sort in every iteration; the higher
complexities of these methods do not actually slow down the overall
algorithm much.

\subsection{The Loss-Complexity Tradeoff}
The asymptotic loss-complexity tradeoff of LDGM quantizers can now be
analyzed heuristically.  For simplicity we assume $K$ to be a
constant, and the time complexity of the quantizer per symbol can then
be simplified to $\asymptO(L\cdot R\db)$.  We analyze the \emph{extra
  loss}, denoted by $1/\kappa$, compared to the $2^K$-ary
random-coding loss, and $n$ is assumed to be large enough that the
small-$n$ loss does not dominate this extra loss.

Now the extra loss $1/\kappa$ consists mainly of two parts, namely the
\emph{monotonicity threshold loss} due to the gap between $K\Icthr$
and its ideal value $R$, and the \emph{delta-area loss} due to the
finiteness of the iteration count $L$.  We have observed in
Table~\ref{tab:impact-db-beq} that the monotonicity threshold loss
diminishes exponentially fast with the increase of $\db$ for BEQ, and
this is apparently true for MSE quantization as well; more precisely,
the loss appears to be diminishing exponentially with the average
\node{c}-degree $R\db/K$, therefore in order to reduce this loss to
$\asymptO(1/\kappa)$, $R\db$ must be on the order of $\log \kappa$.
As for the delta-area loss, \eqref{eq:mse-Ai-emp} suggests that it is
proportional to the delta-area $\Aignore$, and since $L\Aignore$ is
almost a code-independent constant in our simulations when $\Info{c}
\le \IcthrorL$, $\Aignore$ is in turn inversely proportional to the
iteration count $L$, therefore $L$ on the order of $\kappa$ is
necessary to make this loss $\asymptO(1/\kappa)$.  The overall
complexity per symbol necessary for $\asymptO(1/\kappa)$ extra loss is
thus $\asymptO(\kappa \log\kappa)$ according to these heuristic
arguments.  Note that this is similar to previous results and
conjectures on the tradeoff between gap-to-capacity and complexity for
LDPC channel codes; see \cite{perf-cplx-per-it-ldpc-info-theo} and
references therein.

In comparison, the complexity needed to achieve $1/\kappa$ loss with
TCQ appears from Table~\ref{tab:tcq-perf} to be exponential in
$\kappa$, and current achievability results in
\cite{on-red-trellis-src-coding} also achieves this
$\asymptO(\e^{\kappa})$ complexity only.  It thus seem unlikely that a
similar $\asymptO(\kappa \log\kappa)$ complexity can be achieved with
TCQ.

\subsection{Strengths of LDGM Quantizers versus TCQ}
From the numerical results and heuristical analysis above, we conclude
that the proposed LDGM quantizers are superior to TCQ in terms of the
loss-complexity tradeoff, when the block length $n$ is large and
near-zero shaping losses are desired.  On the other hand, TCQ does
perform better for $n$ smaller than $10^3$--$10^4$, and a simple
4-state TCQ may also suffice in undemanding applications where its
0.5371-dB shaping loss is acceptable.

Till now we have talked about the complexity at the encoder
(quantization) side only.  In shaping applications, particularly DPC,
the advantage of LDGM quantizers is more evident at the decoder side,
which according to \eqref{eq:dpc-rx} must usually iteratively separate
the superposition of a channel codeword $\seq{u}$ and a quantizer
codeword $\seq{a}$ \cite{near-cap-dpc-tcq-ira}.  When TCQ is used and
when the operating SNR is close to threshold, the BCJR algorithm must
be run in full many times on the trellis, making the decoder-side
complexity much higher than the encoder side.  When LDGM-based
quantizers are used, on the other hand, the inner iterations of the
channel decoder (usually LDPC) and those on the LDGM quantization code
can be interleaved, and in practice the total complexity is usually no
higher than at the encoder side, both comparable to an ordinary LDPC
decoder.

It is also worth noting that increasing the number of states in TCQ
increases both time and memory complexity, whereas a larger $L_0$ in
the LDGM quantizer increases only the encoder-side time complexity,
not the memory complexity.  This is, however, partially offset by the
LDGM quantizer's need of larger block lengths.

\section{Conclusion}
\label{sec:conclusion}
In this paper we have designed LDGM-based codes and corresponding
iterative quantization algorithms for the MSE quantization problem of
$\SetR^n$.  The optimization of the degree distributions is
formulated, via the introduction of the TTD, as the maximization of a
monotonicity threshold that can be determined using density evolution
methods and optimized by linear programming.  The finite number of
iterations $L$ is then accounted for by optimizing the pace of
decimation and using a modified criterion in degree distribution
optimization.

As shown by the simulation results, the proposed quantizers can
achieve much lower shaping losses than TCQ at similar complexity.  The
methods employed in the analysis of the decimation process, in
particular the typical decimator synchronized to the TTD, may also
prove useful elsewhere.

The proposed LDGM-based quantizers are useful in lossy source coding
and shaping, but in practice their good performance is most important
in dirty-paper coding in the low-SNR regime.  According to our
preliminary investigations, a superimposed structure similar to
\cite{sup-coding-side-info-chan,close-to-cap-dpc,near-cap-dpc-tcq-ira}
can be used directly, where the transmitted signal has the form
\eqref{eq:dpc-tx}, consisting of an LDPC codeword (usually modulated
into a 4-PAM or higher signal) containing the desired information,
pre-subtracted known interference, plus a codeword from the LDGM
quantizer to minimize the overall transmission power.  The design of
the LDPC code, such that the LDPC and LDGM parts can be correctly
separated at the receiver, appears to be straightforward although more
work is necessary in the details.  The scheme is then expected to give
better performance than existing TCQ-based schemes at the same level
of computational complexity.  Alternatively, in
\cite{ld-ach-wz-gp-bounds} a ``nested'' structure for the binary
symmetric Gelfand-Pinsker problem has been proposed, in which the
codewords of an LDGM quantization code are divided into cosets
according to linear equations on $\seq{b}$ and the known interference
is quantized into a codeword chosen from one coset that corresponds to
the information to be conveyed.  In
\cite{info-embedding-codes-graphs-it-enc-dec}, a similar construction
is proposed for the binary erasure case.  It is not difficult to
extend this scheme to DPC on Gaussian channels, and code design,
though much more complex, is still possible.  However, as in BEQ, our
BP-based quantizer will generally leave some hard constraints related
the transmitted information unsatisfied, and the necessary overhead to
correct such errors may make such nested codes less attractive than
the superpositional structure above.  More investigation is necessary in
this aspect.

On the quantizer itself, the currently achieved long-block shaping
losses are already quite good, and we have been able to account for
the losses, through theoretical analysis or heuristic arguments, with
the random-coding loss, the nonideality of the monotonicity threshold,
the delta-area loss due to finite iteration count $L$, and the loss
due to finite block length $n$.  In future work, it would be useful to
rigorously investigate the correctness of these heuristics.  Our
analysis is also limited to the typical decimator with recovery; as we
have shown in Section~\ref{sec:sim-typ}, the greedy decimator used in
practice can have significantly different performance when the code is
not well optimized in terms of $\Icthr$ or when $\Info{c}$ is far from
$\Icthr$, therefore an analysis of the GD would be interesting.

Further improvement in MSE performance may come from appropriate use
of the recovery algorithm, a better optimized strategy for controlling
the decimation process (see Section~\ref{sec:dec-control}), and a more
refined degree distribution optimization method based on the results
of quantized DE.  In addition, there is still plenty of room for
improvement in small-$n$ performance.  We have found that better edge
assignment algorithms, such as progressive edge growth (PEG)
\cite{prog-edge-growth-tanner}, could noticeably improve LDGM
quantizers' shaping losses for small $n$, though the improvement is
not large, partly due to the change in EXIT curves caused by such
algorithms.  Larger gains may result from applying the PEG method
more carefully, or from the use of non-binary or generalized LDGM
codes, which may be viewed as a combination of TCQ and LDGM
techniques.

\IEEEtriggeratref{24}
\bibliographystyle{IEEEtran.bst}
\bibliography{IEEEabrv,ldgm-vq}

\begin{thebibliography}{10}
\providecommand{\url}[1]{#1}
\csname url@samestyle\endcsname
\providecommand{\newblock}{\relax}
\providecommand{\bibinfo}[2]{#2}
\providecommand{\BIBentrySTDinterwordspacing}{\spaceskip=0pt\relax}
\providecommand{\BIBentryALTinterwordstretchfactor}{4}
\providecommand{\BIBentryALTinterwordspacing}{\spaceskip=\fontdimen2\font plus
\BIBentryALTinterwordstretchfactor\fontdimen3\font minus
  \fontdimen4\font\relax}
\providecommand{\BIBforeignlanguage}[2]{{%
\expandafter\ifx\csname l@#1\endcsname\relax
\typeout{** WARNING: IEEEtran.bst: No hyphenation pattern has been}%
\typeout{** loaded for the language `#1'. Using the pattern for}%
\typeout{** the default language instead.}%
\else
\language=\csname l@#1\endcsname
\fi
#2}}
\providecommand{\BIBdecl}{\relax}
\BIBdecl

\bibitem{ldgm-vq-globecom07}
Q.~C. Wang and C.~He, ``Approaching 1.53-{dB} shaping gain with {LDGM}
  quantization codes,'' in \emph{Proc. {GLOBECOM} 2007}, Washington, DC, Nov.
  2007.

\bibitem{lattices-good-everything}
U.~Erez, S.~Litsyn, and R.~Zamir, ``Lattices which are good for (almost)
  everything,'' \emph{{IEEE} Trans. Inf. Theory}, vol.~51, no.~10, pp.
  3401--3416, Oct. 2005.

\bibitem{lattice-trellis-quant-highrate}
M.~V. Eyuboglu, G.~D. Forney~Jr, M.~Codex, and M.~A. Mansfield, ``{Lattice and
  trellis quantization with lattice-and trellis-bounded codebooks---High-rate
  theory for memoryless sources},'' \emph{{IEEE} Trans. Inf. Theory}, vol.~39,
  no.~1, pp. 46--59, Jan. 1993.

\bibitem{trellis-shaping}
G.~D. Forney~Jr, M.~Codex, and M.~A. Mansfield, ``Trellis shaping,''
  \emph{{IEEE} Trans. Inf. Theory}, vol.~38, no. 2 Part 2, pp. 281--300, Mar.
  1992.

\bibitem{writing-on-dirty-paper}
M.~H.~M. Costa, ``Writing on dirty paper,'' \emph{{IEEE} Trans. Inf. Theory},
  vol.~29, no.~3, pp. 439--441, May 1983.

\bibitem{sup-coding-side-info-chan}
A.~Bennatan, D.~Burshtein, G.~Caire, and S.~Shamai, ``Superposition coding for
  side-information channels,'' \emph{{IEEE} Trans. Inf. Theory}, vol.~52,
  no.~5, pp. 1872--1889, May 2006.

\bibitem{close-to-cap-dpc}
U.~Erez and S.~Brink, ``A close-to-capacity dirty paper coding scheme,''
  \emph{{IEEE} Trans. Inf. Theory}, vol.~51, no.~10, pp. 3417--3432, Oct. 2005.

\bibitem{trellis-conv-prec-tx-intf-presub}
W.~Yu, D.~P. Varodayan, and J.~M. Cioffi, ``Trellis and convolutional precoding
  for transmitter-based interference presubtraction,'' \emph{{IEEE} Trans.
  Commun.}, vol.~53, no.~7, pp. 1220--1230, Jul. 2005.

\bibitem{near-cap-dpc-tcq-ira}
Y.~Sun, A.~D. Liveris, V.~Stankovic, and Z.~Xiong, ``Near-capacity dirty-paper
  code designs based on {TCQ} and {IRA} codes,'' in \emph{Proc. {ISIT} 2005},
  Aug. 2005, pp. 184--188.

\bibitem{img-data-hiding-cap-appr-dpc}
Y.~Yang, Y.~Sun, V.~Stankovic, and Z.~Xiong, ``Image data-hiding based on
  capacity-approaching dirty-paper coding,'' in \emph{Proceedings of SPIE},
  vol. 6072, 2006, pp. 429--439.

\bibitem{vector-perturbation-mod-precoding1}
C.~B. Peel, B.~M. Hochwald, and A.~L. Swindlehurst, ``A vector-perturbation
  technique for near-capacity multiantenna multiuser communication--{Part I}:
  channel inversion and regularization,'' \emph{{IEEE} Trans. Commun.},
  vol.~53, no.~1, pp. 195--202, Jan. 2005.

\bibitem{tcq-memoryless-gauss-markov}
M.~W. Marcellin and T.~R. Fischer, ``Trellis coded quantization of memoryless
  and {Gauss-Markov} sources,'' \emph{{IEEE} Trans. Commun.}, vol.~38, no.~1,
  pp. 82--93, Jan. 1990.

\bibitem{turbo-trellis-vq}
V.~Chappelier, C.~Guillemot, and S.~Marinkovic, ``{Turbo trellis-coded
  quantization},'' in \emph{Proc. 5th Intl. Symp. Turbo Codes}, Brest, France,
  Sep. 2003.

\bibitem{it-quant-codes-graphs}
E.~Martinian and J.~S. Yedidia, ``Iterative quantization using codes on
  graphs,'' in \emph{Proc. 41st Annual Allerton Conf.}, Aug. 2004,
  arXiv:cs.IT/0408008.

\bibitem{analysis-ldgm-loss-compression}
E.~Martinian and M.~J. Wainwright, ``Analysis of {LDGM} and compound codes for
  lossy compression and binning,'' in \emph{Workshop on Information Theory and
  its Applications}, Feb. 2006, {arXiv:cs.IT/0602046}.

\bibitem{ld-ach-wz-gp-bounds}
------, ``Low-density constructions can achieve the {Wyner-Ziv} and
  {Gelfand-Pinsker} bounds,'' in \emph{Proc. {ISIT} 2006}, Seattle, WA, Jul.
  2006, pp. 484--488, {arXiv:cs.IT/0605091}.

\bibitem{lossy-src-enc-msgpass-dec-ldgm}
M.~J. Wainwright and E.~Maneva, ``Lossy source encoding via message-passing and
  decimation over generalized codewords of {LDGM} codes,'' in \emph{Proc.
  {ISIT} 2005}, Aug. 2005, pp. 1493--1497, {arXiv:cs.IT/0508068}.

\bibitem{binary-quant-bp-dec-ldgm}
T.~Filler and J.~Fridrich, ``Binary quantization using {Belief Propagation}
  with decimation over factor graphs of {LDGM} codes,'' in \emph{Proc. 45th
  Annual Allerton Conf.}, Oct. 2007, {arXiv:0710.0192v1 [cs.IT]}.

\bibitem{maxwell-constr}
C.~Measson, A.~Montanari, and R.~Urbanke, ``{Maxwell} construction: The hidden
  bridge between iterative and maximum a posteriori decoding,'' Jun. 2005,
  {arXiv:cs.IT/0506083}.

\bibitem{factor-graphs-sum-product}
F.~R. Kschischang, B.~J. Frey, and H.~A. Loeliger, ``Factor graphs and the
  sum-product algorithm,'' \emph{{IEEE} Trans. Inf. Theory}, vol.~47, no.~2,
  pp. 498--519, Feb. 2001.

\bibitem{design-cap-approaching-ldpc}
T.~J. Richardson, M.~A. Shokrollahi, and R.~L. Urbanke, ``Design of
  capacity-approaching irregular low-density parity-check codes,'' \emph{{IEEE}
  Trans. Inf. Theory}, vol.~47, no.~2, pp. 619--637, Feb. 2001.

\bibitem{exit-model-ec-prop}
A.~Ashikhmin, G.~Kramer, and S.~Brink, ``Extrinsic information transfer
  functions: model and erasure channel properties,'' \emph{{IEEE} Trans. Inf.
  Theory}, vol.~50, no.~11, pp. 2657--2673, Nov. 2004.

\bibitem{cap-ldpc-msgpassing-dec}
T.~J. Richardson and R.~L. Urbanke, ``The capacity of low-density parity-check
  codes under message-passing decoding,'' \emph{{IEEE} Trans. Inf. Theory},
  vol.~47, no.~2, pp. 599--618, Feb. 2001.

\bibitem{on-design-ldpc-45e-4-dB}
S.~Y. Chung, G.~D. Forney~Jr, T.~J. Richardson, and R.~Urbanke, ``On the design
  of low-density parity-check codes within 0.0045 {dB} of the {Shannon}
  limit,'' \emph{IEEE Commun. Lett.}, vol.~5, no.~2, pp. 58--60, Feb. 2001.

\bibitem{gen-area-theorem-conseq}
C.~Measson, A.~Montanari, T.~Richardson, and R.~Urbanke, ``The generalized area
  theorem and some of its consequences,'' Nov. 2005, {arXiv:cs.IT/0511039}.

\bibitem{bounds-info-combining}
I.~Land, S.~Huettinger, P.~A. Hoeher, and J.~B. Huber, ``Bounds on information
  combining,'' \emph{{IEEE} Trans. Inf. Theory}, vol.~51, no.~2, pp. 612--619,
  Feb. 2005.

\bibitem{extremes-info-combining}
I.~Sutskover, S.~Shamai, and J.~Ziv, ``Extremes of information combining,''
  \emph{{IEEE} Trans. Inf. Theory}, vol.~51, no.~4, pp. 1313--1325, Apr. 2005.

\bibitem{design-ldpc-mod-det}
S.~ten Brink, G.~Kramer, and A.~Ashikhmin, ``Design of low-density parity-check
  codes for modulation and detection,'' \emph{{IEEE} Trans. Commun.}, vol.~52,
  no.~4, pp. 670--678, Apr. 2004.

\bibitem{on-dec-ldpc-bec}
H.~Pishro-Nik and F.~Fekri, ``On decoding of low-density parity-check codes
  over the binary erasure channel,'' \emph{{IEEE} Trans. Inf. Theory}, vol.~50,
  no.~3, pp. 439--454, Mar. 2004.

\bibitem{bicm}
G.~Caire, G.~Taricco, and E.~Biglieri, ``Bit-interleaved coded modulation,''
  \emph{{IEEE} Trans. Inf. Theory}, vol.~44, no.~3, pp. 927--946, May 1998.

\bibitem{bicm-id}
X.~Li and J.~Ritcey, ``{Bit-interleaved coded modulation with iterative
  decoding},'' in \emph{Proc. {ICC'99}}, vol.~2, 1999.

\bibitem{chan-coding-multi-level-phase}
G.~Ungerboeck, ``Channel coding with multilevel/phase signals,'' \emph{{IEEE}
  Trans. Inf. Theory}, vol.~28, no.~1, pp. 55--67, Jan. 1982.

\bibitem{quant-sig-sp-gen-fg-codes}
J.~S. Yedidia and E.~Martinian, ``Quantizing signals using sparse generator
  factor graph codes,'' U.S. Patent 6\,771\,197, Aug. 3, 2004.

\bibitem{perf-cplx-per-it-ldpc-info-theo}
I.~Sason and G.~Wiechman, ``Performance versus complexity per iteration for
  low-density parity-check codes: An information-theoretic approach,'' in
  \emph{Proc. 4th Intl. Symp. Turbo Codes and Related Topics}, Munich, Germany,
  Apr. 2006, {arXiv:cs.IT/0512075}.

\bibitem{on-red-trellis-src-coding}
G.~Zhou and Z.~Zhang, ``On the redundancy of trellis lossy source coding,''
  \emph{{IEEE} Trans. Inf. Theory}, vol.~48, no.~1, pp. 205--218, Jan. 2002.

\bibitem{info-embedding-codes-graphs-it-enc-dec}
V.~Chandar, E.~Martinian, and G.~W. Wornell, ``Information embedding codes on
  graphs with iterative encoding and decoding,'' in \emph{Proc. {ISIT} 2006},
  Jul. 2006, pp. 866--870.

\bibitem{prog-edge-growth-tanner}
X.~Y. Hu, E.~Eleftheriou, and D.~M. Arnold, ``Regular and irregular progressive
  edge-growth {Tanner} graphs,'' \emph{{IEEE} Trans. Inf. Theory}, vol.~51,
  no.~1, pp. 386--398, Jan. 2005.

\end{thebibliography}



\end{document}